\documentclass[12pt,a4paper]{article}
\usepackage[latin1]{inputenc}
\usepackage{setspace}
\usepackage{amssymb,amsmath,amsfonts,mathrsfs}
\usepackage{a4wide}
\usepackage{cite}

\usepackage{tikz}
\usetikzlibrary{trees}
\usetikzlibrary{snakes}
\usetikzlibrary{arrows}
\usepackage{color}
\usepackage{array}

\usepackage{mdframed}
\usepackage{esint}
\usepackage[hidelinks]{hyperref}

\def \be  	{\begin{equation}}
\def \ee  	{\end{equation}}
\def \ba  	{\begin{eqnarray}}
\def \ea  	{\end{eqnarray}}

\def \Li	{ {\rm Li}}

\def\i		{{\bf  i} }

\def \twist 			{ {\tau} }

\usepackage{mathrsfs}
\DeclareMathAlphabet{\mathpzc}{OT1}{pzc}{m}{it}
\usepackage[mathscr]{euscript}

\def\su2{{SU(2)}}

\def\[{\left[}
\def\]{\right]}

\def\({\left(}
\def\){\right)}
\def\[{\left[}
\def\]{\right]}

\def\<{\langle}
\def\>{\rangle}

\def\i2{\frac{i}{2}}

\def\spi{\relax{\rm \pi\kern-0.5em /}}
\def\sA{\relax{\rm A\kern-0.5em /}}
\def\sp{\relax{\rm p\kern-0.5em /}}
\def\sd{\relax{\rm \d\kern-0.5em /}}
\def\sk{\relax{\rm k\kern-0.5em /}}
\def\sn{\relax{\rm n\kern-0.5em /}}
\def\sl{\relax{\rm l\kern-0.5em /}}
\def\sP{\relax{\rm P\kern-0.7em /}}
\def\sBethe{\relax{\rm \Bethe\kern-0.5em /}}

\def\cO{{\cal O}}

\def\2F1{\,_2{\rm F}_1}

\def\beq 	{\begin{equation}}
\def\eeq	{\end{equation}}
\def\bea 	{\begin{eqnarray*}}
\def\eea	{\end{eqnarray*}}

\def\newsigma { {\tilde U} }
\def\newtau { {\tilde V} }

\begin{document}

\thispagestyle{empty}


~\\[-1.75cm]

\begin{center}
{\Large\bf
			Large $p$ explorations. \\[.3cm]      From SUGRA to big STRINGS in Mellin space
}\\
\vskip 1truecm
{\bf 				
			Francesco~Aprile${}^{1}$ and  Pedro Vieira${}^{2,3}$
}

	\vskip 0.4truecm
	
	{\it
		${}^{1}$ Dipartimento di Fisica, Universit\`a di Milano-Bicocca \& INFN, 
		Sezione di Milano-Bicocca, I-20126 Milano,\\
		\vskip 0.7truecm }

	{\it
		${}^{2}$ Perimeter Institute for Theoretical Physics, 31 Caroline St N Waterloo, \\
		Ontario N2L 2Y5, Canada, 
		}
		
	{\it 
		${}^{3}$ Instituto de Fisica Teorica, UNESP, \\
		ICTP South American Institute for Fundamental Research, \\ 
		Rua Dr Bento Teobaldo Ferraz 271, 01140-070, S\~ao Paulo, Brazil \\
		 }

\end{center}

\vskip 1truecm 
\centerline{\bf Abstract} 
We explore a new way of probing scattering of closed strings in $AdS_5\times S^5$, 
which we call `the large $p$ limit'. It consists of studying four-point correlators of single-particle  
operators in $\mathcal{N}=4$ SYM  at large $N$ and large 't Hooft coupling $\lambda$, 
by looking at the regime in which the dual KK modes become short massive strings. 
In this regime the charge of the single-particle  operators is order $\lambda^{1/4}$ 
and the dual KK modes are in between fields and strings. Starting from SUGRA we compute the large $p$ 
limit of the correlators by introducing an improved $AdS_5\times S^5$ Mellin space amplitude,   
and we show that the correlator is dominated by a saddle point.  Our results are consistent 
with the picture of four geodesics shooting from the boundary of $AdS_5\times S^5$ towards 
a common bulk point, where they scatter as if they were in flat space. The Mandelstam invariants 
are put in correspondence with the Mellin variables and in turn with certain combinations of cross ratios.
At the saddle point the dynamics of the correlator is directly related to the bulk Mellin amplitude,  
which in the process of taking large $p$ becomes the flat space \emph{ten-dimensional} S-matrix.
We thus learn how to embed the full type IIB S-matrix in the $AdS_5\times S^5$ Mellin amplitude, 
and how to stratify the latter in a large $p$ expansion. 
We compute the large $p$ limit of all genus zero data currently available, pointing out 
additional hidden simplicity of known results. We then show that the genus zero resummation at large $p$ naturally leads 
to the Gross-Mende phase for the minimal area surface around the bulk point. 
At one-loop, we first uncover a novel and finite 
Mellin amplitude, and then we show that the large $p$ limit beautifully asymptotes the 
gravitational S-matrix. 

 \noindent

\newpage
\setcounter{page}{1}\setcounter{footnote}{0}
\tableofcontents
\newpage

\section{Introduction}

Half-BPS single trace gauge invariant operators 
\beq
T_p \equiv \text{Tr} \left(\, \vec{Y}\! \cdot \phi(x)\right)^p \label{operators}
\eeq
are probably the simplest conceivable operators in $\mathcal{N}=4$ SYM. A dream is to compute their 
correlation functions at any value of the 't Hooft coupling $\lambda$ and for any number of colours $N$. 
This would amount to computing a full closed string scattering amplitude in $AdS_5\times S^5$ for any 
string tension ($\sqrt{\lambda}$) and any string coupling ($1/N$). For four operators and at large $N$ -- corresponding to the planar theory -- 
this would be the AdS analogue of the flat space Virasoro-Shapiro amplitude. 
Despite the fact that finding this amplitude, and more generally gaining full control over the genus expansion of the theory, 
is still a formidable open problem, the recent years  have witness an increasing number of important
 insights into this program. New ideas have come from the bootstrap approach \cite{Susskind:1998vk,Polchinski:1999ry,Giddings:1999jq,Heemskerk:2009pn,
 Penedones:2010ue,Goncalves:2014ffa,Rastelli:2016nze,Alday:2017xua,Aprile:2017bgs,Aprile:2017xsp,Aprile:2017qoy,Aprile:2018efk,Aprile:2019rep}
 integrability \cite{Basso:2017khq,Coronado:2018ypq,Bargheer:2019exp,Belitsky:2019fan,
 Belitsky:2020qrm,Belitsky:2020qir} and localization   \cite{Binder:2019jwn,Chester:2019pvm,Chester:2019jas,Chester:2020dja}.

Here we propose to consider the corner of parameters space in which $p$ is large 
as a way to explore closed string scattering in $AdS_5\times S^5$.
Indeed, from the large $p$ limit we will learn an important lesson about 
how four-point correlation functions behave in general.

We consider the limit where $N$ is the largest parameter, 
and $\lambda$ is also large. 
Depending on how large $p$ is we have various different physical regimes:
\beq\label{regimes_intro}
\begin{array} {ll}
p = O(1) & \text{SUGRA} \\
p = O(\lambda^{1/4}) & \text{short massive strings} \\
p = O(\lambda^{1/2}) & \text{big classical strings} 
\end{array}
\eeq
We will start with the correlators in SUGRA where $p=O(1)$ and take $p \gg 1 $ in those results. 
We will see that they simplify dramatically but still retain a lot of non trivial information. 
In particular, in the domain $p=O(\lambda^{1/4})$ we make contact with the low energy limit of the short 
massive string \cite{Minahan}, and since we are dealing  with small point-like or near point-like 
objects, compared to the ambient space,  we shall see that in this regime the strings scatter as in flat space. 

When $p$ increases further, outside the short string domain, the strings start to open up 
and we reach the high energy scattering region dominated by the flat space result of Gross-Mende \cite{Gross:1987kza}. 
As $p = O(\sqrt{\lambda})$ the angular momenta of the strings is so huge that their centripetal forces 
open up the strings in the full $AdS_5\times S^5$.  At this point integrability should kick in and allow us to compute the 
relevant minimal areas generalising to  $AdS_5\times S^5$ the computation of Gross-Mende.
We make a few comments about this short/long string transition in the conclusions.


\begin{figure}[t]
\begin{minipage}{\textwidth}
	 \begin{minipage}{\textwidth}
    	\centering
	%
	%
	%
	%
	%
	%
    	\begin{minipage}{.3\textwidth}
      				\begin{tikzpicture}

				\def\xuno {0}
				\def\yuno {0}
				\def\radgrande {2cm}

				\def\xptuno {0.4*\radgrande}
				\def\yptuno {0.1*\radgrande}
				\def\xptdue {-0.4*\radgrande}
				\def\yptdue {0cm}
	
				\def\rad{.075cm}

				\def\psxuno {\radgrande*0.5}  		
				\def\psyuno {\radgrande*0.866}     	%
				
				\def\psxdue {-\radgrande*0.9}  		
				\def\psydue {\radgrande*0.43}    		 %
				
				\def\psxtre {-\radgrande*0.866}  	
				\def\psytre {-\radgrande*0.5}     	%

				\def\psxqua {\radgrande*0.58}  		
				\def\psyqua {-\radgrande*0.8}     	%
				\definecolor{mycolor}{RGB}{0,140,235}


				\draw[]  (-2.15cm,0)   node[below, thick] 				{$$};

				\draw[]  (\xuno-.1,1.25*\radgrande)   node[thick,red,font=\footnotesize] 				{\it p=O(1) SUGRA};


				\draw[fill=gray!15,draw=black] (\xuno,\yuno) circle (\radgrande);

				\draw[draw=mycolor,snake=bumps] (\psxuno,\psyuno) -- (\xptuno,\yptuno);
				\draw[draw=mycolor,snake=bumps] (\psxdue,\psydue) -- (\xptdue,\yptdue);
				\draw[draw=mycolor,snake=bumps] (\psxtre,\psytre) -- (\xptdue,\yptdue);	
				\draw[draw=mycolor,snake=bumps] (\psxqua,\psyqua) -- (\xptuno,\yptuno);
		
				\draw[snake=coil,segment amplitude=5pt,red,thick] (\xptuno,\yptuno)-- (\xptdue,\yptdue);

				\draw[fill=mycolor,draw=mycolor] (\psxuno,\psyuno) circle (\rad);
				\draw[fill=mycolor,draw=mycolor] (\psxdue,\psydue) circle (\rad);
				\draw[fill=mycolor,draw=mycolor] (\psxtre,\psytre) circle (\rad);		
				\draw[fill=mycolor,draw=mycolor] (\psxqua,\psyqua) circle (\rad);
		
				\draw[fill=red!90!white,draw=red!90!white] (\xptuno,\yptuno) circle (.075cm);
				\draw[fill=red!90!white,draw=red!90!white] (\xptdue,\yptdue) circle (.075cm);		
				
			\end{tikzpicture}
    \end{minipage}
	%
	%
	%
	%
	%
	%
    	\begin{minipage}{.3\textwidth}
      				\begin{tikzpicture}

	
				\def\xuno {0}
				\def\yuno {0}
				\def\radgrande {2cm}

				\def\xptuno {0.0*\radgrande}
				\def\yptuno {0.0*\radgrande}

				\def\rad{.075cm}

				\def\psxuno {\radgrande*0.5}  		
				\def\psyuno {\radgrande*0.866}     	%
				
				\def\psxdue {-\radgrande*0.9}  		
				\def\psydue {\radgrande*0.43}    		 %
				
				\def\psxtre {-\radgrande*0.866}  	
				\def\psytre {-\radgrande*0.5}     	%

				\def\psxqua {\radgrande*0.58}  		
				\def\psyqua {-\radgrande*0.8}     	%
				\definecolor{mycolor}{RGB}{0,140,235}

				\draw[]  (-2.25cm,0)   node[below, thick] 				{$$};
				
				\draw[]  (\xuno,1.26*\radgrande)   node[thick,red,font=\footnotesize] 				{\it\ 1\ $\ll$\,p\,$\ll\lambda^{1/4}$};


				\draw[fill=gray!15,draw=black] (\xuno,\yuno) circle (\radgrande);

				\draw[draw=mycolor,thick] (\psxuno,\psyuno) -- (\xptuno,\yptuno);
				\draw[draw=mycolor,thick] (\psxdue,\psydue) -- (\xptuno,\yptuno);
				\draw[draw=mycolor,thick] (\psxtre,\psytre) -- (\xptuno,\yptuno);
				\draw[draw=mycolor,thick] (\psxqua,\psyqua) -- (\xptuno,\yptuno);

				\draw[fill=mycolor,draw=mycolor] (\psxuno,\psyuno) circle (\rad);
				\draw[fill=mycolor,draw=mycolor] (\psxdue,\psydue) circle (\rad);
				\draw[fill=mycolor,draw=mycolor] (\psxtre,\psytre) circle (\rad);		
				\draw[fill=mycolor,draw=mycolor] (\psxqua,\psyqua) circle (\rad);
		
				\draw[fill=red!90!white,draw=red!90!white] (\xptuno,\yptuno) circle (.075cm);
				\draw[] (\xptuno-.1cm,\yptuno-.1cm) node[below,color=red!90!white]    {$P$}; 
				
			\end{tikzpicture}    
\end{minipage}
	%
	%
	%
	%
	%
	%
    	\begin{minipage}{.3\textwidth}
      				\begin{tikzpicture}

	
				\def\xuno {0}
				\def\yuno {0}
				\def\radgrande {2cm}

				\def\xptuno {0.0*\radgrande}
				\def\yptuno {0.0*\radgrande}

				\def\rad{.075cm}

				\def\psxuno {\radgrande*0.5}  		
				\def\psyuno {\radgrande*0.866}     	%
				
				\def\psxdue {-\radgrande*0.9}  		
				\def\psydue {\radgrande*0.43}    		 %
				
				\def\psxtre {-\radgrande*0.866}  	
				\def\psytre {-\radgrande*0.5}     	%

				\def\psxqua {\radgrande*0.58}  		
				\def\psyqua {-\radgrande*0.8}     	%
				\definecolor{mycolor}{RGB}{0,140,235}
				\definecolor{othercolor}{RGB}{10,190,130}

				\draw[]  (-2.35cm,0)   node[below, thick] 				{$$};
				
				\draw[]  (\xuno,1.26*\radgrande)   node[thick,red,font=\footnotesize] 				{\it p\,$/\lambda^{1/4}\equiv\ $x\ $\ll$ 1};


				\draw[fill=gray!15,draw=black] (\xuno,\yuno) circle (\radgrande);

				\draw[draw=mycolor,thick] (\psxuno,\psyuno) -- (\xptuno,\yptuno);
				\draw[draw=mycolor,thick] (\psxdue,\psydue) -- (\xptuno,\yptuno);
				\draw[draw=mycolor,thick] (\psxtre,\psytre) -- (\xptuno,\yptuno);
				\draw[draw=mycolor,thick] (\psxqua,\psyqua) -- (\xptuno,\yptuno);

				\draw[fill=mycolor,draw=mycolor] (\psxuno,\psyuno) circle (\rad);
				\draw[fill=mycolor,draw=mycolor] (\psxdue,\psydue) circle (\rad);
				\draw[fill=mycolor,draw=mycolor] (\psxtre,\psytre) circle (\rad);		
				\draw[fill=mycolor,draw=mycolor] (\psxqua,\psyqua) circle (\rad);

				\def\radgrande{.6cm}
				\def\psxuno {1.3*\radgrande*0.5}  		
				\def\psyuno {1.3*\radgrande*0.866}     	%
				
				\def\psxdue {-1.2*\radgrande*0.9}  		
				\def\psydue {1.2*\radgrande*0.43}    		 %
				
				\def\psxtre {-1.3*\radgrande*0.866}  	
				\def\psytre {-1.3*\radgrande*0.5}     	%

				\def\psxqua {1.1*\radgrande*0.58}  		
				\def\psyqua {-1.1*\radgrande*0.8}     	%

				\draw[snake=bumps,segment amplitude=5pt,very thick,othercolor] (\psxuno,\psyuno) -- (\psxdue,\psydue);
				\draw[snake=bumps,segment amplitude=5pt,very thick,othercolor] (\psxdue,\psydue) -- (\psxtre,\psytre);
				\draw[snake=bumps,segment amplitude=5pt,very thick,othercolor] (\psxtre,\psytre) -- (\psxqua,\psyqua);
				\draw[snake=bumps,segment amplitude=5pt,very thick,othercolor] (\psxqua,\psyqua) -- (\psxuno,\psyuno);

								\draw[thick,othercolor] (\xptuno,\yptuno) -- (\psxuno,\psyuno);
								\draw[thick,othercolor] (\xptuno,\yptuno) -- (\psxdue,\psydue);
								\draw[thick,othercolor] (\xptuno,\yptuno) -- (\psxtre,\psytre);
								\draw[thick,othercolor] (\xptuno,\yptuno) -- (\psxqua,\psyqua);
								
								\draw[fill=red!90!white,draw=red!90!white] (\xptuno,\yptuno) circle (.075cm);
								\draw[] (\psxuno,\psyuno-.5cm)   node[right] {\it $\mathcal{S}_{flat}$}; 
								\draw[] (\psxuno+.35cm,\psyuno-.25cm)   node[right,font=\scriptsize] {\it lowE};

			\end{tikzpicture}    
\end{minipage}%
\end{minipage}
%
 %
 %
 %
 %
 %
 %
 %
 %
\\[.2cm]\!
\begin{minipage}{\textwidth}
    	\centering
	%
	%
	%
	%
	%
	%
    	\begin{minipage}{.3\textwidth}
      				\begin{tikzpicture}

				\def\xuno {0}
				\def\yuno {0}
				\def\radgrande {2cm}

				\def\xptuno {0.0*\radgrande}
				\def\yptuno {0.0*\radgrande}

				\def\rad{.075cm}

				\def\psxuno {\radgrande*0.5}  		
				\def\psyuno {\radgrande*0.866}     	%
				
				\def\psxdue {-\radgrande*0.9}  		
				\def\psydue {\radgrande*0.43}    		 %
				
				\def\psxtre {-\radgrande*0.866}  	
				\def\psytre {-\radgrande*0.5}     	%

				\def\psxqua {\radgrande*0.58}  		
				\def\psyqua {-\radgrande*0.8}     	%
				\definecolor{mycolor}{RGB}{0,140,235}

				\draw[]  (-2.25cm,0)   node[below, thick] 				{$$};
				
				\draw[]  (\xuno,1.26*\radgrande)   node[thick,red,font=\footnotesize] 				{\it p\,$/\lambda^{1/4}\equiv\ $x\ =\ O(1)};


				\draw[fill=gray!15,draw=black] (\xuno,\yuno) circle (\radgrande);

				\draw[draw=mycolor,thick] (\psxuno,\psyuno) -- (\xptuno,\yptuno);
				\draw[draw=mycolor,thick] (\psxdue,\psydue) -- (\xptuno,\yptuno);
				\draw[draw=mycolor,thick] (\psxtre,\psytre) -- (\xptuno,\yptuno);
				\draw[draw=mycolor,thick] (\psxqua,\psyqua) -- (\xptuno,\yptuno);

				\draw[fill=mycolor,draw=mycolor] (\psxuno,\psyuno) circle (\rad);
				\draw[fill=mycolor,draw=mycolor] (\psxdue,\psydue) circle (\rad);
				\draw[fill=mycolor,draw=mycolor] (\psxtre,\psytre) circle (\rad);		
				\draw[fill=mycolor,draw=mycolor] (\psxqua,\psyqua) circle (\rad);

				\definecolor{othercolor}{RGB}{10,190,130}
				\draw[shading=radial, inner color=othercolor] (\xptuno,\yptuno) circle (.4cm);
				\draw[] (\xuno+.4cm,\yuno) node[right] {$\mathcal{S}_{flat}$};
				
			\end{tikzpicture}    
\end{minipage}
	%
	%
	%
	%
	%
	%
	%
     	\begin{minipage}{.3\textwidth}
      				\begin{tikzpicture}

	
				\def\xuno {0}
				\def\yuno {0}
				\def\radgrande {2cm}

				\def\xptuno {0.4*\radgrande}
				\def\yptuno {0.1*\radgrande}
				\def\xptdue {-0.4*\radgrande}
				\def\yptdue {0cm}
	
				\def\rad{.075cm}

				\def\psxuno {\radgrande*0.5}  		
				\def\psyuno {\radgrande*0.866}     	%
				
				\def\psxdue {-\radgrande*0.9}  		
				\def\psydue {\radgrande*0.43}    		 %
				
				\def\psxtre {-\radgrande*0.866}  	
				\def\psytre {-\radgrande*0.5}     	%

				\def\psxqua {\radgrande*0.58}  		
				\def\psyqua {-\radgrande*0.8}     	%
				\definecolor{mycolor}{RGB}{0,140,235}
				\definecolor{othercolor}{RGB}{10,190,130}


				\draw[]  (-2.15cm,0)   node[below, thick] 				{$$};

				\draw[]  (\xuno,1.26*\radgrande)   node[thick,red,font=\footnotesize] 				{\it p\,$/\lambda^{1/4}\equiv\ $x\ $\gg$\ 1};

				
				\draw[fill=gray!15,draw=black] (\xuno,\yuno) circle (\radgrande);
								\filldraw[othercolor ,draw=gray!15] (\psxuno-.2,\psyuno+.1) -- (\psxdue+.2,\psydue+.1) -- (\psxtre+.2,\psytre) -- (\psxqua-.2,\psyqua-.1) -- cycle;

				\draw[draw=mycolor,fill=gray!15] (\psxuno,\psyuno+.1cm) .. controls (\xuno+.05cm,\yuno-.2cm) and (\xuno-.05cm,\yuno-.1cm).. (\psxdue,\psydue+.05cm);
				\draw[draw=mycolor,fill=gray!15] (\psxdue,\psydue-.05cm) .. controls (\xuno+.3cm,\yuno-.2cm) and (\xuno+.3cm,\yuno-.2cm).. (\psxtre,\psytre+.05cm);
				\draw[draw=mycolor,fill=gray!15] (\psxtre,\psytre-.05cm) .. controls (\xuno+.3cm,\yuno-.1cm) and (\xuno+.3cm,\yuno-.2cm).. (\psxqua,\psyqua-.1cm);
				\draw[draw=mycolor,fill=gray!15] (\psxqua,\psyqua+.1cm) .. controls (\xuno+.05cm,\yuno-.05cm) and (\xuno+.05cm,\yuno+.05cm).. (\psxuno,\psyuno-.1cm);

				\draw[fill=mycolor,draw=mycolor] (\psxuno,\psyuno) circle (\rad);
				\draw[fill=mycolor,draw=mycolor] (\psxdue,\psydue) circle (\rad);
				\draw[fill=mycolor,draw=mycolor] (\psxtre,\psytre) circle (\rad);		
				\draw[fill=mycolor,draw=mycolor] (\psxqua,\psyqua) circle (\rad);

				\draw[thick] (\xuno-.6cm,\yuno-.6cm) rectangle  (\xuno+.5cm,\yuno+.5cm);
				
				\draw[] (\xuno+.6cm,\yuno-.62cm) node[below, font=\scriptsize] {\it Gross-Mende};
				\draw[] (\xuno-.2cm,\yuno-.8cm) node[below] {$\mathcal{S}_{flat}$};

			\end{tikzpicture}
    \end{minipage}	%
	%
	%
	%
	%
	%
      	\begin{minipage}{.3\textwidth}
      				\begin{tikzpicture}

	
				\def\xuno {0}
				\def\yuno {0}
				\def\radgrande {2cm}

				\def\xptuno {0.4*\radgrande}
				\def\yptuno {0.1*\radgrande}
				\def\xptdue {-0.4*\radgrande}
				\def\yptdue {0cm}
	
				\def\rad{.075cm}

				\def\psxuno {\radgrande*0.5}  		
				\def\psyuno {\radgrande*0.866}     	%
				
				\def\psxdue {-\radgrande*0.9}  		
				\def\psydue {\radgrande*0.43}    		 %
				
				\def\psxtre {-\radgrande*0.866}  	
				\def\psytre {-\radgrande*0.5}     	%

				\def\psxqua {\radgrande*0.58}  		
				\def\psyqua {-\radgrande*0.8}     	%
				\definecolor{mycolor}{RGB}{0,140,235}
				\definecolor{othercolor}{RGB}{10,190,130}


				\draw[]  (-2.35cm,0)   node[below, thick] 				{$$};

				\draw[]  (\xuno+.1cm,1.26*\radgrande)   node[thick,red,font=\footnotesize] 				{\it p=O$(\lambda^{1/2})$};


				\draw[fill=gray!15,draw=black] (\xuno,\yuno) circle (\radgrande);
								\filldraw[othercolor ,draw=gray!15] (\psxuno-.2,\psyuno-.2) -- (\psxdue+.2,\psydue-.1) -- (\psxtre,\psytre) -- (\psxqua,\psyqua+.2) -- cycle;

				\draw[draw=mycolor,fill=gray!15] (\psxuno,\psyuno) .. controls (\xuno+.6cm,\yuno+.4cm) and (\xuno-.6cm,\yuno+.4cm).. (\psxdue,\psydue);
				\draw[draw=mycolor,fill=gray!15] (\psxdue,\psydue) .. controls (\xuno-.6cm,\yuno+.4cm) and (\xuno-.6cm,\yuno-.4cm).. (\psxtre,\psytre);
				\draw[draw=mycolor,fill=gray!15] (\psxtre,\psytre) .. controls (\xuno-.6cm,\yuno-.4cm) and (\xuno+.6cm,\yuno-.4cm).. (\psxqua,\psyqua);
				\draw[draw=mycolor,fill=gray!15] (\psxqua,\psyqua) .. controls (\xuno+.6cm,\yuno-.4cm) and (\xuno+.6cm,\yuno+.4cm).. (\psxuno,\psyuno);

				\draw[fill=mycolor,draw=mycolor] (\psxuno,\psyuno) circle (\rad);
				\draw[fill=mycolor,draw=mycolor] (\psxdue,\psydue) circle (\rad);
				\draw[fill=mycolor,draw=mycolor] (\psxtre,\psytre) circle (\rad);		
				\draw[fill=mycolor,draw=mycolor] (\psxqua,\psyqua) circle (\rad);

				

			\end{tikzpicture}
    \end{minipage}	
\end{minipage}  
\end{minipage}
\caption{From AdS SUGRA ($p= O(1)$) to AdS string minimal areas $p=O(\sqrt{\lambda})$. 
We go from SUGRA taking first $p$ large, then we dress the amplitude by low energy 
flat space string theory ($p/\lambda^{1/4} \ll 1$), full flat space string theory~($p/\lambda^{1/4} \sim 1$) 
and high energy flat space string theory ($p/\lambda^{1/4} \gg 1$). The latter governed 
by string minimal surfaces as studied by Gross and Mende \cite{Gross:1987kza}.}
\label{regimes}
 \end{figure}
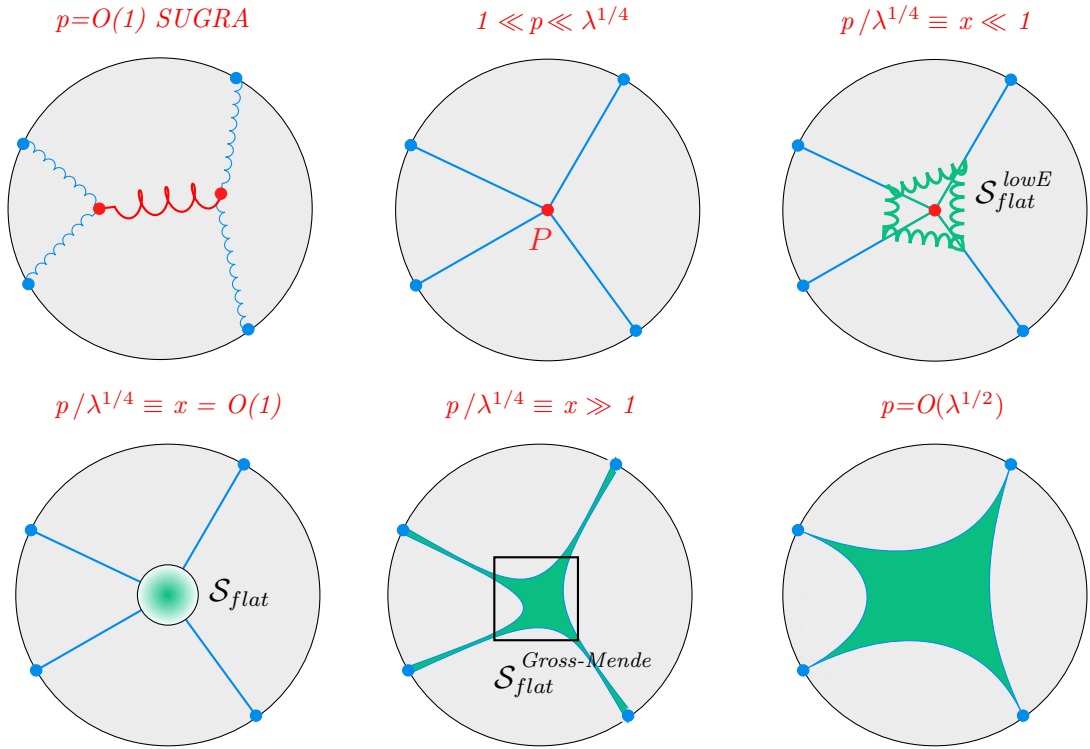

The bulk of the paper is mostly about the transition between the SUGRA/short string transition
and about the insights extracted from this analysis at fixed $p$. 
To analyse that, we focus on the dynamical part of the correlator 
\beq \label{decomposition}
\< \cO_{p_1}(x_1)\dots \cO_{p_4}(x_4)\>_\text{connected} = 
					(\texttt{free theory}) + (\texttt{kinematics}) \times \mathcal{A}_{\vec{p}}(U,V,\tilde U,\tilde V) 
\eeq
which is what we call the \textit{amplitude factor} $\mathcal{A}_{\vec{p}}$. 
The superconformal splitting of the correlator \cite{Eden:2000bk}, and the kinematic elements, 
are standard and detailed in appendix \ref{appendixNotation}. 
Let us highlight here a few important elements about this starting point:

{\bf The operators ${\cal O}_p$} are the single particle operators (SPOs) defined in \cite{Aprile:2018efk}, 
and are the half-BPS operators properly dual to the Kaluza Klein modes on the $AdS_5\times S^5$ background.
SPOs coincide with the single trace operators $T_p$ in (\ref{operators}) the
strict large $N$ limit, but in general are given by an admixture of single and multi-trace operators of the form
\begin{align}
\mathcal{O}_{p}= T_{p} + \sum_{q_1+q_2=p} C_{q_1 q_2}(N) \ T_{q_1} T_{q_2} +  \sum_{q_1+q_2+q_3=p} C_{q_1 q_2 q_3}(N) \ T_{q_1} T_{q_2} T_{q_3}+\ldots 
\end{align}
The various terms correspond to all possible partitions of the charge $p$. At finite $N$, 
the coefficient $C_{ \{q_i\} }$ are obtained by imposing that the two point functions 
of $\mathcal{O}_{p}$ with any other multi-trace operator, they all vanish. 
These coefficients, and other surprising properties of the SPOs, 
are determined for arbitrary charge in \cite{newpaperAprile}.
The difference between SPO and single-trace operators is crucial 
for bootstrapping correctly higher genus corrections \cite{Aprile:2019rep}.

{\bf The amplitude factor $\mathcal{A}_{\vec{p}}$\,} is only a 
function of the conformal cross-ratios $U,V$, and of the R-charge cross ratios $\tilde U$ and $\tilde V$.  
We sometimes refer to these cross ratios as AdS and Sphere cross-ratios respectively. 
The amplitude $\mathcal{A}_{\vec{p}}$ is usually written as a double integral \cite{Penedones:2010ue} and a double sum \cite{Rastelli:2016nze,Rastelli:2017udc}:
The integral is a Mellin transform, i.e.~a Fourier transform with respect to the (logarithm of the) 
space-time cross-ratios $U,V$. The double sum is a discrete Mellin transform w.r.t.~the R-charge cross-ratios $\tilde U,\tilde V$. 

For our study of the large $p$ limit, it will be convenient to transform the discrete sum 
into a double integral a la Sommerfield, and write the amplitude $\mathcal{A}$ as a four-fold 
integral in which all Mellin variables are treated equally. 
The result is what we call the $AdS_5\times S^5$ Mellin representation
\footnote{The contour of integration is a standard Mellin-Barnes contour, 
i.e.~a straight line parallel to the imaginary axis separating left poles from right poles, 
in all complex planes, $s$, $t$, $\tilde s$ and $\tilde t$. }
\beq 
\mathcal{A}_{\vec{p}}(U,V,\newsigma,\newtau)= 
					\iint \! ds dt \iint \! d\tilde s d\tilde t \  \, U^s V^t \newsigma^{\tilde s} \newtau^{\tilde t} \times 
					\,\Gamma_{\otimes} \times \mathcal{M}_{\vec{p}}(s,t,\tilde s,\tilde t\, )  \label{rep4ints}
\eeq 
where now 
\beq\label{amplitude}
\begin{array}{ccc}
\displaystyle
{\Gamma}_{\otimes}=\mathfrak{S}\  
\!\frac{ \Gamma[-s]^2\Gamma[-t]^2 \Gamma[-{u} ]^2  }{ \Gamma[1+\tilde s\,]^2 \Gamma[1+\tilde t\,]^2 \Gamma[ 1+\tilde u\,]^2 }\qquad\ &;&
\displaystyle
\quad \mathfrak{S}=\pi^2 \frac{\ (-)^{\tilde t}(-)^{\tilde u} }{\sin(\pi \tilde t\, )\sin(\pi \tilde u)} 
\\[.7cm]
u\equiv -s-t-(p+2)\qquad\  &;&\quad\tilde u \equiv - \tilde s - \tilde t + (p-2) 
\end{array}
\eeq
This formula for $\Gamma_{\otimes}$ refers to the balanced configuration $p_{i=1,2,3,4}=p$.
The general case of unequal charges is discussed in appendix \ref{appendixNotation}.
The symbol ${\Gamma}_{\otimes}$ stresses that this expression contains 
gamma functions for the product space $AdS_5\times S^5$.\footnote{Notice also that
the sphere part contains the tree level normalisation found in \cite{Aprile:2018efk}. }

When the external charges are large, poles to the left of the original contour are pushed away and
we can move the contour to a region where the integration variables are also large. 
The arguments in $\Gamma_{\otimes}$ then become large and 
we can replace each gamma function by its Stirling asymptotics
\begin{align}
\Gamma(x) \simeq \sqrt{2 \pi } e^{-x} x^{x-\frac{1}{2}} \,.
\end{align}
Collecting the various exponential contributions
we find a simple classical action ${S}_{cl}$, and perform the integrations by saddle point. 
We thus get 
\beq
\lim_{p\rightarrow \infty} \mathcal{A}_{\vec{p}}(U,V,\newsigma,\newtau) =
				\Bigg[ \exp( -p{S}_{cl}) \times |\texttt{det}\big(\texttt{Hessian}(S_{cl})\big)|^{-\frac{1}{2}} \times  
				\mathcal{M}_{\vec{p}}\Bigg]_{s_{cl}, t_{cl} , \tilde s_{cl},\tilde t_{cl}} \label{factors}
\eeq
where the r.h.s is evaluated at the saddle point $ s_{cl}, t_{cl} , \tilde s_{cl}, \tilde t_{cl}$. 
The saddle point is fixed in terms of the space-time and R-charge cross-ratios 
by extremizing the classical action.  On the $AdS_5$  we find 
\begin{align}\label{solu_pppp_AdS5}
-s_{cl}= p \frac{   \sqrt{U}  }{ 1+  \sqrt{U} +  \sqrt{V} }\quad;\quad 
-t_{cl} = p  \frac{  \sqrt{V}  }{ 1+  \sqrt{U} +  \sqrt{V} }
\end{align}
and on the $S^5$
\begin{align}\label{solu_pppp_S5}
\tilde s_{cl}= p \frac{   \sqrt{\newsigma}  }{ 1+  \sqrt{\newsigma} +  \sqrt{\newtau} }\qquad;\qquad 
\tilde t_{cl} = p \frac{  \sqrt{\newtau}  }{ 1+  \sqrt{\newsigma} +  \sqrt{\newtau} }
\end{align}

The large $p$ limit of the four-point correlator is thus given by the simple formula\\

\begin{mdframed} 
\begin{align}
\!\!\!
\lim_{p\rightarrow \infty} \mathcal{A}_{\vec{p}}(U,V,\newsigma,\newtau)=
\frac{ 4\pi^4 }{ (U V \newsigma \newtau)^{\frac{1}{4} } }
\frac{  (1+\sqrt{\newsigma}+\sqrt{\newtau} )^{2(p-2) +\frac{3}{2} } }{ (1+\sqrt{U}+\sqrt{V} )^{2(p+2) - \frac{3}{2} }   } \!
 \times\!
 \Big(p^4 \lim_{p\rightarrow \infty} \mathcal{M}_{\vec{p}}( s_{cl} ,t_{cl}, \tilde s_{cl},\tilde t_{cl})  \Big) 
 \label{saddle_AdS_S}
\end{align}
\end{mdframed}

The combination of the $AdS_5\times S^5$ saddle is neat. 
Some terms of the form $e^{\pm 2p} p^{\mp 2p}$ present separately on the factorised $AdS_5$ and $S^5$ saddles cancel out,
and the functions of the cross ratios combine nicely (see  appendix \ref{appendixsaddleAdS} for more details.)
We are assuming that at each order in perturbation theory, which here means both $1/N^2$ and $\lambda^{-1/2}$, 
the Mellin amplitude does not exponentiate for large values of the variables, and therefore does not shift the saddle point.

Provided $p\ll \lambda^{1/2}$ we expect our computation to reproduce the behaviour of four 
strings propagating as point-like geodesics in $AdS_5\times S^5$ until a small interaction region~\cite{Susskind:1998vk,
Polchinski:1999ry,Giddings:1999jq,Penedones:2010ue,Fitzpatrick:2011jn,Fitzpatrick:2011hu,Goncalves:2014ffa,Minahan}. 
We will provide further evidence for this picture in section~\ref{geodi_sec} where we will map the first factor in (\ref{saddle_AdS_S}) to the contribution to four heavy geodesics propagating in the bulk. At the intersection point the strings 
would now scatter as in flat space.  This picture then predicts that $ \mathcal{M}_{\vec{p}}( s_{cl} ,t_{cl}, \tilde s_{cl},\tilde t_{cl})$
should be further related to a scattering amplitude with a precise identification of Mandelstam invariants.

Indeed, the four variable function $\mathcal{M}_{\vec{p}}( s_{cl} ,t_{cl}, \tilde s_{cl},\tilde t_{cl})$ 
in the large $p$ limit reduces to a special two-variable function: the ten-dimensional flat space S-matrix of IIB string theory. 
This is our next non trivial result: \\

\begin{mdframed}
\begin{align}\label{largep_limit_relation}
\!\!
\lim_{p\rightarrow \infty} \mathcal{M}_{\vec{p}}^{AdS_5\times S^5}\Big( s_{cl} ,t_{cl}, \tilde s_{cl},\tilde t_{cl};\frac{1}{\sqrt{\lambda}};\frac{1}{N^2} \Big)= 
 \frac{1}{ {\bf s}_{cl} {\bf t}_{cl}  {\bf u}_{cl} } \mathcal{S}^{flat}\left(\frac{4\Sigma  }{R^2} \,{\bf s}_{cl}, \frac{4\Sigma }{R^2}\, {\bf t}_{cl} ;\alpha';g_s\right) \\ 
\notag
\end{align}
\end{mdframed}
where $\Sigma=\frac12(p_1+p_2+p_3+p_4)$,  $R^4=4\pi g_s N \alpha'^2$ is the radius,  $\lambda=4\pi g_s N$ is the the 't Hooft coupling, 
and the bold font variables are 
\begin{align}\label{10d_variables}
{\bf s}_{cl}=s_{cl}+\tilde s_{cl}\quad;\quad
{\bf t}_{cl}=t_{cl}+\tilde{t}_{cl} \quad; \quad\ \ 
{\bf s}_{cl}+{\bf t}_{cl}+{\bf u}_{cl}=0 \,.
\end{align}
The connection between the $AdS_5\times S^5$ Mellin amplitude and the ten-dimensional flat space 
amplitude here is complete, and depends on all variables $s$, $t$, $\tilde s$, $\tilde t$ and $p$.  
Notice that being the S-matrix defined in terms of the Mandelstam invariants, here we have found 
a non trivial identification of those with geometric data, namely a pure function of the cross ratios, 
and also a proper identification of the effective coupling for the low energy approximation. 
In particular, since $\Sigma {\bf s}_{cl}\sim p^2$, as far as $\alpha'$ goes, the effective couplings is
\begin{align}
\frac{p^2\alpha'}{R^2}=\frac{p^2}{\sqrt{\lambda} } \,.
\end{align}
The validity of the  low energy expansion  is precisely 
the distinction between SUGRA, short strings and big strings that we made in  \eqref{regimes_intro}.

The compatibility of the $AdS_5\times S^5$ amplitude at fixed $p$, with the 
 the ten-dimensional flat space amplitudes in the large $p$ limit, gives us with a new 
 set of optimal constraints for bootstrapping correlators at fixed $p$.  
Firstly, the uplift of \eqref{10d_variables} to bold font $AdS_5\times S^5$ variables 
is dictated by the hidden conformal symmetry at tree level \cite{Caron-Huot:2018kta}, and reads 
\begin{align}\label{ten_dim_STU_p}
{\bf s}=s+\tilde s\quad;\quad 
{\bf t}=t+\tilde{t} \ \quad; \quad\ \ {\bf s}+{\bf t}+{\bf u}=-4 \,.
\end{align}
In particular, the constraint on the r.h.s.~goes over the flat space ${\bf s}_{cl}+{\bf t}_{cl}+{\bf u}_{cl}=0$ 
in the large $p$ limit, since the bold font variables become very large.
%
Secondly, even though $\mathcal{M}_{\vec{p}}(s,t,\tilde{s},\tilde{t})$ 
depends on all variables separately, it does it in a particular way. 
Schematically, we  expect 
\begin{align}\label{guess_transcendental}
\mathcal{M}^{\ell-loop}_{\vec{p}} \ \sim \sum \ \texttt{Transcendetal}({\bf s},{\bf t}) \times 
						\texttt{Rational}_{\vec{p}}(s,t,\tilde{s},\tilde{t})
\end{align}
where the $\texttt{Transcendetal}$ functions only depend on the bold font variables. 
These are meromorphic functions in the complex plane such that poles are in 
correspondence with the OPE. In addition, we now find that these transcendental functions have 
to asymptote those appearing in the ten-dimensional flat space amplitude \cite{Alday:2017vkk}.
Moreover, at each order in the $\alpha'$ expansion of the loop amplitude, the rational functions in 
$s,t,\tilde{s},\tilde{t}, \vec{p}$, acquire a stratification under the large $p$ scaling,
since this is the natural scaling of the saddle point in \eqref{solu_pppp_AdS5}-\eqref{solu_pppp_S5}.

We will demonstrate our claims by showing that the large $p$ limit of 
the $AdS_5\times S^5$ Virasoro-Shapiro amplitude \emph{equals} 
the flat space Virasoro-Shapiro amplitude, written in the variables ${\bf s,t,u}$, 
and that the subleading in $p$ contributions are polynomials of 
lower degree in $s,t,\tilde{s},\tilde{t},\vec{p}$, again stratified in powers of $p$. 
Then, we will  uncover a {\it novel Mellin representation at one-loop}
which encompasses the position space correlators constructed in~\cite{Aprile:2019rep}. 
Our organisation of the amplitude is thus very natural, and in a sense complements a 
different approach involving infinite series \cite{Alday:2018kkw,Alday:2019nin,Bissi:2020wtv}, 
where the transcendental structure of the amplitude is less manifest.

\section{Intuition from geodesics}\label{geodi_sec}

In this section we explain the bulk AdS/CFT picture underlying our large $p$ limit. In particular, the
relation between the position space correlator and its Mellin amplitude, given in \eqref{saddle_AdS_S}, 
and also for the relation between the Mellin amplitude and the flat space S-matrix, given in \eqref{largep_limit_relation}. 

Let us begin by interpreting the prefactor, 
\beq
\texttt{prefactor}=\frac{ 4\pi^4 }{ (U V \newsigma \newtau)^{ \frac{1}{4} } }
\frac{  (1+\sqrt{\newsigma}+\sqrt{\newtau} )^{ +\frac{3}{2}} }{ (\,1+\sqrt{U}+\sqrt{V} \,)^{ -\frac{3}{2} }   }  \times 
	\underbrace{	\(\frac{  1+\sqrt{\newsigma}+\sqrt{\newtau} }{ 1+\sqrt{U}+\sqrt{V}    }\)^{\!\!2p }  }_{\texttt{leading exponential  factor}} \label{prefactor}
\eeq
which we encountered\footnote{Note that (\ref{prefactor}) is the prefactor in (\ref{saddle_AdS_S}) if 
it were not for the $p\pm 2$ shifts. These shifts follows from the superconformal split of the correlator, through the partial 
non renormalisation theorem, which in \eqref{decomposition} fixes the space time dependence 
of $\texttt{kinematics}$, and therefore fixes the crossing properties of the dynamical amplitude. 
As far as our bosonic estimates is concerned, we can ignore them without loss of generality. }  
in (\ref{saddle_AdS_S}) as arising from the propagation of four very energetic particles all the way 
to a common interaction region in $AdS_5\times S^5$.
The idea of our interpretation is similar to that of a three point correlator of three heavy short string states, obtained by Minahan \cite{Minahan}. 
In our case, the four operators create four particles at the boundary of $AdS_5\times S^5$, which will meet at a point $(P,\tilde P)$ in the bulk, 
whose location we integrate over as in
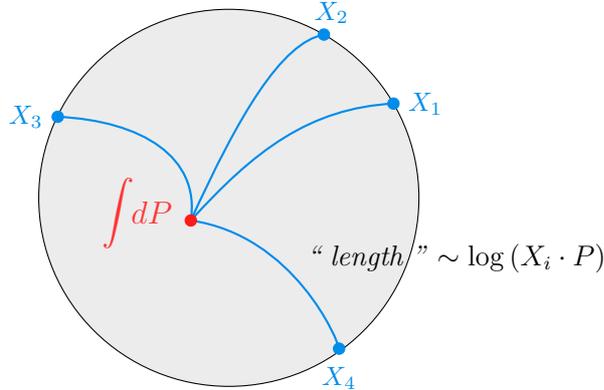
\be\label{geodesic_compu_pic}
\begin{tikzpicture}
\def\xuno {0}
\def\yuno {0}
\def\radgrande {2.5cm}

\def\xpt {-.5cm}
\def\ypt {-.3cm}

\def\rad{.075cm}
\def\psxuno {\radgrande*0.866}  	
\def\psyuno {\radgrande*0.5}     	%

\def\psxdue {\radgrande*0.5}  		
\def\psydue {\radgrande*0.866}     	%

\def\psxtre {-\radgrande*0.9}  		
\def\psytre {\radgrande*0.43}    		 %

\def\psxqua {\radgrande*0.58}  		
\def\psyqua {-\radgrande*0.8}     	%

\def\radpiccolo{1cm}

\def\psxunobis {\radpiccolo*0.866}  	
\def\psyunobis {\radpiccolo*0.5}     	%

\def\psxduebis {\radpiccolo*0.5}  		
\def\psyduebis {\radpiccolo*0.866}     	%

\def\psxtrebis {-\radpiccolo*0.9}  		
\def\psytrebis {\radpiccolo*0.43}    		 %

\def\psxquabis {\radpiccolo*0.58}  		
\def\psyquabis {-\radpiccolo*0.8}     	%

\definecolor{mycolor}{RGB}{0,140,235}

		\draw[]  (-3.5cm,0)   node[below, thick] 				{$$};


		\draw[fill=gray!15,draw=black] (\xuno,\yuno) circle (\radgrande);

		\draw[fill=mycolor,draw=mycolor] (\psxuno,\psyuno) circle (\rad);
		\draw[fill=mycolor,draw=mycolor] (\psxdue,\psydue) circle (\rad);
		\draw[fill=mycolor,draw=mycolor] (\psxtre,\psytre) circle (\rad);		
		\draw[fill=mycolor,draw=mycolor] (\psxqua,\psyqua) circle (\rad);

		\draw[draw=mycolor, thick] (\psxuno,\psyuno)  .. controls (-1cm+0.95*\psxuno,0.95*\psyuno)   and  (-.75cm+0.5*\psxuno,0.5*\psyuno) .. 	 (\xpt,\ypt);
		\draw[draw=mycolor,thick] (\psxdue,\psydue)  .. controls (-.5cm+0.95*\psxdue,0.95*\psydue)   and  (-.5cm+0.5*\psxdue,0.5*\psydue) ..  	(\xpt,\ypt);
		\draw[draw=mycolor,thick] (\psxtre,\psytre)  .. controls (1cm+0.95*\psxtre,0.95*\psytre)   and  (.75cm+0.5*\psxtre,0.5*\psytre) ..  		(\xpt,\ypt);
		\draw[draw=mycolor,thick] (\psxqua,\psyqua)  .. controls (.5cm+0.65*\psxqua,0.95*\psyqua)   and  (.5cm+0.25*\psxqua,0.25*\psyqua) .. 	 (\xpt,\ypt);

		\draw[] (\psxuno+.05cm,\psyuno)	 	node[right,mycolor,font=\footnotesize] 				{$X_1$};
		\draw[] (\psxdue+.1cm,\psydue) 	 	node[above,mycolor,font=\footnotesize] 				{$X_2$};
		\draw[] (\psxtre-.05cm,\psytre)			 node[left,mycolor,font=\footnotesize] 				{$X_3$};
		\draw[] (\psxqua,\psyqua-.1cm)  	 	node[below,mycolor,font=\footnotesize] 				{$X_4$};		
		
		\draw[fill=red!90!white,draw=red!90!white] (\xpt,\ypt) circle (.075cm);
		\draw  (\xpt-.1cm,\ypt+.1cm)  			 	node[left,red!80!white] 						{$dP $}; 
		\draw  (\xpt-.6cm,\ypt+.1cm)  			 	node[left,red!80!white,scale=1] 				{$\displaystyle \int$};

	
		\draw[]  (\xuno+.35*\radgrande,\yuno-.8cm)   node[right,scale=.9] 				{\it `` length " $\sim  \log\left( X_i\cdot P\right)$};

\end{tikzpicture}
\ee

The propagation factor is given by the product of four bulk-to-boundary propagators. 
To specify those, we will use embedding coordinates.  
For $AdS_5$ we consider the hyperboloid, therefore $P\cdot P=-R^2$ is a bulk point, 
and $X_i\cdot X_i=0$ lies on the conformal boundary.  For the sphere $\tilde P\cdot \tilde P=+R^2$ 
belongs to $S^5$, but $Y_i$ is the null six-dimensional polarisation vector on which the $SU(4)$ R-symmetry acts.  

The correlator is thus described by the integral 
\begin{align}\label{geodesic_compu}
\int d^6 P\! \int d^6 \tilde P\! \int \!d\lambda \!\int \!d\tilde \lambda  \, 
\prod_{i=1}^4 \(\frac{Y_i \cdot \tilde P}{X_i \cdot P}\)^{\!\!p_i} 
e^{ -  \lambda (P\cdot P +R^2)- \tilde\lambda (\tilde P\cdot \tilde P -R^2)  }
\end{align}
where  $\lambda$ and $\tilde \lambda$ are Lagrange multipliers
which enforce $P$ and $\tilde P$ to belong to $AdS_5\times S^5$. 
For large external charge $p_{i=1,2,3,4}$, the particles are heavy/very energetic 
and the integration over the interaction point $(P,\tilde P)$,  is localized by saddle point.\footnote{In 
unpublished work from 2017, while studying the flat space limit of general multi-point functions 
of gapped QFT's in AdS, Shota Komatsu has solved very similar kind of problems; see also \cite{ShotaBaltZhao} for beautiful upcoming work on various elements of the flat space limit. 
We thank Shota for several insightful discussions on this point.}
In sum, we should extremize the classical action $S_{cl}=S_{AdS_5}-S_{S^5}$ where 
\beq
S_{AdS_5}= \sum_{i=1}^4 p_i \log(X_i \cdot P)-\lambda \,(P\cdot P +R^2) \label{AdSAction}
\eeq
and a similar expression holds for the sphere part, which can be obtained by the 
substitution $X_i\rightarrow Y_i$, $P\rightarrow \tilde P$, $\lambda\rightarrow \tilde \lambda$, 
and $R\rightarrow i R$. Notice that the first term (\ref{AdSAction}) is indeed the sum of geodesic 
lengths of four geodesics joining the boundary points $X_i$ with the interaction point $P$, 
as the picture above suggested.

Consider first the eqm obtained by varying with respect to $P$, 
\beq
\frac{\partial S_{AdS_5}}{\partial P_A}=0\qquad{\rm iff} \qquad \sum_{i} p_i \frac{X_i^A}{X_i \cdot P} -{2}\lambda \, P^A  =0  \label{SP}
\eeq
Dotting with $P$ leads to $\lambda R^2=-\frac{1}{2}\sum_i p_i$. 
Dotting with $X_j$ yields a set of four equations for the four variables $X_i \cdot P$, 
\beq
\sum_{i\neq j} p_i\frac{X_i\cdot X_j}{X_i \cdot P} \, =- \frac{(p_1+p_2+p_3+p_4) }{R^2} X_j\cdot P \,. 
\eeq
For equal charges, we can solve these easily and we find
\begin{align}
 X_1\cdot P =&\ f \times \left[ \frac{(X_1\cdot X_2)(X_1\cdot X_3)(X_1\cdot X_4)}{(X_2\cdot X_3)(X_2\cdot X_4)(X_3\cdot X_4)} \right]^{\! \frac{1}{4}}  \notag\\
 &\ f= \left[ \frac{R^2}{4}\left( \sqrt{X_3\cdot X_2} \sqrt{X_4\cdot X_1}+\sqrt{X_3\cdot X_1} \sqrt{X_4\cdot
   X_2}+\sqrt{X_2\cdot X_1} \sqrt{X_4\cdot X_3}\right)\right]^{\! \frac{1}{2}} \notag
\end{align}
with similar expressions for the other $X_i\cdot P$. All of them have $f$ as a common factor, 
and the combinatorics w.r.t.~the index $i$ is simple to guess.
Now we plug this solution into the action to estimate its leading exponential behaviour. 
We find a remarkably simple formula
\beq
S_{AdS_5}^\text{classical} = p \log\left(\prod_{i=1}^{4} X_i \cdot P\right)=-p \log f^4 \notag
\eeq
If we also factor out $p \log (X_1\cdot X_3) (X_2\cdot X_4)$ we nicely reproduce the 
expected coordinate prefactor in the \texttt{kinematics} factor sitting in front of the 
amplitude plus a cross-ratio dependent expression:
\beq
e^{S_{AdS_5}^\text{classical}}=\frac{1}{(X_1\cdot X_3)^p (X_2\cdot X_4)^p}  \frac{R^4}{16}
\({1+\sqrt{U}+\sqrt{V}}\)^{\!\!-2p}  \notag
\eeq
Adding the sphere simply completes this expression in the obvious way, and cancel the radius dependence. Finally
\beq
e^{S_{AdS_5}^\text{classical}-S_{Sphere}^\text{classical}}=\frac{(Y_1\cdot Y_3)^p (Y_2\cdot Y_4)^p}{(X_1\cdot X_3)^p (X_2\cdot X_4)^p} 
\(\frac{1+\sqrt{\newsigma}+\sqrt{\newtau}}{1+\sqrt{U}+\sqrt{V}}\)^{\!\!2p}  \notag
\eeq
We have thus reproduced the leading exponential factor in (\ref{prefactor}).

Let us turn now to the non-exponential factor in (\ref{prefactor}). 
In the saddle point approximation, that we just found above,
we expect the non-exponential factor to arise from integrating out quadratic fluctuations. 
For the $AdS_5$ part, the quadratic action around the saddle can be written as
\beq
\frac{\partial^2 }{\partial P_A\partial P_B} S_{AdS_5} =
p \sum_{i=1}^4 \left[-\frac{X_{i}^AX_{i}^B}{(X_i \cdot P)^2}+\frac{1}{R^2} \delta_{AB} \right] \notag
\eeq
To compute the determinant is convenient to go to the basis of the $X_i$. 
In total, the gaussian integration produces
\beq
\frac{ (2\pi)^2}{p^2} \sqrt{  \frac{ \det\limits_{j,k}\left[ X_j\cdot X_k \right]}{  \det\limits_{j,k} \left[ \sum_{i=1}^4 \frac{X_j\cdot X_{i}X_{i}\cdot X_k}{(X_i \cdot P)^2}+ X_j\cdot X_k \right] }   }
=\frac{\pi^2R^4}{16p^2} \frac{ (1+\sqrt{U}+\sqrt{V})^{\frac{3}{2} } }{ (UV)^{\frac{1}{4} } } \notag
\eeq
where $\det\left[ X_j\cdot X_k \right]$ takes into account the non-orthonormality 
of the basis we are using. Combining with the sphere contribution we beautifully match 
the full prefactor in (\ref{prefactor}) up to a simple numerical factor.\footnote{The numerical factor is $R^8/2^6/(2p)^4$.}

Now we would like to explain with a simple argument our second claim in \eqref{largep_limit_relation}, 
namely, how the relation between  the large $p$ limit of the Mellin amplitude and the flat 
space S-matrix comes about in our picture. 
The idea -- nicely described in \cite{Minahan} -- is to measure the slope of each geodesic at the intersection point $(P,\tilde P)$, 
and extract the corresponding momenta as it enters the interaction region.  
At this point, note that the saddle point equations for the two factors of $AdS_5$ and $S^5$ 
are nothing but ``momentum conservation" in the corresponding embedding spaces, if we define
\begin{align}
{M}_i^A\equiv p_i \(\frac{X_i^A}{X_i \cdot P} +\frac{P^A}{R^2}\)\qquad;\qquad  \sum_{i}\, {M}_i^A =0\\
\tilde{ {M}}_i^{\tilde A} \equiv p_i \(\frac{Y_i^{ \tilde A} }{Y_i \cdot \tilde P} -\frac{ \tilde P^{\tilde A} }{R^2} \)\qquad;\qquad \sum_{i}\, \tilde{ {M}}_i^{\tilde A} =0
\end{align}
Notice that ${M}_i^A$ is orthogonal to $P$ and ${ {M}}_i^{\tilde A}$ is orthogonal to $\tilde P$. 
Moreover 
\begin{align}
{M}_i\cdot {M}_i= +\frac{p_i^2}{R^2}\qquad ;\qquad \tilde{ {M}}_i\cdot \tilde{ {M}}_i=-\frac{p_i^2}{ R^2}
\end{align}
We can then assemble four twelve-dimensional vectors,
\begin{align}
K_i= {({M}_i^A, \tilde{{M}}_i^{\tilde A} )}{}\qquad;\qquad i=1,2,3,4
\end{align}
which by constructions are null and tangent to the 10d geodesics at the intersection point. 
They are the 10d Lorentzian momenta. 
By explicit computation we find\footnote{We are using a convention where $2K_1\cdot K_2 = -(k_1+k_2)^2$ when going from embedding to 10d Lorentzian momenta, see e.g. (13) in \cite{Penedones:2010ue}.} 
\begin{align}
\!\! 
{2 K_1\cdot K_2}{} \Bigg|_{p_{i}=p} 
=\ \frac{8p}{R^2} \left[ \frac{{ p} \sqrt{U}}{\sqrt{U}+\sqrt{V}+1}- \frac{ {p}\sqrt{\newsigma}}{\sqrt{\newsigma}+\sqrt{\newtau}+1}\right] 
=\frac{ 4\Sigma}{R^2} {\bf s}_{cl} \Bigg|_{p_{i}=p}
\label{finally}
\end{align}
and similarly for the others thus perfectly reinforcing the identifications of the previous section. 
Relation \eqref{finally} is a very important dictionary type relation; it gives us the link between physical Mandelstam invariants, Mellin variables and space-time cross-ratios.

We can now go back to our picture   \eqref{geodesic_compu_pic} 
and turn on interactions, locally at the intersection point. This should simply 
decorate the integrand with the 10d flat space S-matrix. From the identification of 
physical Mandelstam invariants and Mellin variables given above, we can then argue that 
the large $p$ limit of the  $AdS_5\times S^5$ Mellin amplitude is expected to behave as anticipated {in (\ref{largep_limit_relation}) above. } 

In the next two sections we demonstrate this relation in a very precise way. 
Firstly by looking at the genus zero amplitude, and then by looking at one-loop supergravity.

\section{Genus zero amplitude}\label{tree_level_sec}

\subsection{Tree Level SUGRA}

The tree level Mellin amplitude for arbitrary charges was discovered in the seminal paper of~\cite{Rastelli:2016nze,Rastelli:2017udc},
and it was shown to possess a hidden ten dimensional conformal symmetry by Caron-Huot and Trinh in~\cite{Caron-Huot:2018kta}. 
This hidden symmetry explains the beautiful pattern of anomalous dimensions uncovered in~\cite{Aprile:2018efk}.
In our notation, the tree level Mellin amplitude in (\ref{rep4ints}) for arbitrary charges $\vec{p}$ 
looks extremely simple. It is just\footnote{The bold font variables are still given by \eqref{ten_dim_STU_p} with $p\rightarrow p_3$, 
while $\Gamma_{\otimes}$ upgrades to (\ref{GammaUnbalanced}) 
}
\begin{align}\label{amplitude_RZ}
\mathcal{M}^{(1)}_{p_1p_2p_3p_4}(s,t,\tilde{s}, \tilde{t}\,)=-\frac{ 1 }{ (1+{\bf s})(1+{\bf t}) (1+ {\bf u} ) }   \,.
\end{align}
where ${\bf s}+{\bf t}+{\bf u}=-4$. 
The hidden conformal symmetry is beautifully manifest in the bold font variables, 
i.e.~the four variable function $\mathcal{M}^{(1)}_{\vec{p}}(s,t,\tilde s,\tilde t)$ 
is shown here to depend only on ${\bf s}$, ${\bf t}$ and ${\bf u}$. 

We can now start from the flat space limit 
a la Penedones, in which $s$ and $t$ are large but $\tilde s$ and $\tilde t$ are kept finite. 
We will find trivially that
\begin{align}\label{step_1penedones}
\lim_{s,t,\rightarrow \infty} \mathcal{M}^{(1)}_{p_1p_2p_3p_4}(s,t,\tilde{s}, \tilde{t})\rightarrow \frac{1}{stu}
\end{align}
where $u$ participates in the limit as $u\rightarrow-s-t$. 
This limit is very asymmetric from the point of view of $AdS_5\times S^5$, 
because in reality all four variables $s,t,$ and $\tilde{s}, \tilde{t}$ participate on equal 
footing in the $AdS_5\times S^5$ Mellin amplitude.  In fact, the flat space limit a la Penedones 
and the large $p$ limit are similar, in the sense that in both cases one considers large Mellin variables, 
but the first one `forgets' about the sphere,  because $\tilde s$ and $\tilde t$ are kept finite. 
If we want to restore the sphere what we should do is to covariantise \eqref{step_1penedones}
and write
\begin{align}
\lim_{p\rightarrow\infty}
\mathcal{M}^{(1)}_{p_1p_2p_3p_4}(s,t,\tilde{s}, \tilde{t})\rightarrow \left( \frac{1}{{\bf s}{\bf t}{\bf u} } \right)\bigg|_{{\bf s}={\bf s}_{cl}, {\bf t}={\bf t}_{cl}}
\end{align}
%
%
Notice that we can first covariantise with ${\bf u}=-{\bf s}-{\bf t}-4$, and then localise the amplitude 
on the flat space saddle point ${\bf s}_{cl}+{\bf t}_{cl}+{\bf u}_{cl}=0$. This aspect of our formalism is tied 
with the hidden symmetry~\cite{Caron-Huot:2018kta}, which  
partly born out from the idea of relating the $p_1p_2p_3p_4$ amplitude to the simplest $2222$ correlator. 
In particular, the variables ${\bf s,t,u},$ satisfy the same constraint as the $AdS_5$ Mellin variables $s,t,u,$ for the $2222$ correlator, 
and this is why we can always covariantise as above. The $AdS_5\times S^5$ Mellin representation allows 
us to make manifest both the covariantisation for large bold font Mellin variables, and the hidden 
symmetry of the correlator. We shall see that the  covariatisation principle holds always, both in the full 
genus zero amplitude, and at one-loop.  


\subsection{Large $p$ limit $=$ 10d Virasoro-Shapiro }\label{VSlargepequal10d}

In this section we consider $\alpha'$ genus zero corrections to tree level supergravity. 

Starting from the flat space formula of Penedones \cite{Penedones:2010ue}, 
(which is valid for fixed $su(4)$ channel) we obtain the integral relation
\begin{align}\label{alapenedones_sec}
\mathcal{M}( s,t)= 
				C \int_0^{\infty}\!\!dz\ z^{\Sigma-\frac{d}{2}+3} e^{-z} \mathcal{A}_{VS}^{flat}\left( \frac{4z}{R^2} s,\frac{4z}{R^2} t  \right)\qquad;\qquad s,t\gg 1
\end{align}
where $C$ is a normalisation and $\Sigma=\frac{1}{2} \sum_i p_i$. 
The $\alpha'$ expansion of the VS amplitude is polynomial, 
thus one approach would be to make a polynomial ansatz for $\mathcal{M}(s,t)$, and match
order by order the $\alpha'$ expansion of the r.h.s. This computation boils down to $\Gamma$ 
function integrals multiplying the expansion of the VS amplitude, and thus 
fixes the leading polynomial term of the $AdS_5\times S^5$ amplitude. 
We have done this in appendix~\ref{VS_sugra_string}, where we also explain in more 
details how the covariantisation principle works. 
Here instead we will take a quicker route which leads to the same result. 

Note that \eqref{alapenedones_sec} defines a Mellin amplitude which is also 
a function of $\vec{p}$, because even though the flat space VS amplitude has no 
knowledge of $\vec{p}$, the remaining integrand in $z$ depends on $\Sigma$.
Then, by taking $p_{i} \rightarrow \infty$ the integral localises on a saddle point. 
The effective action is $\Sigma \log z -z$  and the saddle is $z=\Sigma$. 
This saddle point evaluation is almost like our master formula \eqref{largep_limit_relation}, 
but for the fact that we forgot the sphere. However, we learned in the previous 
section how to restore the dependence on $\tilde s$ and $\tilde t$. This is as simple 
as covariantising the result in $s$ and $t$, to a function of ${\bf s}$ and ${\bf t}$. 

Summarising, the large $p$ limit of the genus zero 
$AdS_5\times S^5$ amplitude is the flat space VS amplitude
evaluated at $4\Sigma{\bf s}_{cl}/R^2$ and $4\Sigma{\bf t}_{cl}/R^2$, 
which are the covariantised variables appearing in \eqref{alapenedones_sec}, with $z=\Sigma$, 
evaluated at the saddle point since we are taking $p$ large in the first place.
This is the same identification of Mandelstam invariants we obtained from our picture 
of four point-like strings interacting at the bulk point in the previous section.

Putting together our two computations to fix the details, and using the explicit expression for the Virasoro-Shapiro amplitude recalled in appendix \ref{B1}, we arrive at the prediction 
\begin{mdframed}
\begin{align}
\!\!\lim_{p\rightarrow \infty} \mathcal{M}(s_{cl},t_{cl},\tilde s_{cl},\tilde t_{cl}) \Big|_{genus\,=\,0} \!&=\! 
\frac{1}{ {\bf s}_{cl} {\bf t}_{cl} {\bf u}_{cl}}  
\frac{ \Gamma[1- \frac{\Sigma}{\sqrt{\lambda}} {\bf s}_{cl} ]\Gamma[1-\frac{\Sigma}{\sqrt{\lambda}} {\bf t}_{cl} ]\Gamma[1-\frac{\Sigma}{\sqrt{\lambda}}{\bf u}_{cl} ] }{
\Gamma[1+\frac{\Sigma}{\sqrt{\lambda}} {\bf s}_{cl} ] \Gamma[1+\frac{\Sigma}{\sqrt{\lambda}}{\bf t}_{cl} ]\Gamma[1+\frac{\Sigma}{\sqrt{\lambda}} {\bf u}_{cl} ]}  \notag
\\[.2cm]
\label{largepVS_formulasection}
\end{align}
\end{mdframed}

From the resummed amplitude we can appreciate the three different regimes we mentioned in the Introduction.
Begin by noticing that the `coupling' entering the argument of the $\Gamma$ function is
\begin{align}
 \epsilon= \frac{\Sigma}{\sqrt{\lambda}}\qquad;\qquad  \Sigma=\frac{p_1+p_2+p_3+p_4}{2}\qquad;\qquad |\epsilon|\ll 1
\end{align}
where $|\epsilon|\ll 1$ defines our setting. 
At the saddle point both $p_{i=1,2,3,4}$ and the Mellin variables are taken to be large in the same way. 
For concreteness, let's say ${\bf s}/p$ fixed with $p\rightarrow \infty$. 
This ratio is a function which we might call $\texttt{geometry}$, since it depend only on the cross ratios. 
The effective coupling controlling the behaviour of the amplitude is  thus 
\begin{align}
\epsilon_{eff}= \frac{p^2}{\sqrt{\lambda} }\ ;\quad 
\end{align}
and the three regimes are 
\begin{align}
\begin{array}{c|c|c} 
\quad p\ll \lambda^{1/4}\quad &{\rm SUGRA}  & \Gamma[1\pm \epsilon\, {\bf s}]\rightarrow 1 \\[.2cm] 
\hline
\rule{0pt}{.6cm}
\quad p/\lambda^{1/4}=O(1)\quad &\ {\rm short\, massive\, strings} \  &\Gamma[1\pm \epsilon\, {\bf s}] \text{ held fixed}\\[.2cm]
\hline
\rule{0pt}{.6cm}
\quad \lambda^{1/4} \ll p\ll\lambda^{1/2}\quad &{\rm flat\, classical\, strings}  &\ \Gamma[1\pm\epsilon\, {\bf s}]^{} \rightarrow \exp[{\pm\epsilon\, {\bf s}\log{\bf s}}]\ \\
\end{array}\notag
\end{align}
The SUGRA regime is improved by taking $p/\lambda^{1/4}=x$ fixed with $x\ll 1$. 
This corresponds to taking into account higher derivative corrections to the SUGRA action, coming from string theory, 
which we see here as an expansion in small $\epsilon\, {\bf s}$.
Instead, in the regime of flat classical strings we cross to the high-energy regime, where $\epsilon\, {\bf s}$ is large, 
the amplitude exponentiates
\begin{align}
&
\!\!\lim_{p\rightarrow \infty} \mathcal{M}(s_{cl},t_{cl},\tilde s_{cl},\tilde t_{cl}) \Big|_{genus\,=\,0} = \notag\\
&
\qquad
\frac{1}{{\bf s}_{cl} {\bf t}_{cl} {\bf u}_{cl } }
\exp{\left[ -\epsilon \left( {\bf s}_{cl} \log(-{\bf s}_{cl}^2)+ {\bf t}_{cl} \log(-{\bf t}_{cl}^2) + {\bf u}_{cl} \log(-{\bf u}_{cl}^2) \right)\right]}(1+\ldots )
\end{align}
The exponential contribution is precisely the Gross-Mende amplitude \cite{Gross:1987kza}.  
This has the interpretation of the saddle point action for a minimal area surface 
in flat space contributing to the four-point scattering amplitude.  If before we had two 
decoupled geodesic saddles, separately on $AdS_5$ and $S^5$, coming from $\Gamma_{\otimes}$, 
the effect of reaching a transition to a stringy-like 
object is to couple the two with flat space interactions mediated by the Gross-Mende amplitude. Still, as long as $\epsilon$ is small, even a classical string such as that in Gross-Mende is still only probing a small portion of space around the bulk point where the dynamics is like flat space, see figure \ref{regimes}.
If we further increase $p\rightarrow \sqrt{\lambda}$, $\epsilon$ is no longer small and our approximation breaks down.  
Physically it is clear why:  We would still be describing strings fluctuations 
in a flat space approximation, rather than the full $AdS_5\times S^5$.  
We will comment more about the transition from flat to big strings in $AdS_5\times S^5$ in the discussion section \ref{discussion_sec}.

\subsection{Towards the full amplitude: $\zeta_3$ and $\zeta_5$}

The $AdS_5\times S^5$ Virasoro-Shapiro amplitude is there, waiting to be discovered, 
and only the first two terms in the $\alpha'$ expansion are known in full generality 
\cite{Drummond:2019odu,Drummond:2020dwr}. These have been constructed by complementing 
the results of \cite{Alday:2018pdi, Alday:2018kkw}, which used the flat space limit a la Penedones, 
with the information coming from the unmixing problem \cite{Aprile:2017xsp}, pushed to 
order $\alpha'^3$ and $\alpha'^5$, and localisation results from \cite{Binder:2019jwn,Chester:2019jas,Chester:2020dja}

At each order in the $\alpha'$ expansion, the $AdS_5\times S^5$ Virasoro-Shapiro amplitude 
is a polynomial in the Mellin variables, but differently from its flat space limit, it is a sum of 
polynomials of different degree. The large $p$ limit suggests how to organise this sum. 
The idea is simply to stratify the amplitude according to the scaling
\begin{align}
s\rightarrow p s\quad;\quad t\rightarrow p t\quad ;\quad \tilde s \rightarrow ps\quad;\quad \tilde t \rightarrow p \tilde t\qquad;\qquad p\rightarrow\infty
\end{align}
which is indeed the scaling of the saddle point solutions \eqref{solu_pppp_AdS5} and \eqref{solu_pppp_S5}. 
Then, each stratum of the amplitude, written in the four fold representation, 
must be a function of crossing symmetric polynomials in the letters 
\begin{align}
{\bf s, t,u}\qquad;\qquad \tilde{s},\tilde{t},\tilde{u}\equiv-\tilde s-\tilde t+p_3-2\qquad ;\qquad p_1,p_2,p_3,p_4 \label{utilde}
\end{align}

To demonstrate our observation we will now rewrite the results of \cite{Drummond:2019odu,Drummond:2020dwr}
according to our discussion. We work directly with the most general case of arbitrary external charges, 
and find that 
\begin{align}
\zeta_3^{-1} \mathcal{V}^{(1,0)}= &\ \ \ \, (\Sigma-1)_3 \times 2 \label{zeta_3} \\[.2cm]
\zeta_5^{-1} \mathcal{V}^{(1,2)}=  &\ \ \ \, (\Sigma-1)_5 \ \big( {\bf s}^2 + {\bf t}^2 + {\bf u}^2 \big) \label{zeta_5} \\[.1cm]
						    &+(\Sigma-1)_4 \ \big(-10 (\tilde s\,{\bf s} + \tilde t\, {\bf t} + \tilde u\, {\bf u}) -5(  c_s {\bf s} + c_t {\bf t} + c_u {\bf u } )\big) \notag\\[.1cm]
						    &+(\Sigma-1)_3\  \big( +20( \tilde{s}^2+\tilde{t}^2+\tilde{u}^2) +\tfrac{5}{2} (  c_s^2 + c_t^2+ c_u^2 )+ 20 ( \tilde{s}\, c_s + \tilde{t}\, c_t + \tilde{u}\, c_u) \big) \notag\\[.1cm]
						    &+(\Sigma-1)_3\  \big( +\tfrac{33}{2} -\tfrac{27}{2}\Sigma^2 \big)  \notag
\end{align}
For convenience of the reader we repeat 
\begin{align}
c_s=\tfrac{ p_1+p_2-p_3-p_4}{2}\quad;\quad 
c_t=\tfrac{p_1+p_4-p_2-p_3}{2} \quad;\quad 
c_u=\tfrac{ p_2+p_4-p_3-p_1}{2}\quad;\quad  
\Sigma=\tfrac{p_1+p_2+p_3+p_4}{2} \label{cs}
\end{align}

Let us focus on the lines \eqref{zeta_3} and \eqref{zeta_5} first. 
The Pochhammers in $\Sigma$ come from the (inverse) $\Gamma$ function 
integral built in the flat space limit a la Penedones, (as we show in \eqref{term_by_term_norm} 
in appendix \ref{VS_sugra_string}).  In particular, from the latter one finds the 
leading terms $\zeta_3(\Sigma-1)_3$ and $\zeta_5(\Sigma-1)_5 (s^2+t^2+u^2)$. 
This was the starting point in \cite{Drummond:2019odu,Drummond:2020dwr}. 
We see now that their result secretly covariantises, as we argued in the previous section. 
Notice that away from the saddle point solution, the covariantisation is in terms of bold font variables such that
${\bf s}+{\bf t}+{\bf u}=-4$. Taking the large $p$ limit amounts to turn the Pochhammers into powers, thus in the large $p$ limit 
we find the flat space VS amplitude in ${\bf s}$ and ${\bf t}$, expanded to the given order. For example,
\begin{align}
\lim_{p\rightarrow \infty}  \mathcal{V}^{(1,2)}(s_{cl},t_{cl},\tilde s_{cl}, \tilde t_{cl})=\zeta_5\, \Sigma^5 \, \big( {\bf s}_{cl}^2 + {\bf t}_{cl}^2 + {\bf u}_{cl}^2 \big)
\end{align}

The $\zeta_5$ contribution is the first non trivial case where polynomials of lower 
degree in the Mellin variables are turned on, due to $AdS_5\times S^5$ effects. 
The large $p$ stratification is manifest in the polynomials which accompany the 
various factors $(\Sigma-1)_{k\leq 4}$. Indeed, if we consider the $(\Sigma-1)_4$ term, 
this can only multiply a polynomial which is linear in ${\bf s}$ and ${\bf t}$ and 
at most linear in the remaining variables, $\tilde{s}$ $\tilde{t}$ and $p_{i=1,2,3,4}$, 
otherwise it would contribute to the large $p$ limit.  The concrete result from \cite{Drummond:2020dwr} 
becomes amazingly simple. A similar observation holds for the $(\Sigma-1)_3$ contribution.\footnote{Localisation only enters to fix  
the actual values of $\tfrac{33}{2} -\tfrac{27}{2}\Sigma^2+\tfrac{5}{2} (  c_s^2 + c_t^2+ c_u^2 )$. We thanks J.Drummond and M.Santagata for discussion on this point.}
In addition, the overall degree of $\mathcal{V}^{(1,2)}$ w.r.t.~$\tilde{s}$ and $\tilde{t}$ 
is expected to be at most second order, i.e. it should not exceed
the degree of the top term.\footnote{In the superblock basis this is related to the truncation 
of the 10d spin in the spectrum, i.e.~the fact that only rep $[aba]$ with $a=0,1,2$ appear at this order in the OPE. }

\section{One-Loop SUGRA}\label{one_loop_sec}

The next level of computations we can explore with our large $p$ limit is that of one-loop SUGRA. 
Even though there are no Witten diagrams computations of such kind, a number 
of four-point one-loop amplitudes, fully consistent with the dual CFT picture,
have been bootstrapped in \cite{Aprile:2017bgs,Aprile:2017xsp,Aprile:2017qoy,Aprile:2018efk,Aprile:2019rep}. 
These provide the basis for our explorations as we now explain.

Understanding the anatomy of the four-point one-loop amplitudes goes through the 
analysis of the spectrum of two-particle operators in $\mathcal{N}=4$ SYM \cite{Aprile:2018efk,Aprile:2019rep}. 
Long two-particle operators are exchanged in the leading logarithmic discontinuity, 
more generally at any loop order $\ell$, and determine the $\log^{\ell+1}u$ coefficient 
function of the full amplitude, in terms of their tree level anomalous dimensions 
(to the power $\ell+1$) and their leading three point couplings with the external 
single-particle operators. To complete the logarithmic discontinuity into the 
full one-loop amplitude,  the entire two-particle data, both long and protected, 
is needed. Luckily for us this can all be extracted from the tree level amplitude in 
combination with $1/N^2$ free theory, and propagated in the one-loop amplitude 
according to the OPE. In this way the full one-loop amplitude is fixed, up to stringy ambiguities.

In principle all Kaluza Klein one-loop amplitudes are available from the position space 
algorithm of \cite{Aprile:2019rep}.  However, the large $p$ limit is hard to study in position space. 
Thus we will have to Mellinize those results. The strategy we follow is to use
the hidden conformal symmetry \cite{Caron-Huot:2018kta} to write the leading 
logarithmic discontinuity of a generic correlator in one go. Then, we Mellinize the full position 
space results and take the large $p$ limit. 

The consistency of the large $p$ limit  with the flat space scattering amplitude suggests 
from the very beginning that the transcendental content of the one-loop Mellin amplitude at fixed $p$ 
is made of meromorphic functions which match those of the 10d box in the limit.
We will show that this pattern is the very much the same as for the position space amplitude. 
On one hand, this is the only way position and Mellin space could match precisely, 
on the other hand this computation is highly non trivial. 
 

Some readers might wish to jump directly to section \ref{grav_smatrix_comparison} at this point. There
 we collect the various more technical results obtained in 
sections \ref{main_txt_log_disc} and \ref{big_mellinoneloop_section} and 
we discuss directly the correspondence of the large $p$ limit with the 10d S-matrix.

\subsection{Logarithmic discontinuities}\label{main_txt_log_disc}

In the formalism of \cite{Caron-Huot:2018kta}, 
the $\log^2 u$ discontinuity is obtained by acting with an eight-order differential operator, 
see $\Delta^{(8)}$ in appendix \ref{app_delta8}, on the following prepotential,
\begin{align}\label{dpqrs}
\mathcal{P}^{\ell}_{p_1p_2p_3p_4}(U,V,\newsigma,\newtau)&=\sum_{{\it T}} \left(\frac{\newsigma}{U}\right)^{\!\!\tilde s+2} 
\left(\frac{\newtau}{V}\right)^{\!\!\tilde t}  \widehat{\mathcal{D}}^{}_{\tilde s,\tilde t }(p_1p_2p_3p_4 )\ \mathcal{P}^{\ell}_{2222}(U,V)
\end{align}
By the hidden conformal symmetry the operators 
$\widehat{\mathcal{D}}^{}_{\tilde s,\tilde t }(p_1p_2p_3p_4 )$ do not depend on the loop order, 
i.e. they are the same operator at tree level. We have found the closed form 
expression for these operators:\footnote{Recall (\ref{utilde}) for the definition of $\tilde u$ and (\ref{cs}) for $c_s,c_t,c_u$.} 
\begin{align}
\widehat{\mathcal{D}}^{}_{\tilde s,\tilde t }(\vec{p}\, )=\!&\
\frac{( { U }\partial_{ U } -3-\tilde s )_{\tilde s} }{ \tilde s! }  \frac{ ( { U }\partial_{ U } -3-\tilde s-c_s )_{\tilde s+c_s} }{  (-)^{c_s}( \tilde s+c_s )! }  \times
\label{mostgeneralDpqrs}\\[.2cm]
&
\   \frac{({ V }\partial_{ V }+1 -\tilde t )_{\tilde t} }{\tilde t!} \frac{ ({ V }\partial_{ V } +1-\tilde t-c_t )_{\tilde t+c_t } }{ (-)^{c_t}(\tilde t+c_t )!} \times
\, \frac{  ({ U }\partial_{ U }+{ V }\partial_{ V })_{\tilde u}  }{\tilde u!} \frac{ ({ U }\partial_{ U }+{ V }\partial_{ V })_{\tilde u + c_u} }{ (\tilde u + c_u)!} \notag \,.
\end{align} 
The one loop function $\mathcal{P}_{2222}(u,v)$  has a very simple form, 
given in \cite{Caron-Huot:2018kta} in position space. (It can be found 
in appendix \ref{app_delta8} in our notation.)
Our construction of the one-loop Mellin amplitudes instead begins 
with the following Mellin representation for $\mathcal{P}_{2222}$:
\begin{align}
\!\!\!
\mathcal{P}_{2222}(U,V)&= -\iint ds dt\ (-U)^{s+4}V^t\ \frac{\Gamma[-s]}{\Gamma[s+1]} \Gamma[-t]^2 \Gamma[-{u}]^2 \times \mathcal{N}_{2222}(s,t)
\label{MellinP_2222}
\end{align}
where  {${u}=-s-t-4$} for this correlator, and the Mellin amplitude is
\begin{align}
&\mathcal{N}_{2222}(s,t)=\label{N2222}\\
&
\quad
(\psi^{(0)}(-t)-\psi^{(0)}(s+1)  ) \mathcal{T}^{(4)}(s,t)  +  (  \psi^{(0)}(-{u})-\psi^{(0)}(s+1) ) \mathcal{T}^{(4)}(s,{u}) + \mathcal{T}^{(3)}(s,t) \notag
\end{align}
with the functions $\mathcal{T}^{(4)}$ and $\mathcal{T}^{(3)}(s,t)=\mathcal{T}^{(3)}(s,u)$ given by\footnote{The (top weight) function $\mathcal{T}^{(4)}[s,t]$ 
is the Mellin transform of ${2}\times \mathcal{P}_{2222}^{2,2^-}/(x_1-x_2)^{7}$. 
It follows from the Mellin representation of $(x_1-x_2)^{-7}$, after including the numerator, monomial by monomial
by shifting the contour of integration, as suggested in \cite{Aprile:2019rep}. 
This procedure works at any loop order for the top-weight amplitude. } 
\begin{align}
\label{calAformula}
\mathcal{T}^{(4)}(s,t)&= \frac{1}{120} \frac{ (t-1)t}{(s+1)(s+2)( s+t+1)(s+t+2)(s+t+3) }\\[.2cm]
\mathcal{T}^{(3)}(s,t)&=  \frac{1}{240} \frac{(s+3)}{(s+1)}\frac{ 16 + 8(t+{u})-4(t-{u})^2 + {u} t (t+{u}) -2 (t^3+{u}^3)}{(s+1)(s+2)(t+1)(t+2)({u}+1)({u}+2) }
\end{align}
The contour 
of integration in \eqref{MellinP_2222} is a straight line Mellin-Barnes contour.\footnote{It 
can be deformed to encircle poles at $s\ge 0$  and $t\ge 0$, so to match
the Taylor expansion in position space. Notice there is actually no simple pole 
at $t=-1$!, or at $s=-1$.} 

Turning the operators $\widehat{\mathcal{D}}^{}_{\tilde s,\tilde t }(p_1p_2p_3p_4 )$ 
into Mellin space, we arrive at the counterpart of the formula \eqref{dpqrs} in Mellin space. 
This computation, which looks complicated, can actually be done with pencil and paper, 
and we do so in appendix \ref{app_delta8}.  The result is very illuminating
\begin{align}\label{Mellin_p1234}
&\!\! \mathcal{P}_{p_1p_2p_3p_4}(U,V,\tilde U,\tilde V)=   \\
&
\quad(U\newsigma)^{2} \!\! 
\iint { \newsigma^{\tilde s} \newtau^{\tilde t} }{}  \Bigg[  \!
(-)^{max(0,c_s)}\!\! \iint ({ -U })^{ s }  { V }^{t }\, \mathcal{N}_{2222}({\bf s},{\bf t} ) 
\, \left[\frac{\sin(\pi(-s+max(0,c_s)) )}{\pi} \Gamma_{\otimes} \right] \Bigg] \notag
\end{align}
where the $sin$ flips one of the two $s$-dependent $\Gamma$ functions in $\Gamma_{\otimes}$, 
in order to have just simple poles, as it is the case for the leading  logarithmic discontinuity.
From \eqref{Mellin_p1234} we read off the Mellin amplitude of the prepotential for arbitrary charges, 
\begin{align}
&
\mathcal{N}_{p_1p_2p_3p_4}(s,t,\tilde s,\tilde t) =\mathcal{N}_{2222}({\bf s},{\bf t} )\label{ddisc_mellin_1234}
\end{align}
where the r.h.s. is evaluated precisely at the bold font 10d variables defined in \eqref{ten_dim_STU_p}.

In position space $\mathcal{P}_{p_1p_2p_3p_4}$ is simpler than the corresponding 
leading logarithmic discontinuity, but in Mellin spaceall the complexity has collapsed 
to just the $\Gamma_{\otimes}$ factor, and the rest is carried by 
$\mathcal{N}_{p_1p_2p_3p_4}(s,t,\tilde s,\tilde t) =\mathcal{N}_{2222}({\bf s},{\bf t} )$, 
namely the $\mathcal{N}_{2222}(s,t)$  prepotential promoted to be a function of the bold font 
variables ${\bf s}$, and ${\bf t}$. The fact that $\mathcal{N}_{p_1p_2p_3p_4}$ depends only 
on bold font variables perhaps was to be expected,  since the prepotential still enjoys 
the hidden conformal symmetry. 

The double discontinuity of $\mathcal{A}^{(2)}_{p_1p_2p_3p_4}$ is finally obtained by 
acting with the eight-order differential operator $\Delta^{(8)}$ on $\mathcal{P}_{p_1p_2p_3p_4}$. 
The way to Mellinise $\Delta^{(8)}$ is explained in appendix \ref{app_delta8}, roughly what happens 
is that we write $\Delta^{(8)}=\sum \Omega_{mn\tilde m \tilde n}U^m V^n \newsigma^{\tilde m}\newtau^{\tilde n}$, 
and by shifting the contour we turn the monomials $U^n V^m \newsigma^i \newtau^j$ into a shift operator.
This procedure is straightforward but less illuminating, since
it involves some lengthy polynomials in the Mellin variables.  
Ancillary files are attached to appendix \ref{app_delta8}
to help the reader.

\subsection{One-Loop Mellin amplitudes}\label{big_mellinoneloop_section}

\setcounter{tocdepth}{1}

The total one-loop amplitude for a balanced configuration of charge, $p_{i=1,2,3,4}=p$ has the form 
\begin{align}
\label{oneLmellin_integrand}
\mathcal{A}^{(2)}_{pppp}=
\sum_{\tilde s,\tilde t}\iint \frac{ \Gamma[-s]^2\Gamma[-t]^2 \Gamma[-{u} ]^2  }{ 
						\Gamma[\tilde s+1]^2 \Gamma[\tilde t+1]^2 \Gamma[ \tilde u+1]^2 }\ 
						U^s V^t \newsigma^{\tilde s} \newtau^{\tilde t} \ 
						\underbrace{ \Big(  \mathcal{W}_{pppp}^{w=4}+ 
										\mathcal{W}_{pppp}^{w=3} + 
										\mathcal{W}_{pppp}^{w=2} \Big) }_{ \mathcal{M}^{(2)}_{pppp}(s,t,\tilde s, \tilde t)}
\end{align}
It consists of three sub-amplitudes $\mathcal{W}^{w=4,3,2}$ distinguished by the transcendental weight $w$ they \emph{first} contribute to.  
Each of these functions can be written as a sum of products of \emph{reduced transcendental kernels} multiplying rational functions. 
The full transcendental kernels include of course $\Gamma_{\otimes}$, which we wrote explicitly in \eqref{oneLmellin_integrand}.

We will now introduce the various $\mathcal{W}_{pppp}^{w=4,3,2}$ and take the large $p$ limit. 
Doing so we will  remark how the organisation of the integrand in \eqref{oneLmellin_integrand} is
in perfect agreement with the position space results of \cite{Aprile:2019rep}.  
The case of unequal external charge will be discussed in \cite{to_be_decided}.

\subsubsection*{4.2.1\,\,\,\,\, Top-weight 4}

The structure of the top-weight 4 integrand is 
\begin{mdframed}
\begin{align}
\label{amplitudeW4}
\!\!
\mathcal{W}_{pppp}^{w=4}=
\mathcal{K}[\mathbf{s},\mathbf{t}] \mathpzc{W}_{pppp}^{(4)}(s,t,\tilde s,\tilde t)+ 
\mathcal{K}[\mathbf{s},\mathbf{u}] \mathpzc{W}_{pppp}^{(4)}(s,{u},\tilde s,\tilde u)+
\mathcal{K}[\mathbf{u},\mathbf{t}] \mathpzc{W}_{pppp}^{(4)}({u},t,\tilde u,\tilde t)
\end{align}~
\end{mdframed}
where $\mathcal{K}$ is the reduced transcendental kernel of the double box 
(see below \eqref{doubleboxkernel}) and $\mathpzc{W}^{(4)}_{pppp}(s,t,\tilde s ,\tilde t)$ 
is a rational function with the symmetry  
\begin{align}
\label{symm_W4}
\mathpzc{W}_{pppp}^{(4)}(s,t,\tilde s ,\tilde t)=\mathpzc{W}_{pppp}^{(4)}(t,s,\tilde t, \tilde s)\ .
\end{align}

The reduced kernel $\mathcal{K}$ of the double box \cite{Usyukina:1992jd,Isaev:2003tk} is 
\begin{align}\label{doubleboxkernel}
\mathcal{K}[z,w]= \psi^{(1)}(-z)+\psi^{(1)}(-w) - ( \psi^{(0)}(-z) + \psi^{(0)}(-w) )^2 - \pi^2\ .
\end{align}
This function has been studied in \cite{Allendes:2012mr}, and upon inspection only contains simple poles in $s$ and $t$.
The Mellin transform of the double box is indeed just given by the product of 
$\Gamma[-s]^2 \Gamma[-t]^2 \Gamma[s+t+1]^2$ and the reduced kernel $\mathcal{K}[z,w]$.\footnote{$\mathcal{K}[z,w]$ 
can be bootstrapped from the small $U$ and $V$ expansion of the ladder integrals, 
since these are pure transcendentals over $(x_1-x_2)$.  It is simple to see that 
$\psi(-z)\psi(-w)$ is not enough, but can only be corrected by a linear combinations 
of weight $2$ terms, which at most bring in simple poles, i.e. 
$\psi^{(0)}(-z)\psi^{(0)}(-z) -\psi^{(1)}(-z)$, the analogous for $w$, and $\pi^2$. }

In position space, the top weight four part of the one-loop correlators are given by 
the double box in the three independent orientations, each orientation multiplied by a rational 
functions of the cross ratios, fixed by the double logarithmic discontinuity. The structure of the top-weight 4 
integrand in \eqref{amplitudeW4} perfectly reflects the position space understanding of the correlator. 
In particular, $\mathpzc{W}_p^{(4)}(s,t,\tilde s ,\tilde t)$ is determined 
by the knowledge of $\mathcal{T}^{(4)}({\bf s},{\bf t})$ in \eqref{calAformula} 
and \eqref{ddisc_mellin_1234}, and the action of $\Delta^{(8)}$ as described in the previous section.

We match the $\log^2 U \log^2 V$ projection of the correlator with that of the double logarithmic discontinuity. 
This is equivalent to looking at the residue of the triple poles in $s$ and $t$ in \eqref{oneLmellin_integrand}, 
and triple poles in $t$ of the double logarithmic discontinuity given in \eqref{ddisc_mellin_1234}.
From the explicit Mellin integrals 
%
%
we can present the final result in a simple fashion as a sum over (twenty-five) shifts,
\begin{align}
\mathpzc{W}_{pppp}^{w=4}(s,t,\tilde s,\tilde t) &= 
\sum_{\substack{ -4 \leq a \leq 0 \\ -3-a \leq b\leq 3} }{\delta}_{pppp}(a,b)  \mathcal{T}^{(4)}[ {\bf s}+a, {\bf t}+b  ] 
\end{align}
where the coefficients $\delta_{pppp}(a,b)$ depend on $p$ and the Mellin variables, 
and are attached in an ancillary file. For finite $p$ the rational function 
$\mathpzc{W}_{pppp}^{w=4}(s,t,\tilde s,\tilde t)$ is a four variable function, 
because of the ${\delta}_{pppp}(a,b)$.

Coming to the large $p$ limit, we have to consider the limit of $\mathcal{W}_{pppp}^{w=4}(s,t,\tilde s, \tilde t)$ 
when all the variables are large and scale linearly with $p$. This limit is obviously factorised into 
that of $\mathcal{K}$ and $\mathpzc{W}^{(4)}_{pppp}$, and the split of the full kernel into 
$\Gamma_{\otimes}$ and the reduced kernel is crucial for this analysis.

The reduced kernel $\mathcal{K}$ in the large $p$ limit is dominated by the asymptotic of $\psi^{(0)}$, 
which is leading compared to the asymptotic of $\psi^{(1)}$,
\begin{align}
\lim_{p\rightarrow\infty} \mathcal{K}[ pz, pw ]&=  -(\log(-pz)-\log(-pw))^2 -\pi^2=\ -\log^2((-z)/(-w))-\pi^2
\end{align}
This combination is `dimensionless' and thus does not have additional $p$-dependence. 

The limit on $\mathpzc{W}^{(4)}_{pppp}$ has more features. Without loss of generality, 
we focus on one of the three orientation. The result can be put in the following suggestive form,
\begin{align}\label{limit_suggestive_calW4}
\lim_{p\rightarrow \infty} &\mathpzc{W}_{pppp}^{(4)}(s_{cl},t_{cl},\tilde s_{cl} ,\tilde t_{cl})= 
\left( \sum_{\substack{ -4 \leq a \leq 0 \\ -3-a \leq b\leq 3} } \lim_{p\rightarrow \infty }{\delta}_{pppp}(a,b)  \right) 
\left( \lim_{p\rightarrow \infty}\mathcal{T}^{(4)}( {\bf s}_{cl},{\bf t}_{cl}) \right)
\end{align}
The expression \eqref{limit_suggestive_calW4} is correct because as we now 
show both the sum over $\delta_{pppp}(a,b)$ and $\mathcal{T}^{(4)}({\bf s},{\bf t})$ have a 
non vanishing leading contribution. If we were to be taking the large $s$ and $t$ limit, 
rather than the large $p$ limit, we would not be finding such a simple relation.

The large $p$ limit gives
\begin{align}
\lim_{p\rightarrow \infty} \mathpzc{W}_{pppp}^{(4)}(s_{cl},t_{cl},\tilde s_{cl},\tilde t_{cl})&= 
\ +\frac{1}{120}\times \Big( 16p^4 (s_{cl}+\tilde s_{cl} )^4\, \Big)\times\,
 \frac{  ({\bf t}_{cl})^2 }{ ({\bf s}_{cl})^2 (-{\bf u}_{cl})^3 } \notag\\
&=\ -\frac{1}{120} \times \frac{1}{ {\bf s}_{cl} {\bf t}_{cl} {\bf u}_{cl} } \times 
(\Sigma^4 {\bf s}_{cl}^4)\times  \frac{  ({\bf t}_{cl})^3 }{ ({\bf s}_{cl}) ({\bf u}_{cl})^2 } 
\label{weight4amp_largep}
\end{align}
It can be immediately seen here that the contribution from ${\delta}(a,b)$ 
scales like $p^8$, as it is natural to expect from $\Delta^{(8)}$ 
when all variables are scaled together. But the nicest 
aspect of this calculation is the following.

The sum over the shifts $a$ and $b$ in \eqref{limit_suggestive_calW4} reorganises 
to give a simple multiplicative $(2p)^4({\bf s}_{cl})^4$ factor. This is very remarkable because 
all terms in the sum are actually non trivial!  When we include $p_{21}\neq 0$ and $p_{34}\neq 0$, 
as described in details in appendix \ref{app_delta8},  the result is indeed $\Sigma^4 {\bf s}_{cl}^4$, 
as written in \eqref{weight4amp_largep}. In this case the combinatorics that leads to such 
a simple result is even more surprising, and in fact let us point out that $\Sigma$ is not explicit 
in any of the terms involved in the sum. Individually, these depend on $c_s$, $c_t$, $c_u$ and $p_3$.

The action of $\Delta^{(8)}$ restores the extra symmetry of the $\log^2 U \log^2 V$ coefficient function, 
which we highlighted in \eqref{symm_W4}, that is lost at the level of the preamplitude $\mathcal{T}^{(4)}$. 
We see in \eqref{weight4amp_largep} that at leading order in the large $p$ expansion 
this symmetry becomes a symmetry of the ten-dimensional variables, ${\bf s}_{cl}\leftrightarrow {\bf t}_{cl}$.

\subsubsection*{4.2.2\,\,\,\,\, Top-weight 3}

The structure of the top-weight 3 amplitude is simpler. The reduced kernel is simply 
a polygamma function, and the full $\mathcal{W}_{pppp}^{w=3}$ is
\begin{mdframed}
\begin{align}
\mathcal{W}_{pppp}^{w=3}=
\psi(-{\bf s}) \mathpzc{W}_{pppp}^{(3)}(s,t,\tilde s,\tilde t) +  \psi(-{\bf t}) \mathpzc{W}_{pppp}^{(3)}(t,{u},\tilde t,\tilde u)+ \psi(-{\bf u}) \mathpzc{W}_{pppp}^{(3)}({u},t,\tilde u,\tilde t),\ \notag\\
\end{align}
\end{mdframed}
where the rational function $\mathpzc{W}_{pppp}^{(3)}(s,t,\tilde s,\tilde t)$ has the symmetry
\begin{align}
\label{symm_W3}
\mathpzc{W}_{pppp}^{(3)}(s,t,\tilde s,\tilde t)=\mathpzc{W}_{pppp}^{(3)}(s,u,\tilde s,\tilde u)
\end{align}

In position space, transcendental functions with weight $w=4^{-},3^{\pm}, 2^+$ 
have coefficient functions fully determined by the double discontinuity in the various orientations. 
In the same way, $\mathcal{W}_{pppp}^{w=3}$ is determined by $\mathcal{T}^{(4)}$ 
and $\mathcal{T}^{(3)}$, since $\mathcal{W}_{pppp}^{w=3}$ first contributes to the 
$\log^2 U \log^1 V$ projection of the correlator, which is again fixed by the double discontinuity. 
To see this consider the residue of triple poles in $s$ and double poles in $t$ of 
the amplitude \eqref{oneLmellin_integrand}, and compare with the residue of the double 
poles in $t$ of the double logarithmic discontinuity \eqref{ddisc_mellin_1234}. 

The matching procedure just described follows the same logic as for 
$\mathcal{W}_{pppp}^{w=4}$ but for a minor modification: Shifting the contour in order to 
absorb $\Delta^{(8)}_{}(p)$ produces a shift of the arguments of the polygammas
 in \eqref{ddisc_mellin_1234}. These shifts do not affect the determination of 
 $\mathcal{W}^{w=4}_{pppp}$ itself, given in the previous section, 
 but now they must be included correctly. Taking care of this subtlety we find 
\begin{align}
&
\mathpzc{W}_{pppp}^{(3)}(s,t,\tilde s, \tilde t)=\label{contrA_to_W3}
\sum_{\substack{ -4 \leq a \leq 0 \\ -3-a \leq b\leq 3} }  {\delta}_{pppp}(a,b)\Big( 2 \mathcal{T}^{(3)}[ {\bf s}+a, {\bf t}+b  ] + \\[.1cm]
&\rule{1.2cm}{0pt}
r_{pppp}(a,-b, {\bf s}, -{\bf t})  \mathcal{T}^{(4)}[ {\bf s}+a, {\bf t}+b  ] + 
r_{pppp}(a,a+b, {\bf s},- {\bf u})  \mathcal{T}^{(4)}[ {\bf s}+a, {\bf u}-a-b  ]\Big)\notag
\end{align}
where $r_{pppp}(\alpha,\beta, {\bf s}, {\bf t})$ is the rational function due to the 
shifts in the arguments of the polygammas.\footnote{This is just
$$
r_{pppp}(\alpha,\beta, w, z)=
-\sum_{n=0}^{|\alpha|-1} \frac{1} { {{ \rm sgn}[\alpha] } (w+\frac{{{\rm sgn}[\alpha] }+1 }{2})+n } + \sum_{m=0}^{|\beta|-1} \frac{1} { {{\rm sgn}[\beta] }(z+\frac{ {{\rm sgn}[\beta] } -1}{2})+m }
$$
}

Coming to the large $p$ limit, we focus on one of the three orientations without loss of generality. 
The reduced transcendental kernel is just a polygamma $\psi^{(0)}$ 
which asymptotes to a logarithm, as before.  With a same argument as in 
\eqref{limit_suggestive_calW4}, the limit of the rational function can be put 
in the suggestive form, 
\begin{align}
\lim_{p\rightarrow \infty} &\mathpzc{W}_{pppp}^{w=3}(s_{cl},t_{cl},\tilde s_{cl},\tilde t_{cl})=
 \left( \sum_{\substack{ -4 \leq a \leq 0 \\ -3-a \leq b\leq 3} } \lim_{p\rightarrow \infty }{\delta}_{pppp}(a,b) \right) 
 \left( \lim_{p\rightarrow \infty} 2\mathcal{T}^{(3)}[ {\bf s}_{cl},{\bf t}_{cl}] \right)
 \label{limit_W3ppp_rational}
 \end{align}
Notice also that contributions to $\mathpzc{W}_{pppp}^{(3)}$ coming from $\mathcal{T}^{(4)}$ 
are subleading compared to $\mathcal{T}^{(3)}$, since they come with one extra power of 
$p$ in the denominator, due to $r_{pppp}$. More explicitly, we obtain the result
\begin{align}
 \lim_{p\rightarrow \infty} \mathpzc{W}_{pppp}^{(3)}(s_{cl},t_{cl},\tilde s_{cl},\tilde t_{cl})= 
 \frac{1}{120}\times \frac{\Sigma^4}{{\bf s}_{cl} {\bf t}_{cl} {\bf u}_{cl} } \times 
 \frac{   7 ({\bf t}_{cl})^2({\bf s}_{cl})^4 + 7  ({\bf u}_{cl})^2 ({\bf s}_{cl})^4 -3 ({\bf s}_{cl})^6 }{  2 ({\bf t}_{cl}) ( {\bf u}_{cl}) }
\end{align}
The r.h.s.~shows the symmetry \eqref{symm_W3} at the level of 10d variables,  
${\bf t}_{cl}\leftrightarrow {\bf u}_{cl}$.

\subsubsection*{4.2.3\,\,\,\,\, Top-weight 2}

The top-weight 2 amplitude is just a rational function, since the full kernel in this case coincides with 
$\Gamma_{\otimes}$, i.e. $\mathcal{W}^{w=2}_{pppp}=\mathpzc{W}^{(2)}_{pppp}$. 
For illustration we quote from \cite{to_be_decided} the result for $p=2$, 
\def\tt  {\mathpzc{I}}
\begin{align}
\mathpzc{W}^{(2)}_{2222}(s,t)&=\frac{num( \tt_i) }{(s+1)(s+4)(s+5)(t+1)(t+4)(t+5)(u+1)(u+4)(u+5)}+\beta^{(1)}
\end{align}
where $\tt_i=s^i+t^i+{u}^i$ with ${u}=-s-t-4$ for this correlator, and 
\begin{align}
num=& 
+\tfrac{3}{25} \tt_{5} \tt_5 \notag\\[.2cm]
&
- \tfrac{68 849 782 843}{3 072 000} \tt_4 \tt_5 	 -  \tfrac{524 233}{2 250} \tt_3\tt_6 			 +\tfrac{16 495 714 123}{1 024 000}  \tt_2 \tt_7 \notag\\[.2cm]
&
- \tfrac{63 132 297 731}{768 000} \tt_4 \tt_4 	 - \tfrac{6 745 323 479}{72 000} \tt_3 \tt_5 		+\tfrac{63 132 758 531}{768 000} \tt_2 \tt_6 \notag\\[.2cm]
&
- \tfrac{558 138 949 661}{1 152 000} \tt_3 \tt_4	 -  \tfrac{879 855 811}{38 400} \tt_2 \tt_5  		 \notag\\[.2cm]
&
-\tfrac{11 005 992 151}{18 000} \tt_3\tt_3		 - \tfrac{191 551 927 561}{288 000} \tt_2 \tt_4 	 \notag\\[.2cm]
&
- \tfrac{21 075 629 959}{14 400} \tt_2\tt_3 			\notag\\[.2cm]
&
- \tfrac{11 046 578 393}{18 000} \tt_2 \tt_2 		
\end{align}
This is not a unique decomposition, however it makes crossing invariance 
of $\mathpzc{W}^{(2)}_{2222}$ manifest. The (only) constant ambiguity $\beta$ 
can be fixed by using results for the integrated correlator, which are provided 
independently by localisation \cite{Chester:2019pvm}.

What happens for $\mathpzc{W}^{(2)}_{p_1p_2p_3p_4}$, for fixed values of the external charge,  
is quite non trivial and deserves a separate chapter \cite{to_be_decided}, which we summarize here below.

The position space algorithm of \cite{Aprile:2019rep}, after imposing crossing 
and absence of (euclidean) unphysical $x_1= x_2$ poles, returns a correlator with 
free parameters in the coefficient functions relative to transcendental weights $w=2^{-},1^{+},0$. 
These free coefficients provide the same degrees of freedom of $\mathpzc{W}^{(2)}_{p_1p_2p_3p_4}$. 
In general, fixing the full correlator (up to stringy-type ambiguities) requires
\begin{itemize}
\item[1)] Knowledge about the precise definition of the external single particle operators, 
\item[2)] One-loop OPE predictions in- and below-window for the long sector,
\item[3)] One-loop OPE predictions for protected semi-short multiplets at the unitarity bound, in order to perform correctly multiplet recombination. 
\end{itemize}

There is a non trivial exchange of information in items 2) and 3) between 
free theory and the dynamical amplitude, versus the spectrum of supergravity. 
These considerations motivated \cite{Aprile:2019rep}  to introduce the \emph{minimal one-loop function}, 
which is nicely understood to descend from the double discontinuity, upon 
modifying the tree level Mellin amplitude by an \emph{upgraded} tree level function.   
This upgraded tree level function takes care of some trivial multiplet recombination, 
for example the multiplet recombination to cancel stringy states in supergravity, 
and ascribe all the new features of the one-loop dynamics
to the minimal one-loop function. The upgraded tree level function
extends consistently $\mathcal{M}^{(1)}_{p_1p_2p_3p_4}$
to a non-planar function, and it is an exact function of $N$. Re-expanding the 
latter at one-loop gives a contribution to $\mathpzc{W}^{(2)}_{p_1p_2p_3p_4}$.

One-loop stringy-type ambiguities contribute only to finite spin and 
make their appearance only in $\mathpzc{W}^{(2)}_{p_1p_2p_3p_4}$.
Generically they are described by an ansatz of the form
\begin{align}\label{stringy_mellin}
\mathpzc{W}^{(2)}_{p_1p_2p_3p_4}(s,t,\newsigma,\newtau)\Big|_{ambiguities}= 
\sum_{ \textit{T}} \left(\beta^{(1)}_{\tilde s,\tilde t;\vec{p}} + \beta^{(s)}_{\tilde s,\tilde t;\vec{p}}\, s + \beta^{(t)}_{\tilde s,\tilde t;\vec{p}}\, t \right)\newsigma^{\tilde s} \newtau^{\tilde t}
\end{align}
where $\beta^{type}$ are arbitrary constants. As far as the counting goes we 
can think of terms labelled by $\beta^{(1)}$ as corresponding to $R^4$ in the bulk 10d effective action, 
and terms labelled by $\beta^{(s)}$ and $\beta^{(t)}$ as corresponding to $\partial^2 R^4$. 
If a correlator is symmetric under a crossing transformation the same symmetry needs to 
be imposed on \eqref{stringy_mellin}. For example, equal charges correlators are fully crossing invariant.

As detailed already in \cite{Aprile:2019rep}, the position space bootstrap for 
the minimal one-loop function, after taking into account items 1), 2) and 3),
returns a function with finitely many free parameters, which case by case can be shown 
to be in correspondence with the parameters $\beta$ in \eqref{stringy_mellin}.
We do not know a closed and explicit expression for $\mathpzc{W}^{(2)}_{pppp}$, yet.  
However, we will now argue that we can deduce the large $p$ limit of $\mathpzc{W}^{(2)}_{pppp}$ 
by covariantising the large $s$ and $t$ limit at fixed $p$. We already seen this idea at work in the case of the VS, 
here the procedure is slightly more involved since we are dealing with rational functions rather than polynomials.

Given $\mathpzc{W}^{(2)}_{pppp}$ for $p=2,3,4$ \cite{to_be_decided}, 
we notice that the corresponding large $s$ and $t$ limit is controlled by a very simple 
function,\footnote{We have used the results in \cite{Aprile:2019rep}, cfr. Section 4.4 and 4.5. 
The limit here compares with the limit taken on the minimal one-loop 
functions attached in \cite{Aprile:2019rep} by a redefinition of the ambiguities, which is always possible.}
\begin{align}
\lim_{s,t\rightarrow \infty}  \mathpzc{W}^{(2)}_{2222}(s,t)&=\frac{3(s^2+ st + t^2 )^2}{ st(s+t) } \label{weight2_largest_2222}\\[.2cm]
\lim_{s,t\rightarrow \infty}  \mathpzc{W}^{(2)}_{3333}(s,t,\newsigma,\newtau)&=\frac{14(s^2+ st + t^2 )^2}{st(s+t) }(1+\newsigma+\newtau)+ 
 \beta^{(s)}_{3333}(\newsigma s +\newtau t - (s+t) ) 
\label{weight2_largest_3333}\\[.2cm]
\lim_{s,t\rightarrow \infty}  \mathpzc{W}^{(2)}_{4444}(s,t,\newsigma,\newtau)&=
\frac{42(s^2+ st + t^2 )^2}{st(s+t) }(1+\newsigma^2+\newtau^2+4\newsigma+4\newtau+4\newsigma\newtau) \\ 
&\ +
\beta^{(s)}_{4444} (\newsigma^2 s +\newtau^2 t - (s+t) )+ 
\beta^{(t)}_{4444}(\newsigma\newtau(s+t)-\newsigma t -\newtau s ) 
\label{weight2_largest_4444}
\end{align} 

Importantly, the large $s$ and $t$ limit scale linearly with $s$, and thus depends 
only on non constant stringy ambiguities. In particular, it  does not see the presence of the 
upgraded tree level function, which at most scales like $1/s^2$  \cite{Aprile:2019rep}. 

A pattern similar to that in \eqref{weight2_largest_2222}-\eqref{weight2_largest_4444} 
shows up in   in $\mathpzc{W}^{(4)}_{2222,3333,4444}$, and $\mathpzc{W}^{(3)}_{2222,3333,44444}$
if we repeat the large $s$ and $t$ analysis at fixed external charges. 
Indeed, the following general result holds \footnote{This prefactor can be written as the 
prefactor coming from the flat space limit a la Penedones:
\begin{align}
\tfrac{p(p+1)(2p-1)(2p+1)}{30}=  \tfrac{ (\Sigma-1)_4}{120}\Big|_{p_{i=1,2,3,4}=p} \qquad;\qquad\Sigma=\tfrac{p_1+p_2+p_3+p_4}{2}
\end{align}
}
\begin{align}
\lim_{s,t\rightarrow \infty } \mathpzc{W}^{w=4,3}_{pppp}(s,t,\tilde s,\tilde t)&= 
\frac{p(p+1)(2p-1)(2p+1)}{30}  f_{}^{w=4,3}(s,t) \sum_{\textit{T}} \frac{\newsigma^{\tilde s} \newtau^{\tilde t}}{ \tilde s!^2\tilde t!^2\tilde u!^2}\\
f^{(4)}&=\frac{s^2 t^2}{(s+t)^3 } \quad \quad \\[.2cm]
f^{(3)}&=\frac{   7 t^2 s^3 + 7 (s+t)^2 s^3 -3 s^5 }{  2 t^2  (s+t)^2   }
\end{align} 
This same pattern goes over the top-weight 2 function and we can match \eqref{weight2_largest_2222}-\eqref{weight2_largest_4444} with
\begin{align}
\lim_{s,t\rightarrow \infty } \mathpzc{W}^{(2)}_{pppp}(s,t,i,j)&=
\frac{p(p+1)(2p-1)(2p+1)}{30}  f^{(2)}(s,t) \sum_{\textit{T}} \frac{ \newsigma^{\tilde s} \newtau^{\tilde t}}{ \tilde s!^2\tilde t!^2\tilde u!^2} \\
f^{(2)}&= \frac{(s^2+ st + t^2 )^2}{st(s+t) } 
\end{align}  
For $\mathpzc{W}^{(4)}_{pppp}$, and $\mathpzc{W}^{(3)}_{pppp}$ we know how to 
compare the large $s$ and $t$ limit with the large $p$ limit.
Covariantising we find the large $p$ limit of $\mathpzc{W}^{(2)}_{pppp}$ to be
\begin{align}
\lim_{ p\rightarrow \infty} \mathpzc{W}^{(2)}_{pppp}(s_{cl},t_{cl},\tilde s_{cl},\tilde t_{cl})=& 
-\frac{1}{120}\times \frac{\Sigma^4}{{\bf s}_{cl}{\bf t}_{cl}{\bf u}_{cl}} \times \frac{( ({\bf s}_{cl} )^2+ ({\bf u}_{cl})^2 + ({\bf t}_{cl})^2 )^2}{4 }
\end{align}

In this formula we dropped the ambiguities, which instead we saw in  
\eqref{weight2_largest_2222}-\eqref{weight2_largest_4444}  contributing to the 
flat space limit a la Penedones.  Notice now that in the position space bootstrap 
there is an abundance of such ambiguities since they are constrained only from crossing and spin truncation. 
This is essentially how the counting in \eqref{stringy_mellin} works. However, all these ambiguities should be 
consistent with the four-fold representation, something that at fixed $p$ cannot be appreciated. 
From this point of view, our previous counting is reduced to only two $AdS_5\times S^5$ ambiguities, namely 
\begin{align}
\mathpzc{W}^{(2)}_{2222}(s,t,\tilde s, \tilde t)\Big|_{ambiguities}= \beta^{R^4} + \beta^{\partial^2 R^4} ( s\tilde s+ t\tilde t +u \tilde u )
\end{align}
In fact, the stringy origin of these ambiguities are the $R^4$ term and its derivatives 
compactified on $AdS_5\times S^5$. It is tempting to set $\beta^{\partial^2 R^4}$ to zero 
but the only argument that can be made would be to try to use 
independent results from localisation to fix it, as in the case of $\beta^{R^4}$ \cite{Chester:2019pvm}.

In the next section we collect all the various contributions to 
$\mathcal{M}^{(2)}_{pppp}$ in the large $p$ limit and we discuss the relation
with the ten-dimensional box integral in flat space.

\subsection{Gravitational S-matrix in the large $p$ limit}\label{grav_smatrix_comparison}

Let us recall the saddle point relation \eqref{saddle_AdS_S}  between 
position and Mellin space representation of the amplitude, in the large $p$ limit, 
\begin{align}\label{saddle_AdS_S_forloop}
\lim_{p\rightarrow \infty} \mathcal{A}_{}(U,V,\newsigma,\newtau)=
\frac{ 4\pi^4 }{ (U V \newsigma \newtau)^{\frac{1}{4} } }\frac{  
					(1+\sqrt{\newsigma}+\sqrt{\newtau} )^{2(p-2) +\frac{3}{2} } }{ (1+\sqrt{U}+\sqrt{V} )^{2(p+2) - \frac{3}{2} }   } \times 
p^4 \lim_{p\rightarrow \infty} \mathcal{M}( s_{cl} ,t_{cl}, \tilde s_{cl},\tilde t_{cl}) 
\end{align}
We will now show how the r.h.s. looks like at one loop.

The result for $\mathcal{M}^{(2)}$, obtained in the previous section, 
once organised by weight $w$, and divided out by the tree level amplitude becomes
\begin{align}
&
\lim_{ p\rightarrow \infty}  \mathcal{M}^{(2)}_{pppp}( s_{cl} ,t_{cl}, \tilde s_{cl},\tilde t_{cl})= 
\frac{1}{ {\bf s}_{cl} {\bf t}_{cl}  {\bf u}_{cl} }\times \Phi({\bf s}_{cl}, {\bf t}_{cl})\label{one_loop_largep_allin}
\end{align}
where
\begin{align}
\!\!
 \Phi({\bf s}_{cl}, {\bf t}_{cl})= \frac{\Sigma^4}{5!} \Bigg[ &\!
+\frac{  ({\bf s}_{cl})^3 ({\bf t}_{cl})^3  }{  ({\bf u}_{cl} )^2   } \ (\,\pi^2 +\log^2((-{\bf s}_{cl})/(-{\bf t}_{cl} )) \  + \, {\rm crossing }\,\, + \notag \\[.2cm]
& 
\!
+\frac{   7 ({\bf t}_{cl} )^2({\bf s}_{cl})^4 + 7  ({\bf u}_{cl})^2 ({\bf s}_{cl})^4-3 ({\bf s}_{cl})^6 }{  2({\bf t}_{cl}) ( {\bf u}_{cl}) } \log(-{\bf s}_{cl}  ) \, +\, {\rm crossing }\,\, + \notag\\[.2cm]
& 
\!
- \frac{( ({\bf s}_{cl})^2 + ({\bf t}_{cl})^2 + ({\bf u}_{cl})^2)^2}{4 }\ +\,{\rm crossing } \ \Bigg] \label{last_formula_box}
\end{align}

We want to compare \eqref{one_loop_largep_allin}-\eqref{last_formula_box} 
with the flat space amplitude of type IIB SUGRA in 10d, from string theory \cite{Green:1982sw,Bern:1998ug}. 
The idea of this computation was pioneered in \cite{Alday:2017vkk}, and 
here we will consider its Mellin space version, uplifted to 10d. Assembling the 
various results we find \footnote{On the stringy side, 
we refer to (4.1) (4.22) and (4.25) of \cite{Green:2008uj}. In their (4.22) 
we seem to find a $8\pi$ instead of $2\pi$. This would match formula (5.28) 
of  \cite{Alday:2017vkk} (with an $s^4I_{box}$). In formula (4.25) of \cite{Green:2008uj},
we used the identity ${\rm Li}_2(z)+{\rm Li}_2(1/z)= -\frac{1}{2}\log^2(-z)-\frac{\pi^2}{6}$.
Notice also that upon shifting $\alpha'\rightarrow \alpha'/\Lambda^2$ we would find that 
$s+t+u=0$ implies $$\log\Lambda^2\left( \tfrac{s^3 t}{u} (u-2t ) +\tfrac{t^3 s}{u} (u-2s ) + {\rm crossing} \right)=0$$
Decomposing $\log(-s)=\frac{1}{2}(\log(-s)-\log(-t)) + \frac{1}{2}(\log(-s)+\log(-t))$, 
and similaly for $\log(-t)$, it is clear that the property above is a property of the 
coefficient function for the symmetric term. }
\begin{align}
\mathcal{S}^{flat}(s,t)= 1+ &\mathcal{S}^{1-loop, flat}(s,t)+\ldots\quad \quad \\
&
\mathcal{S}^{1-loop,flat}(s,t) ={8\pi g_s^2}{}\Big( \mathbb{B}_{}(\alpha' s,\alpha' t)+ \mathbb{B}_{}(\alpha' s,\alpha' u)+ \mathbb{B}_{}(\alpha' t,\alpha' u) \Big) \notag
\end{align}
%
%
where $ \mathbb{B}_{}$ is related to the box function in 10d. In particular,  
\begin{align}\label{SIMON_10dbox}
\mathbb{B}_{}(s,t)=
 \frac{\pi}{5!2^{1+6}} \Bigg[  &
\frac{s^3 t^3}{u^2} \left( \pi^2 +\log^2 \frac{-s}{-t} \right)+ \\
&
\frac{s^3 t}{u} (u-2t )  \log(-s) +\frac{t^3 s}{u} (u-2s )  \log(-t) -s^2t^2 + C_2 st u^2\Bigg]  \notag
\end{align}
The term denoted with $C_2=1$ cancels in the sum over orientations, 
when $s+t+u=0$.\footnote{The 10d box presented in \cite{Alday:2017vkk} 
includes also the quadratic divergence, restored after dim-reg.} 

By looking at $\mathcal{S}^{1-loop,flat}(s,t)$ and  comparing with 
$\lim_{ p\rightarrow \infty}  \mathcal{M}^{(2)}_{pppp}( {\bf s}_{cl} ,{\bf t}_{cl})$ 
we find the relation
\begin{mdframed}
\begin{align}
\lim_{ p\rightarrow \infty}\left( \frac{1}{N^2} \mathcal{M}^{(2)}_{pppp}( s_{cl} ,t_{cl}, \tilde s_{cl},\tilde t_{cl})\right)=  
\frac{1}{ {\bf s}_{cl} {\bf t}_{cl} {\bf u}_{cl} } 
\mathcal{S}^{1-loop, flat}\left( \frac{4\Sigma }{R^2} {\bf s}_{cl}, \frac{4\Sigma }{R^2}{\bf t}_{cl}  ;g_s\right)\\
\notag
\end{align}
\end{mdframed}
with $R^4/\alpha'^2 =4\pi g_s N$. 
This is the same identification of 10d Mandelstam variables we found in the VS amplitude, 
and again agrees with our expectations from the picture of four propagating 
geodesics towards a common bulk point.   Thus we have showed the coincidence 
between the large $p$ limit of the $AdS_5\times S^5$ amplitude and the 
10d flat space scattering amplitude, at one-loop.

The scaling with $p$ of the Mellin amplitude can be generalised to the $\ell$-loop 
order, and leads to the following scheme:
\begin{align}
&
 \lim_{p\rightarrow \infty} \mathcal{M}_{\vec{p}}^{\ell-loop} ( s_{cl} ,t_{cl}, \tilde s_{cl},\tilde t_{cl}) =  
 \frac{1}{ {\bf s}_{cl} {\bf t}_{cl} {\bf u}_{cl}  }
 \times  
 \left( \frac{\Sigma^4 {\bf s}_{cl}^4 }{N^2} \right)^{\!\!\ell}  
 \Bigg[  \notag\\
&
\rule{.5cm}{0pt}
 \sum_{0\leq d\leq 2\ell} \texttt{lim-Transcendetal}_{\ell,d}({\bf s}_{cl} ,{\bf t}_{cl} ) \times 
 					\texttt{Rational}_{\ell,d}\left( \frac{ {\bf t}_{cl} }{ {\bf s}_{cl} }\right)\, +\, {\rm crossing} \Bigg]
 \label{scheme_all_loops}
\end{align}
where $\texttt{lim-Transcendetal}$ are the trascendental functions appearing in 
the flat space scattering amplitude, which come from the limit  of the $\texttt{Transcendetal}$ 
functions in the $AdS_5\times S^5$ Mellin amplitude, when the Mellin variables are large.  
The latter are meromorphic, with poles dictated by the OPE, while the first ones generically 
have also branch cuts, related to the physics of the 10d scattering amplitude. From the 
$AdS_5\times S^5$ point of view these branch cuts arise from the accumulation of the poles 
when the Mellin variables are large. The transcendental weight goes down by two in the process, 
because of $\Gamma_{\otimes}$ which plays a different role, compared to 
the \emph{reduced transcendental kernels}. It is quite interesting to notice that, 
differently from perturbation theory where the transcendental content of the amplitude 
has fixed weight at given loop order,  at strong coupling all lower weights are turned on, 
and there is a correspondence between transcendental functions across dimensions. The rational functions denoted schematically by $\texttt{Rational}$ in \eqref{scheme_all_loops}, 
have  degree zero under the large $p$ limit. For example at  one-loop we have found
\begin{align}
\lim_{p\rightarrow \infty} \mathpzc{W}_{pppp}^{(4)}(s_{cl},t_{cl},\tilde s_{cl},\tilde t_{cl})    \sim  \frac{1}{ {\bf s}_{cl} {\bf t}_{cl} {\bf u}_{cl} } \times p^4 {\bf s}_{cl}^4\times 
\left[ \texttt{Rational}_{1,4}= \frac{  ({\bf t}_{cl})^3 }{ ({\bf s}_{cl}) ({\bf u}_{cl})^2 } \right]
\end{align}
and similarly for $\mathpzc{W}^{(3)}_{pppp}$ and $\mathpzc{W}^{(2)}_{pppp}$.\footnote{In appendix \ref{two_loop_exercise} we provide a 
test of our scheme for the case of the leading logs at two-loop.} 

Finally, note that since both $\Sigma$ and the bold font variables scale with $p$ we see that the genus $\ell$ contribution comes dressed by an effective $(p^8/N^2)^\ell$ coupling. For us, since $N$ is the largest parameter, this is always small. In some recent explorations discussed in the next section, $p \sim N^{1/2}$; this effective coupling would be large and we would need to fully resum the non-planar series.

\section{Discussion}\label{discussion_sec}

In this paper we studied the large $p$ limit of the four-point correlators 
$\langle \mathcal{O}_{p_1} \mathcal{O}_{p_2} \mathcal{O}_{p_3} \mathcal{O}_{p_4}\rangle $ 
of single-particle operators in $\mathcal{N}=4$ SYM in the large $N$ limit 
at strong 't Hooft coupling.  We considered different regimes of $p$ having in 
mind the greater picture illustrated by figure~\ref{regimes}. The idea is to interpolate
from the top left corner (AdS SUGRA) to the bottom right corner (AdS minimal areas), 
and see how the interpolation goes.  In this paper we took a first step and we focused mostly 
on perturbative string theory, pushing SUGRA results up to the flat space Gross-Mende phase, 
where the nature of string theory as an extended object  appears for the first time, but 
does not yet backreact on the $AdS_5\times S^5$ geometry.


When $p=O(\sqrt{\lambda})$ the minimal surfaces are big, occupying a finite 
fraction of the AdS space. Intuitively, the angular momentum of the string, 
which is of order $p$, is comparable to the string tension~$\sqrt{\lambda}$, 
thus the fight between centripetal forces and tension leads to a big macroscopic string. 
Decreasing $p$ the string tension starts to win and shrinks the string. Thus,
as we approach $p=O(\lambda^{1/4})$ we expect the various strings 
to become point-line particles flying from the AdS boundary towards 
a small interaction region in AdS,  as represented in the middle panel in the bottom 
row of Figure \ref{regimes}.  These four almost point-like strings will then enter the 
small interaction region with very large energy.  Interactions there are described by high 
energy (at fixed angle) string theory as studied by Gross-Mende \cite{Gross:1987kza}, and
the leading contribution to the string amplitude will now be given by a \textit{flat space} minimal area.

A direct study of the AdS minimal area is hard, due to the non-linearities on the world-sheet, but  
in recent years an integrability based technology  has been developed precisely for computing the \textit{action} 
of minimal surfaces \textit{without} ever finding the shape of these surfaces. This is fully developed 
for Scattering Amplitudes/Wilson loops \cite{WLAreas} and three point functions of local operators \cite{3ptAreas}.
In the case of four-point functions, the work of~\cite{Caetano:2012ac} has shown how to 
determine  the action of the AdS minimal surfaces when the four points are aligned along 
a single line in the AdS boundary.  Generalising this result would lead to the AdS minimal area 
we are looking for. A complementary question to ask is how the minimal 
area opens from the flat configuration at $p=O(\lambda^{1/4})$ up to occupy a finite fraction of the AdS space when $p=O(\lambda^{1/2})$. 
We hope to report soon on these problems and determine the full minimal areas depicted in the corner of Figure \ref{regimes}. 

In this context we would like to make a few comments on some very interesting correlation function 
developments in $N=4$ SYM  arising from Integrability explorations and commonly known as the \textit{octagon} correlation functions. 
The octagon is a four-point correlator of very large external weights computed by Frank Coronado in perturbation 
theory \cite{Coronado:2018ypq} by using the hexagonalization technology \cite{Fleury:2016ykk,Hexagons}, 
re-derived through a boostrap program \cite{Coronado:2018cxj}, recast as a compact infinite dimensional 
determinant in \cite{det} and compactly described as a solution to a simple set of integro-differential equations 
in \cite{Belitsky:2019fan,Belitsky:2020qrm}. In \cite{Bargheer:2019kxb} it was shown how the octagon also 
computes the full non-planar expansion of particular four point correlation functions of large weights $p$ provided we scale $p 
\sim \sqrt{N}$ (which is of course much larger than anything discussed so far in this paper). At strong coupling 
the octagon was studied in \cite{Bargheer:2019exp,Belitsky:2020qrm,Belitsky:2020qir}. In \cite{Bargheer:2019exp} 
it was shown that the hexagon exponentiates at strong coupling and it was conjectured that the exponent is 
nothing but the associated minimal area. So far this area was not independently computed from the worldsheet. 
In \cite{Belitsky:2020qrm,Belitsky:2020qir} Belitsky and Korchemsky beautifull understood how to systematically 
extract any term in the strong coupling expansion.

Would be fantastic to compare those strong coupling results with our SUGRA extrapolations. As of now that is 
not possible for a few reasons which would be very interesting to overcome. 
The first reason concerns the fact the octagon is related to a correlator whose external weights are very large, 
much larger than anything considered in this paper and in particular much larger than $\lambda^{1/4}$ which 
is as far as we can confidently reach starting from SUGRA. Would be very interesting to compute the associated 
minimal area for the octagon with weights $p \sim \sqrt{\lambda}$ and take $p/\sqrt{\lambda}$ to be very small 
to hopefully interpolate between AdS minimal areas and the Gross-Mende flat space areas alluded to in this work. 
The second reason is more trivial: The octagon corresponds to a specific polarisation of the single-trace correlator 
on the $S^5$, i.e.~it corresponds to the dynamics  of a particular propagator {structure},  thus it is not quite 
the amplitude $\mathcal{A}$ we studied. Would be interesting to repeat the analysis in this work for correlators 
with fixed polarizations instead of fixed $R$-charge cross-ratios. 

Let us nonetheless note that there is a very nice aspect of the SUGRA results which is Taylor made to allow 
for a smooth interpolation with the octagon strong coupling findings. One aspect of the strong coupling expansion is the scaling with 
$\sqrt{\lambda}$ of the correlator, which follows from the simple universal $A_1=1$ result of \cite{Belitsky:2019fan,Belitsky:2020qrm} once we put together
the results 
\begin{align}\label{octagon_explanation}
\langle T_p T_p T_p T_p\rangle\Big|_{\texttt{octagon config}} \sim\ &[\mathbb{O}(U,V)]^2\\
&\ \mathbb{O}(U,V) =(\sqrt{\lambda})^{A_1^2/2} e^{-g A_0(U,V)+B+ O(1/g)} \quad;\quad A_1=1
\end{align}
We match this scaling by considering the universal scaling with $p$ which follows from our master relation \eqref{saddle_AdS_S}, namely
\begin{align}
\!\!\!
\lim_{p\rightarrow \infty} \mathcal{A}_{\vec{p}}(U,V,\newsigma,\newtau)=
\frac{ 4\pi^4 }{ (U V \newsigma \newtau)^{\frac{1}{4} } }\frac{  
(1+\sqrt{\newsigma}+\sqrt{\newtau} )^{2(p-2) +\frac{3}{2} } }{ (1+\sqrt{U}+\sqrt{V} )^{2(p+2) - \frac{3}{2} }   } \!
 \times\!
 \frac{p^4}{ {\bf s}_{cl} {\bf t}_{cl} {\bf u}_{cl} }\bigg[1+O(1/N^2,\lambda^{-1/2}) \bigg]
\end{align}
Notice now that ${\bf s}_{cl}$ and ${\bf t}_{cl}$ in \eqref{solu_pppp_AdS5}-\eqref{solu_pppp_S5}, 
are function of cross ratios, but scale linearly with $p$. Thus the overall scaling of the amplitude goes with $p$. 
In the regime $p=O(\sqrt{\lambda})$ we recover precisely the scaling of the octagon.  

It would also be fascinating to compare our findings with the octagon results of \cite{Bargheer:2019kxb} 
where the octagon shows up as a building block in a completely re-summed large $N$ correlation function. 
One obstacle here pertains to the difference between single-trace operators (used in \cite{Bargheer:2019kxb}) 
and single-particle operators (which we are using here). In particular, our connected correlator is supposed to 
be exponentially small (as arising from the geodesic propagation) at any order of the $1/N$ expansion while 
the octagon correlator studied in~\cite{Bargheer:2019kxb} goes to a constant result as $\lambda \to \infty$. 
We believe this difference is precisely accounted by the change of basis we just mentioned and look forward 
to investigating this further with Till Bargheer and Frank Coronado.


The regimes of $p$ in which $p/\lambda^{1/4}\ll 1$ are regimes in which SUGRA, 
with higher derivative corrections coming from string theory, provides a valid low energy 
descriptions of the physics.  In this framework, our large $p$ limit  
has unveiled a piece of the greater puzzle offered by Figure \ref{regimes}. 
We started from the Mellin description of the correlators \cite{Rastelli:2016nze},  
and we improved that by introducing the $AdS_5\times S^5$ four-fold representation, 
built around $\Gamma_{\otimes}$, such that all Mellin variables (and cross ratios) are treated equally.
We then took the large $p$ limit and we identified two types of contributions. 
From $\Gamma_{\otimes}$ we found the saddle point configuration of four 
geodesics shooting towards a common bulk point.  From the bulk Mellin amplitude 
we got the IIB S-matrix in flat space. Furthermore we understood the identification 
between geometric data and Mandelstam invariants of the approximately ten-dimensional physics at the bulk point.

Then, the large $p$ limit guided us towards a more precise understanding of the full $AdS_5\times S^5$ amplitude.
In particular, we understood how to covariantise the flat space limit a la Penendones, 
and how to stratify the Mellin amplitude in a large $p$ expansion.  We demonstrated these 
two points by studying the Virasoro-Shapiro and the one-loop amplitudes, i.e~all the 
data currently available. For the VS amplitude we pointed out additional simplicity of the known results. 
At one-loop we have constructed a new finite Mellin amplitude, in correspondence 
with the position space bootstrap of \cite{Aprile:2017bgs,Aprile:2017xsp,Aprile:2017qoy,Aprile:2018efk,Aprile:2019rep}.
 Our findings put in place a more general scheme for studying the $AdS_5\times S^5$ 
 correlators in the future.


The equality between the 10d flat space VS amplitude and the large $p$ limit of the 
$AdS_5\times S^5$ genus zero amplitude, which is largely unknown, is the first aspect 
of the large $p$ limit we worked out. Understanding a similar statement at one-loop, 
i.e. going beyond the 10d box in SUGRA by adding one-loop stringy corrections would 
be highly desirable. These corrections  are known in the form of stringy integrals and it 
would be fascinating to connect those integrals with the Mellin space representation of the amplitude, directly.
This line of thoughts goes along a similar reinterpretation of classic stringy computations
under the light of new localisation results \cite{Chester:2019pvm,Chester:2020dja}. 
In this sense, it would be even more intriguing to make contact with the generating 
techniques for closed and open string amplitudes recently developed in \cite{Mafra:2019ddf,Gerken:2019cxz}.

A number of results are now available at tree level on $AdS\times S$ spaces. 
The case of $AdS_3\times S^3$ \cite{Rastelli:2019gtj,Giusto:2019pxc,Giusto:2020neo} 
is likely to enjoy hidden-symmetric properties analogous to $AdS_5\times S^5$, 
and it would be a natural place where to explore the large $p$ limit in the way we did in this paper. 
At one loop the six-dimensional box is even simpler, and it could be the right starting 
point for bootstrapping the $AdS_3\times S^3$ one-loop amplitudes.

Very recently, the authors of  \cite{Alday:2020lbp,Alday:2020dtb} conjectured 
all tree level amplitudes for the maximally symmetric theories in $d=3$ and $d=6$. 
Amplitudes in these dimensions are more likely to be attached to the various propagator structures, 
since it is yet unclear what superconformal symmetry manifests into -- for example long superconformal blocks 
do not show any \texttt{kinematic} common factor -- However, as we discussed in this paper, 
we expect all fundamental properties of flat M-theory to emerge in the large $p$ limit.
As an anticipation of this expectation, \cite{Alday:2020lbp,Alday:2020dtb} pointed out 
the emerging of nice features in the flat space limit a la Penedones, 
specifically in what they call the R-symmetry polynomial in combination with the universal 
$1/stu$ term, which appears to be the same as in $AdS_5\times S^5$.
As we learned in this paper, this is the starting point towards the large $p$ limit.

Kid A and kid B want to dig a big solid tunnel in the beach. Each one starts digging a tunnel entrance. 
They seem so far at first. They dig and dig and  at some point they feel each other's finger tips, 
barely touching as they scratch the sand digging from either side. 
They stand and look through each tunnel entrance to see a beautiful tiny light ray coming from the other side. 
The kids smile. At this point they know they will succeed. It is just a matter of digging carefully 
and steadily. Kid A is Integrability. Kid B is SUGRA.

\section*{Acknowledgements}
We would like thank  Till Bargheer, Nathan Berkovits, Frank Coronado, Vasco Goncalves, Andrea Guerrieri,
Shota Komatsu, Joao Penedones, Amit Sever for several enlightening discussions. 
FA would like to thank James Drummond and Paul Heslop, Hynek Paul and Michele Santagata, 
for collaboration on related topics, and especially James Drummond for pointing 
out ref.~\cite{Allendes:2012mr} in the 2017.
FA acknowledges the ICTP South American Institute for Fundamental Research (ICTP-SAIFR) 
for hospitality and financial support during the period this work started. 
FA is partially supported by the ERC-STG grant 637844- HBQFTNCER. 
Research at the Perimeter Institute is supported in part by the Government of Canada 
through NSERC and by the Province of Ontario through MRI. This work was additionally 
supported by a grant from the Simons Foundation (PV: \#488661) and FAPESP 
grants 2016/01343-7 and 2017/03303-1.

\appendix

\section{Holographic correlators and the large $p$ limit}\label{largeplimit_sec}

\subsection{Conventions on Mellin and Mandelstam variables} \label{appendixNotation}

We will define first the Mellin transform of a generic four-point function,  
\begin{align}
\langle\mathcal{O}_{p_1}(x_1)\mathcal{O}_{p_2}(x_2) \mathcal{O}_{p_3}(x_3) \mathcal{O}_{p_4}(x_4) \rangle =
\oint \mathcal{M}( c_{\alpha\beta}) \prod_{1  \leq i<j\leq 4}^4\frac{ \Gamma[c_{\alpha\beta} ]}{ \ \ (x_{i}-x_{j})^{2c_{\alpha\beta} } }
\end{align}
The Symanzik variables $c_{\alpha\beta}=c_{\beta,\alpha}$ 
(with $c_{\alpha\alpha}=0$) are constrained requiring 
\begin{align}
\sum_{\alpha \neq \beta} c_{\alpha\beta} =p_\beta 
\end{align}
and can be solved by analogy with Mandelstam variables, by 
defining $c_{\alpha\beta}=c_{\alpha\beta}(s,t,\vec{p})$, where the integration 
variables are $s$, and $t$, and a single constraint gives $u$. 
In Mellin space it is then natural to have $-s=c_{12}$ to appear in the 
argument of the $\Gamma$ functions, similarly for $t$ and $u$.

The S-matrix for massless scattering is by construction a function only of the physical Mandelstam invariants, 
and since we are considering a $2\rightarrow 2$ process, we find ${\cal S}^{flat}={\cal S}^{flat}(k_s,k_t)$ where
\begin{align}
k_s=-(k_1+k_2)^2;\qquad k_t=-(k_1+k_4)^2; \qquad k_u=-(k_1+k_3)^2 
\end{align}
The quantities $k_s$ and $k_t$ are the physical $s$ and $t$ Mandelstam invariants, but
these are not the same as the Mellin space variables. 
In the above conventions, the flat space formula of Penedones \cite{Penedones:2010ue}, 
reads, 
\begin{align}\label{alapenedones}
\mathcal{M}( s,t)= C \int_0^{\infty} d\beta \beta^{\Sigma-\frac{d}{2}-1} e^{-\beta} \mathcal{S}_{}^{flat}\left( k_s=\frac{4\beta}{R^2} s,k_t=\frac{4\beta}{R^2} t  \right)
\end{align}
To compare with \cite{Penedones:2010ue} notice that 
$\delta_{ij}^{there}=c_{\alpha\beta}^{here}$ and $s_{ij}^{there}$ scales like $2s^{here}$.

\subsection{Conventions on SUGRA amplitudes} \label{appendixNotation}

Four-point functions of half-BPS operators admit a splitting into free and dynamical contributions, according to the 
partial-non-renormalisation theorem \cite{Eden:2000bk}, as we now review.

Free theory four-point correlators are given by a sum over propagator 
structures weighted by the corresponding color factor $a_{\gamma,k}$. 
Without loss of generality we consider $p_{43}\ge p_{21}\ge 0$, 
and write the free theory contribution as 
\begin{align}
\!\!
\< \cO_{p_1}(x_1)\dots \cO_{p_4}(x_4)\>_{ free}= \mathcal{P}( \{ g_{ij}\} ) 
\sum_{ \substack{ p_{43} \leq \gamma_{12} \\[.1cm] 0 \leq \gamma_{23} \leq \gamma_{12}}  }^{} a_{\gamma_{12},\gamma_{23}} \left( 
\frac{ g_{13} g_{24} }{ g_{12} g_{34} }  \right)^{\!\! \frac12(\gamma_{12}-p_3-p_4)} \left(\frac{ g_{14} g_{23} }{ g_{13} g_{24} }\right)^{\!\!\gamma_{23}} \notag
\end{align}\\[-.2cm]
where
\begin{align}
 \mathcal{P}\!=\! g_{12}^{\frac{ p_1+p_2-p_{3}-p_4 }{2} } g_{14}^{\frac{p_1+p_{4}-p_3-p_{2}}{2} }g_{24}^{\frac{ p_{2}+p_{4}-p_1-p_3 }{2} } (g_{13}g_{24})^{p_3} 
\end{align}
Here $\gamma_{12}$ counts the number of bridges going from $\cO_{p_1}(x_1)\cO_{p_2}(x_2)$ to $\cO_{p_3}(x_1)\cO_{p_4}(x_2)$. 
The first diagram, i.e. $\gamma_{12}=p_{43}$ with $\gamma_{23}=0$, 
is the diagram where all bridges of $\mathcal{O}_{p_3}$ link with $\mathcal{O}_{p_4}$, 
therefore there are $p_{43}$ bridges going from $\mathcal{O}_{p_4}$ t
o the pair $\mathcal{O}_{p_1}\mathcal{O}_{p_2}$. 

Cross ratios will be parametrised as follows, 
\begin{align}
\frac{ g_{13} g_{24} }{ g_{12} g_{34} } = \frac{U}{\sigma} \quad ;\quad 
\frac{ g_{14} g_{23} }{ g_{13} g_{24} }= \frac{\tau}{V} \quad;\quad
\begin{array}{lll}
u=x_1 x_2 &;\quad \sigma=y_1 y_2 \\
v=(1-x_1)(1-x_2) &;\quad \tau=(1-y_1)(1-y_2) 
\end{array}\label{crosseratios}
\end{align}

The variables $\sigma$ and $\tau$ are those coming from analytic superspace 
(see \cite{Doobary:2015gia} for a more recent discussion). We will start from 
these, since it is useful to match conventions, and we will move to our 
$\tilde U=\sigma$ and $\tilde V=\tau$ along the way.

The dynamical contribution inherits from superconformal symmetry a specific structure \cite{Eden:2000bk},
\begin{align}
\< \cO_p(x_1)\dots \cO_p(x_4)\>_{ dynamical} &=  
\prod_{1\leq i,j\leq 2}(x_i-y_j)\times \mathcal{P}(\{ g_{ij} \})\times \mathcal{D}_{\vec{p}}(U,V,\sigma,\tau) \label{correlator}
\end{align}

In the strong 't Hooft coupling regime, the theory lives on a classical $AdS_5\times S^5$
and the quantum corrections are organised as a double expansion in 
$1/N^2$ and $\lambda^{-1/2}$. In particular, the dynamical function scales like a connected correlator.
In this regime we define the amplitude $\mathcal{A}_{\vec{p}}$ of the correlator as
\begin{align}
\mathcal{D}_{\vec{p}}(U,V,\sigma,\tau)&=\frac{p_1p_2p_3p_4}{N^2} \mathcal{A}_{\vec{p}}(u,v,\sigma,\tau)  
\end{align}

The amplitude $\mathcal{A}$ encodes all the non-trivial information 
about the dynamics and can be written as a double integral and a double sum,
\begin{align}
\mathcal{A}_{\vec{p}}(u,v,\sigma,\tau)& \equiv   \iint \! ds dt\, u^s v^t \,\sum_{\textit{T} } \sigma^{-i-j+p_3-2} \tau^j  \times \Gamma(s,t,i,j) \times \mathcal{M}_{\vec{p} }(s,t,i,j)
\end{align}
where
\begin{gather}
\!\!\!\!\Gamma_{}(s,t,i,j) = \frac{ \Gamma[-s]\Gamma[-t]\Gamma[-u]\Gamma[-s+c_s\,]\Gamma[-t+c_t]\Gamma[-u+c_u]}{k!j!i! (k+c_s)! (j+c_t)! (i+c_u)!}\\[.2cm]
c_s=\frac{ p_1+p_2-p_3-p_4}{2}\quad;\quad c_t=\frac{p_1+p_4-p_2-p_3}{2} \quad;\quad c_u=\frac{ p_2+p_4-p_3-p_1}{2}
\end{gather}
The definition of $k$ follows from the relation $i+j+k=p_3-2$. The definition of $u$ is
\begin{align}
s+t+u=-p_3-2 
\end{align} 
The double sum over $i$ and $j$ runs over the set of integers,
\begin{align}
\textit{T}=\{ \ i\ge0\ ;\  j\ge 0 \ ;\ \ i+j \leq \kappa-2\  \}\qquad \kappa=\tfrac12(  min(p_1+p_2,p_3+p_4)-p_{43} )
\end{align}
where $\kappa$ is the degree of extremality. Notice that $\kappa=min(c_s,0)+p_3$, 
therefore $ i+j \leq \kappa-2$ iff $k\ge max(-c_s,0)$, and $k!(k+c_s)!$ is well defined. 
In our conventions $c_u,c_t\ge 0$ but $c_s$ can have both signs. 
In the main text we changed variables from $i$ and $j$ to the new variables
\begin{align}
\tilde s= -i-j+p_3-2\quad;\quad \tilde t = j \quad;\quad \tilde s+\tilde t +\tilde u = p_3-2
\label{change_var}
\end{align}
In these new variables $s$ is aligned with $\tilde s$ and the gamma factor becomes
\begin{align}\label{newGammatimes}
\Gamma(s,t,\tilde s, \tilde t\,) = \frac{  \Gamma[-s]\Gamma[-t]\Gamma[-u]\Gamma[-s+c_s\,]\Gamma[-t+c_t]\Gamma[-u+c_u]}{ 
\Gamma[1+\tilde s] \Gamma[1+\tilde t]\Gamma[1+\tilde u] \Gamma[1+\tilde s + c_s]\Gamma[1+\tilde t + c_t] \Gamma[1+\tilde u + c_u] }
\end{align}
The triangle is now
\begin{align}
\textit{T}=\{ \ \tilde s\ge max(0,-\tfrac12(p_1+p_2-p_3-p_4))\ ;\  \tilde t\ge 0 \ ;\ \ \tilde u\ge 0\  \}
\end{align}
The two cases of $\tilde s$ have to do with the two possible inequalities 
$c_s\ge s\ge 0$ or $\tilde s\ge -c_s\ge 0$, which depend on $max(p_1+p_2,p_3+p_4)$. 
This is the only freedom left in our conventions. 

We obtain the four-fold representation of the amplitude by going from the 
discrete sum over $T$ to a Mellin integral over $\tilde s$ and $\tilde t$. 
The inequalities which defines $T$ are implemented by the positivity of the $\Gamma$ 
functions in the denominator of $\Gamma(s,t,\tilde s,\tilde t)$, and thus we 
only need to pick a domain where to insert simple poles. There are various choices, and here 
we consider
\beq 
\mathcal{A}_{\vec{p}}(U,V,\newsigma,\newtau)= \iint \! ds dt \iint \! d\tilde s d\tilde t \  \, U^s V^t \newsigma^{\tilde s} \newtau^{\tilde t} \times 
\,\Gamma_{\otimes} \times \mathcal{M}_{\vec{p}}(s,t,\tilde s,\tilde t\, )  
\eeq 
where now 
\beq
\begin{array}{c}
\displaystyle
{\Gamma}_{\otimes}= \mathfrak{S} \
\frac{ \Gamma[-s]\Gamma[-t]\Gamma[-u]\Gamma[-s+c_s\,]\Gamma[-t+c_t]\Gamma[-u+c_u]}{ 
\Gamma[1+\tilde s] \Gamma[1+\tilde t]\Gamma[1+\tilde u] \Gamma[1+\tilde s + c_s]\Gamma[1+\tilde t + c_t] \Gamma[1+\tilde u + c_u] } \\[.7cm]
\displaystyle
\quad \mathfrak{S}=\pi^2 \frac{\ (-)^{\tilde t}(-)^{\tilde u} }{\sin(\pi \tilde t\, )\sin(\pi \tilde u)} 
\end{array} \label{GammaUnbalanced} 
\eeq
In the four-fold representation we turned to the more appropriate $\sigma=\tilde U$ and $\tau=\tilde V$.

Below we relate $\Gamma_{\otimes}$ to the OPE expansion. 
{First, notice that double poles of $s$ and $t$ occur respectively for $s\ge max(0,c_s)$, i.e. $U^{max(0,c_s)}$, and $t \ge c_t\ge 0$. }

\subsection{OPE view on the Mellin amplitude}\label{OPE_mellin_app}

Shifting the AdS Mellin variables to align double poles in $s$ and $t$ to lie 
on the positive real axis,  we find that the $\log^1{ u }$ discontinuity of the dynamical 
function has a Taylor expansion in small $u$ and small $v$ starting with leading powers \footnote{The 
l.h.s  normalisation ${ U}^{\frac{p_{43}}{2}+p_3}$  is due to the form of the superblocks, and the prefactor in \eqref{correlator}.}
\begin{align}\label{sample_disc}
{ U^{\frac{p_3+p_4}{2} }\ }\mathcal{D}^{(1)}_{p_1p_2p_3p_4}(U,V,\newsigma,\newtau)\Big|_{\log^1 { u }} =
{ U }^{\frac{ { max }(p_1+p_2,p_3+p_4)}{2} }  { V }^{\frac{p_{43}-p_{21}}{2} }( \mathcal{A}_{0,0}(\newsigma,\newtau) +\ldots  
\end{align}
The value of ${max}(p_1+p_2,p_3+p_4)$ is the \emph{threshold twist}  for exchange of long 
two-particle operators in the common OPE $(\mathcal{O}_{p_1}\times\mathcal{O}_{p_2})\cap(\mathcal{O}_{p_3}\times\mathcal{O}_{p_4})$. 
The power of ${ V }$ follows from crossing.  
To see this threshold twist consider the behaviour of the three point functions of two external 
single-particle operators $\mathcal{O}_{p_i}\mathcal{O}_{p_j}$  with a two-particle operator 
$\mathcal{K}$ of twist $\twist$ in the rep $[aba]$ of $su(4)$, as function of the twist $\twist$ 

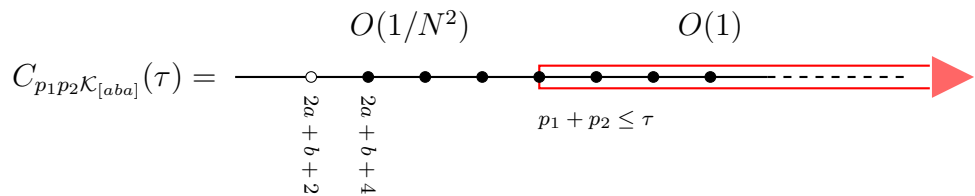
\begin{center} 

             \begin{tikzpicture}
             
             \def \originx {3}
             \def \originy {2}
             \def \lato {7}

            \draw (\originx-.1,	\originy-.05) 	node[left] 	{$C_{{p_1}{p_2} \mathcal{K}_{[aba]}}(\twist)=$};

             \draw             [thick, red]    (\originx+ 1 +.75*4, \originy+.15) 	rectangle 		(\originx+ 1 +.75*11, \originy-.15);
             \filldraw		 [thick, fill=red!60,draw=white]    (\originx+ .9 +.75*11, \originy-.3) -- (\originx+ .9 +.75*11, \originy+.3) -- (\originx+ 1.5 +.75*11, \originy)  -- cycle;
             
              \draw  (\originx+ 1+.75*5, \originy-.3)		node[below,font=\scriptsize] 	{$p_1+p_2\leq \twist$};
              \draw  (\originx+ 1.25+.75*0, \originy-.9)	node[below,font=\scriptsize,rotate=-90] 	{$2a+b+2$};
              \draw  (\originx+ 1.25+.75*1, \originy-.9)	node[below,font=\scriptsize,rotate=-90] 	{$2a+b+4$};

              \draw  (\originx+ 1+.75*7, \originy+.3)	node[above] 	{$O(1)$};
              \draw  (\originx+ .8+.75*2, \originy+.3)		node[above] 	{$O(1/N^2)$};

             \draw[thick]   (\originx,\originy) --  (\originx+\lato,\originy);
             \draw[thick,dashed]   (\originx+\lato,\originy) --  (\originx+\lato+1.8,\originy);

           	  \filldraw [fill=white,draw=black] (\originx+ 1,\originy) circle (2pt);
           	  \foreach \x in {1,...,7}
			\filldraw [fill=black, draw=black] (\originx+ 1 +.75*\x,\originy) circle (2pt); 

         	   \draw [] (\originx+ 1 +.75*12,\originy) node {$\phantom{s}$}; 
            
              \end{tikzpicture}
\end{center}

The OPE has the following implications:

{\it 1}. Two particle operators with twist above the threshold have an $O(1)$ three 
point functions in disconnected free theory. 

{\it 2}. Long two-particle operators acquire a $1/N^2$ anomalous dimension 
\cite{Aprile:2019rep} and have an $O(1)$ three point functions in disconnected free theory. 
Therefore they build up the leading logarithmic discontinuity \eqref{sample_disc}, 
when they are exchanged at tree level, above threshold.

{\it 3}. Other CFT data can be extracted from the window 
$min(p_1+p_2,p_3+p_4)\leq \twist< max(p_1+p_2,p_3+p_4)$. 

{\it 4}. The below-window region is bounded by the unitarity bound, 
which for a given rep $[aba]$ of $su(4)$ is $\twist_{[aba]}=2a+b+2$. 
The first contributions below window are of order $1/N^4$.

In fact, simple poles in $s$ and $t$ in $\mathcal{M}^{(1)}= 1\big/({\bf s}+1)/({\bf t}+1)({\bf u}+1)$ 
are known to contribute precisely at the unitarity bound $\twist_{[aba]}=2a+b+2$, and 
cancel the corresponding contribution coming from connected free theory, in the basis 
of superblock \cite{Dolan:2004iy,Dolan:2006ec,Aprile:2019rep}. We see now that 
all these simple poles correspond to the locus $\mathbf{s}=-1$ and $\mathbf{t}=-1$ 
projected onto the planes $s$ and $t$, for fixed $\tilde{s}$ and $\tilde{t}$. 
More precisely, a point $(\tilde s, \tilde t)$ in the triangle ${\it T}$,
\begin{align}\label{appe_tildestildettildeu}
{\it T}=\{ \tilde t\ge 0 \ ;\ \tilde u \ge 0 \ ;\   \tilde t+\tilde u\le min(0,c_s)+p_3-2\ \}\quad;\quad \tilde s +\tilde t +\tilde u =p_3-2
\end{align}
gives a simple pole in $s$, from $\mathbf{s}=-1$, which corresponds 
to a contribution at twist $\twist=p_3+p_4-2-2\tilde s= p_{43}+2+2(\tilde t+ \tilde u)$. In the second equality, 
we used the relation in \eqref{appe_tildestildettildeu}. Equating this value to $\tau_{[aba]}$ we find that 
all possible contributions at the unitarity bound $\tau=\twist_{[aba]}$ are filled in 
\begin{align}
\tau_{[aba]}=2(\tilde t+\tilde u )+2 +p_{43}\qquad\leftrightarrow\qquad a+\tfrac12(b-p_{43}) =\tilde t + \tilde u\in{\it T}
\end{align}
because $a=0,\ldots \kappa-2$ and $b=p_{43},\ldots,  p_{43}+2(\kappa-2)-2a$, with 
$\kappa= min(0,c_s)+p_3$ the degree of extremality we already encountered. 

\section{Mellin Saddle Point Details}  \label{appendixsaddleAdS}

\subsection*{B.1\,\,\,\,Saddle point on $AdS_5$}

In the supergravity regime the Mellin amplitude does not exponentiate 
and the saddle factorises into $AdS_5$ and the $S^5$ contribution, since it only comes from $\Gamma_{\otimes}$.

Let us begin from $AdS_5$ saddle. Matching the arguments of $\Gamma_{\otimes}$, we define 
\begin{gather}
\label{H_AdS5_Cnm}
\mathcal{H}_{AdS_5}(s,t,\vec{p}\,)= U^{s} V^{t}  \prod_{1 \leq \alpha< \beta \leq 4} \Gamma[c_{\alpha\beta}(s,t,u,{p}_i+2)]
\end{gather}
where $\sum_{\alpha=1}^4 c_{\alpha\beta}(s,t,p_i)= p_\beta$ are Symanzik variables. Written in this fashion, the study 
of the $AdS_5$ saddle is a $\overline{D}$ generalisation of a classic study for hypergeometrics \cite{Mellin_book}.

The large $p$ limit is taken by rescaling $\{s,t,p_{i=1,2,3,4}\}=n\{s^*,t^*,d_{i=1,2,3,4}\}$ and letting $n\rightarrow \infty$.
The limiting behaviour of the Symanzik variables has the following form
\begin{align}
\begin{array}{ccc}
c_{\alpha\beta}(s,t,{p}_i+2)\quad &\rightarrow& \quad n c_{\alpha\beta}(s^*,t^*,{d}_i)+ r_{\alpha\beta}\\[.2cm]
\Gamma[c_{\alpha\beta}(s,t,{p}_i+2) ]\quad &\rightarrow &\quad \sqrt{2\pi}\ e^{-n c_{\alpha\beta} } \, (n c_{\alpha\beta} )^{ n c_{\alpha\beta} + r_{\alpha\beta} -1/2} 
\end{array}
\label{asymptotics_Gamma}
\end{align}
Notice that $u^*=-s^*-t^*-q$ and there are only two non vanishing $r_{\alpha\beta}$, 
with value $+2$, i.e.~those taking into account the difference $-u+u^*$. 
The limit of $\mathcal{H}_{AdS_5}$ is then
\begin{align}
\mathcal{H}_{AdS_5}(s,t,\vec{p}\,)\rightarrow\quad &  
\frac{ (2\pi)^3 }{ n^3} \left( \prod_{\alpha<\beta} n^{r_{\alpha\beta}} \right)   
{ e^{-\frac{ p_1+p_2+p_3+p_4}{2} } n^{\frac{ p_1+p_2+p_3+p_4}{2} } }  \times\label{otra_formula_H_with_G_asympt} \\[.2cm]
&
\quad
\oint \left( n^2 ds^* dt^*\right) \prod_{\alpha<\beta}   {c}_{\alpha\beta}^{ { r_{\alpha\beta} }-\frac{1}{2} }(s^*,t^*,d_i)\,\exp[{- n\, {S}_{AdS_5} }]  \notag
\end{align}  
where the effective action reads,
\begin{align}
 {S}_{AdS_5}= -s^*\log(U) - t^*\log(V) - \sum_{1\leq\alpha<\beta\leq 4} {c}_{\alpha\beta}(s^*,t^*,d_i) \log[  {c}_{\alpha\beta}(s^*,t^*,d_i) ]
\end{align}

The saddle point equations are
\begin{align}
\frac{(-s^*)( +\frac{d_1+d_2-d_3-d_4}{2}-s^*) }{ (-u^*)(+\frac{ d_2+d_4-d_1-d_3}{2}-u^*) }= U\quad;\quad
\frac{(-t^*)( +\frac{d_1-d_2-d_3+d_4}{2}-t^*) }{ (-u^*)(+\frac{ d_2+d_4-d_1-d_3}{2}-u^*) } = V
\label{saddle_eq_AdS5}
\end{align}
Recall that the saddle point should be such that the arguments of the $log$ are positive, 
i.e.~the saddle point should stay away from the accumulation of the gamma function poles in the large $p$ limit.

We can rewrite the Effective Action as
\begin{align}
&
{S}_{AdS_5}= \notag\\
&+ s^* \left(\log\!\left[  \frac{(-s^*)( c_s-s^*) }{ (-u^*)(c_u-u^*) } \right] - \log(U) \right) 
					+ t^* \left(  \log\!\left[\frac{(-t^*)( c_t-t^*) }{ (-u^*)(c_u-u^*) } \right] - \log(V) \right)\notag\\
& -\ \ \, \left( {S}^0_{AdS_5}\equiv \sum_{1 \leq \alpha<\beta\leq 4}c_{\alpha\beta}(0,0,d_i) \log[ c_{\alpha\beta}(s^*,t^*,d_i)] \right)
\end{align}

When the external charges are all equal, we can take the square root of 
equations \eqref{saddle_eq_AdS5}. The solution is then 
\begin{align}\label{solu_pppp_AdS5here}
-s_{cl.}= d \frac{   \sqrt{U}  }{ 1+  \sqrt{U} +  \sqrt{V} }\quad;\quad -t_{cl.} = d  \frac{  \sqrt{V}  }{ 1+  \sqrt{U} +  \sqrt{V} }
\end{align}
In order to compute $\mathcal{H}_{AdS_5}(u,v,pppp)$ at the saddle, 
we then need to evaluate both $\mathcal{S}^0_{AdS_5}$ and the 
quadratic fluctuations $\mathcal{S}_{AdS_5}''$.
For the action on-shell  we find
\begin{align}
-n{S}^0_{AdS_5}(u,v,pppp)=2p\log(d+s_{cl}+t_{cl})={+ }2p\log\left(\frac{ d  }{ 1+ \sqrt{U} + \sqrt{V} } \right)
\end{align}
and for the Hessian 
\begin{align}
\det {S}_{AdS_5}''(s^*,t^*,pppp)=
\det \left(\begin{array}{cc} -\frac{2}{s^*} -\frac{2}{u^*} &   -\frac{2}{u^*} \\[.2cm]  -\frac{2}{u^*} & -\frac{2}{t^*} - \frac{2}{u^*}\end{array}\right)= -\frac{4d}{s^* t^* u^* }
\end{align}
We will have to include $\frac{2\pi}{n} |\det \mathcal{S}_{AdS_5}''|^{-\frac{1}{2}}$.

Finally, the saddle point approximation gives
\begin{align}
\!\!\!
\mathcal{H}_{AdS_5}(u,v,pppp)\Big|_{ saddle\, approx} =
 \frac{(2\pi p)^4}{2 p^2}  \frac{e^{-2p} p^{2p} }{\color{blue} (1+\sqrt{U}+\sqrt{V})^{2(p+2) }  } \Bigg[ \frac{d^{3/2}}{ \sqrt{  (d+s_{cl} +t_{cl}) s_{cl} t_{cl}  } }\Bigg]
 \label{saddle_AdS5_pppp}
\end{align}

In the last line we have made crossing symmetry manifest, i.e.~we extracted a term 
which has the right crossing transformation, {\color{blue} in blue}, 
and put the remaining (crossing invariant) terms together.  The crossing transformations are
\begin{align}
\mathcal{H}_{AdS_5}(U,V,pppp)= \frac{1}{V^{p+2} }\mathcal{H}_{AdS_5}\left(\frac{U}{V},\frac{1}{V},pppp\right)= \mathcal{H}_{AdS_5}(V,U,pppp)
\end{align}
and the crossing invariant term is  
\begin{align}\label{gaussian_fluctuation_AdS}
 \frac{d^{3/2}}{ \sqrt{  (d+s_{cl} +t_{cl}) s_{cl} t_{cl}  } }= \frac{ (1+\sqrt{U}+\sqrt{V})^{3/2} }{ (U V)^{1/4} }
\end{align}
Both $U\leftrightarrow V$ and $U\rightarrow U/V$, $V\rightarrow 1/V$, leave \eqref{gaussian_fluctuation_AdS} invariant. 

By writing $\mathcal{H}_{AdS_5}(U,V,pppp)$ in the form \eqref{saddle_AdS5_pppp} 
we should interpret the contribution above as counting for the total gaussian fluctuation 
around the saddle point.  In fact, \eqref{gaussian_fluctuation_AdS} is the combined result 
from the Hessian and $  \prod_{\alpha<\beta} {c}_{\alpha\beta}^{-1/2 }(s_{cl},t_{cl},d)$.

\subsection*{B.2\,\,\,\,Saddle point on $S^5$}

The study of the saddle on the sphere follows similar steps.  Define from $\Gamma_{\otimes}$, 
\begin{align}
\mathcal{H}_{S^5}(\tilde s,\tilde t,\vec{p}\,)=
{  \newsigma^{\tilde s} \newtau^{\tilde t} }\left[ \mathfrak{S}=\pi^2\frac{(-)^{\tilde u}(-)^{\tilde t} }{ \sin(\pi \tilde u) \sin(\pi \tilde t ) }  \right]\prod_{1 \leq \alpha<\beta\leq 4} \frac{ 1}{ 
 \Gamma[1+C_{\alpha\beta}(\tilde s,\tilde t,p_i-2)] } 
 \label{H_S5_Cnm}
\end{align}
which includes the factor $\mathfrak{S}$ and where again we have 
introduce Symanzik variables $C_{\alpha\beta}$ such that $\sum_{\alpha} C_{\alpha\beta}=p_{\beta}$.

In the large $p$ limit take $\{\tilde s,\tilde t,p_{i=1,2,3,4}\}=n\{\tilde s^*,\tilde t^*,d_{i=1,2,3,4}\}$, 
$\tilde u^*=-\tilde s^*-\tilde t^*+d_3$, and let $n\rightarrow \infty$. We find
\begin{align}
\begin{array}{ccc}
C_{\alpha\beta}(\tilde s,\tilde t,p_i-2)\quad &\rightarrow& \quad nC_{\alpha\beta}(\tilde s^*,\tilde t^*,d_i) + R_{\alpha\beta} \\[.2cm]
1/\Gamma[1+C_{\alpha\beta}(s,t,{p}_i+2) ]\quad &\rightarrow &\quad 1/\sqrt{2\pi}\ e^{+n C_{\alpha\beta} } \, (n C_{\alpha\beta} )^{ -n C_{\alpha\beta} - R_{\alpha\beta} -\frac{1}{2}} 
\end{array}
\label{asymptotics_Gammasphere}
\end{align}
where the two non vanishing $R_{\alpha\beta}$ now have values $-2$. 
Notice also that $\frac{(-)^{\tilde u}(-)^{\tilde t} }{ \sin(\pi \tilde u) \sin(\pi \tilde t ) } \rightarrow 4$. 
From the asymptotics 
\begin{align}
\mathcal{H}_{S^5}(\tilde s,\tilde t,\vec{p}\,)\quad\rightarrow\quad &  \frac{ 4\pi^2 }{ (2\pi)^3 n^3}\left( \prod_{\alpha<\beta} n^{-R_{\alpha\beta}} \right)  
{ e^{+\frac{ p_1+p_2+p_3+p_4}{2} } n^{- \frac{ p_1+p_2+p_3+p_4}{2} } }   \label{} \\[.2cm]
&
\quad
\oint \left( n^2 d\tilde s^* d\tilde t^* \right) \prod_{\alpha<\beta}   {C}_{\alpha\beta}^{ {  -R_{\alpha\beta} }-\frac{1}{2} }(\tilde s^*,\tilde t^*,d_i)\,\exp[{- n\, {S}_{S^5} }]  \notag
\end{align}
The effective action reads
\begin{align}
 {S}_{S^5}= -\tilde s^*\log(\newsigma) - \tilde t^*\log(\newtau) + \sum_{1\leq\alpha<\beta\leq 4} {C}_{\alpha\beta}(\tilde s^*,\tilde t^*,d_i) \log[  {C}_{\alpha\beta}(\tilde s^*,\tilde t^*,d_i) ]
\end{align}

The saddle point equations are
\begin{align}
\frac{(+\tilde s^*)(c_s+\tilde s^*) }{ (+\tilde u^*)(c_u+\tilde u^*) }= \newsigma\quad;\quad
\frac{ (+\tilde t^*)( c_t + \tilde t^*) }{ (+\tilde u)(c_u+\tilde u^*)} =\newtau
\end{align}

We can also rewrite the Effective Action as
\begin{align}
&
{S}_{S_5}= \notag\\
&+ \tilde s^* \left( \log\!\left[  \frac{(+\tilde s^*)( c_s+\tilde s^*) }{ (+\tilde u^*)(c_u+\tilde u^*) } \right] - \log(\newsigma) \right) 
					+ t^* \left(  \log\!\left[\frac{(+t^*)( c_t+\tilde t^*) }{ (+\tilde u^*)(c_u+\tilde u^*) } \right] - \log(\newtau) \right)\notag\\
& +\ \ \, \left( {S}^0_{S_5}\equiv \sum_{1 \leq \alpha<\beta\leq 4}C_{\alpha\beta}(0,0,d_i) \log[ C_{\alpha\beta}(\tilde s^*,\tilde t^*,d_i)] \right)
\end{align}

The explicit solution for the case of equal charges is
\begin{align}
\tilde s_{cl.}= d \frac{   \sqrt{\newsigma}  }{ 1+  \sqrt{\newsigma} +  \sqrt{\newtau} }\qquad;\qquad \tilde t_{cl.} = d  \frac{  \sqrt{\newtau}  }{ 1+  \sqrt{\newsigma} +  \sqrt{\newtau} }
\end{align}
The action on shell then reads
\begin{align}
-n{S}^0_{S_5}(\tilde U,\tilde V,pppp)=-2p\log(d-\tilde s_{cl}-\tilde t_{cl})={\color{red} - }2p\log\left(\frac{ d  }{ 1+ \sqrt{\newsigma} + \sqrt{\newtau} } \right)
\end{align}
and for the Hessian we find
\begin{align}
\det {S}_{S^5}''(S^*_{pppp},T^*_{pppp})= \frac{4d}{\tilde s^* \tilde t^* \tilde u^* }
\end{align}
Considering all the relevant contributions (we will have to 
include $\frac{2\pi}{n} |\det \mathcal{S}_{S_5}''|^{-\frac{1}{2}}$) 
the saddle point approximation gives
\begin{align}
&
\mathcal{H}_{S_5}(\newsigma,\newtau,pppp)\Big|_{ saddle\, approx} = 
\frac{p^2}{2} \frac{ e^{+2p} p^{-2p} }{\color{blue} (1+\sqrt{\newsigma}+\sqrt{\newtau})^{2(2-p) }  } \Bigg[ \frac{d^{3/2}}{ \sqrt{  (d-\tilde s_{cl} -\tilde t_{cl}) \tilde s_{cl} \tilde t_{cl}  } }\Bigg]
 \label{saddle_S5_pppp}
\end{align}

As in the case of $AdS_5$, we have written the last expression by 
making manifest crossing symmetry, which for $\mathcal{H}_{S^5}$ means,
\begin{align}
\mathcal{H}_{S^5}(\newsigma,\newtau,pppp)=\newtau^{p-2} \mathcal{H}_{S^5}\left(\frac{\newsigma}{\newtau},\frac{1}{\newtau},pppp\right)= \mathcal{H}_{S^5}(\newtau,\newsigma,pppp)
\end{align}
The blue colored term provides the correct  crossing transformation, 
and the remaining one is crossing invariant. The latter carries the total 
fluctuations of the saddle configuration.

\section{Virasoro-Shapiro}

\subsection{Low energy expansion and Asymptotics} \label{B1}

The Virasoro-Shapiro amplitude reads \cite{Green:2008uj}
\beq
\mathcal{A}_{VS} = \frac{1}{stu}  
\frac{\Gamma (1-\frac{\alpha' s}{4}) \Gamma (1-\frac{\alpha' t}{4}) \Gamma
   (1-\frac{\alpha'  u}{4})}{\Gamma (1+\frac{\alpha'  s}{4}) \Gamma (1+\frac{\alpha'  t}{4})
   \Gamma (1+ \frac{\alpha' u}{4})} \,, \quad;\quad u=-s-t
\eeq
where $s=-(k_1+k_2)^2, t=-(k_1+k_4)^2$ and $u=-(k_1+k_3)^2$. 
At low energy we expand around $s,t,u=0$ to recover the famous infinite series representation 
\beq
\mathcal{A}_{VS} = \frac{1}{stu} 
\exp\left[ \,  \sum_{n\ge 1} \frac{2\zeta_{2n+1}}{2n+1} \left(\frac{\alpha'}{4}\right)^{\!\!2n+1}(s^{2n+1}+t^{2n+1}+u^{2n+1} )  \right] \label{lowE}
\eeq
in terms of the odd zeta functions $\zeta_{3},\zeta_5,\dots$. 

A totally different expansion corner corresponds to the Gross-Mende (GM) regime of large energy at fixed angle corresponding to all Mandelstam large. In this case, we obtain another beautiful infinite series representation
\begin{align}
\mathcal{A}_{GM}&=\exp\left[ -\frac{\alpha'}{4} s \log(-s^2)-\frac{\alpha'}{4} t \log(-t^2) -\frac{\alpha'}{4} u \log(-u^2) \right]\\[.2cm]
\mathcal{A}_{VS} &= 
+\frac{i}{stu}\mathcal{A}_{GM} \left[\sum_{n\ge 1} \frac{B_{2n}}{n(2n-1)}\left( \frac{4}{\alpha'}\right)^{\!2n-1} \left(\frac{1}{s^{2n-1}}+\frac{1}{t^{2n-1}}+\frac{1}{u^{2n-1}}\right)\right] \label{highE} 
\end{align}
which must also be known to the experts (although we could not find it anywhere). Here $B_n$ are the Bernoulli numbers. 
The amplitude $\mathcal{A}_{GM}$ is the famous high energy result of Gross and Mende \cite{Gross:1987kza}. 
Recall that this result can be understood from the worldsheet as a saddle point calculation which localized the
 integration over the string moduli around its extremum value. Of course, the full integral is not just the saddle but is 
 obtained by including all the quantum fluctuations around that classical configuration. That is what comes with the Bernoulli numbers. 

There are probably interesting games one could play about understanding the resurgence properties of these infinite sum representations and how the various low energy zeta function coefficients resum into the high energy moduli space expansion Bernouli numbers and vice-versa. After all, as explained in the text, we end up resumming the low energy expansion~(\ref{lowE}) when coming from SUGRA while we will probably land on the high energy expansion~(\ref{highE}) when arriving from the large classical strings side. As such, understanding the interpolation between these two regimes is key to taming the full interpolation in Figure \ref{regimes} discussed above. 

Amusingly this interpolation is very reminiscent of the dressing phase interpolation in the $N=4$ SYM spin chain. As explained in \cite{Beisert:2006ez,Beisert:2006ib} the weak coupling expansion there is full of Zeta functions while the strong coupling expansion is populated by Bernouli numbers which turn out to be also given by (analytically continued) zeta functions. Mathematically, the reason for this similarity is clear: both Virasoro-Shapiro and the AdS/CFT dressing phase have Gamma functions as their building blocks \cite{Dorey:2007xn}. Would be very nice to find a physical analogy beyond this technical observation

\subsection{More on the large $p$ limit}\label{VS_sugra_string}

Here we give more details about the flat space limit of the VS amplitude in $AdS_5\times S^5$, 
as we mentioned in section \ref{VSlargepequal10d}.
Recall the usual flat space limit formula (in $d$ dimensions) of Penedones~\cite{Penedones:2010ue},
\begin{align}\label{Penedones_here}
\mathcal{A}_{VS}(s,t)= C \lim_{R\rightarrow \infty} \oint \frac{dz}{2\pi i} z^{-(\Sigma-\frac{d}{2} +4) } e^{z}\ \mathcal{M}\left( \frac{R^2}{4z} s, \frac{R^2}{4z} t ,\tilde s, \tilde t\right)
\end{align}
where $C$ is a normalisation. The exponent $\Sigma-\frac{d}{2}+4$ is correct for taking into account the dynamical correlator. 
We use \eqref{Penedones_here} to fix the top degree monomials in $s$ and $t$ in the $AdS_5$ part of the tree level amplitude, order by order in $1/\sqrt{\lambda}$, 
\begin{align}
\mathcal{M}_{\vec{p}}(s,t,\tilde s, \tilde t)\Big|_{genus=0}=\frac{1}{({\bf s}+1)({\bf t}+1)({\bf u}+1)}   +  \sum_{h=0}^{\infty}\left(\frac{1}{\sqrt{\lambda}}\right)^{\!\! h+3} 
\mathcal{V}^{(1,h)}_{\vec{p}}(s,t,\tilde s,\tilde t) 
\end{align}
where $\mathcal{V}^{(1,h)}$ is a polynomial of degree $h$, 
against the the Virasoro Shapiro amplitude.
\begin{align}
\mathcal{A}_{VS}(s,t)
=-\frac{1}{st(s+t)}\left (1+ \sum_{h=0}^{\infty}   \,\left(\frac{\alpha'}{4}\right)^{\!\!h+3} \mathcal{A}_{VS}^{(h+3)}(s,t) \right)
\end{align}
where the $\mathcal{A}^{(h)}_{VS}$ are known. 

Computing the integral in \eqref{Penedones_here} amounts to solve an inverse $\Gamma$ function integral
and results in some Pochhammers, after diving by the normalisation $C= (\frac{R}{2})^6 \Gamma[\Sigma-\tfrac{d}{2}+1 ]$. 
Then, 
\begin{align}\label{term_by_term_norm}
\lim_{s,t,\rightarrow \infty }\left(\frac{1}{\sqrt{\lambda}}\right)^{\!\!h+3}\mathcal{V}^{(1,h)}(s,t,\tilde s, \tilde t)= 
\left(\frac{\alpha'}{R^2} \right)^{\!\! h+3} \frac{ (\Sigma-1)_{h+3} \, \mathcal{A}^{(h+3)}_{VS}({s, t}) }{ { s\, t\, (-s-t)}}
\end{align}
Notice that the r.h.s. will always be polynomial. A well known feature of the Virasoro-Shapiro amplitude. 

Switching to the large $p$ limit we restore the dependence on $\tilde s$ and $\tilde t$ introducing the bold font variables, 
and because $\mathcal{A}_{VS}^{(h)}$ is a homogeneous polynomial, we arrive at 
\begin{align}
\lim_{p\rightarrow \infty}
\mathcal{V}^{(1,h)}(s_{},t_{},\tilde s_{},\tilde t_{}) =  
& 
 \left[ -\frac{\mathcal{A}_{VS}^{(h+3)}(\Sigma { s}_{} ,\Sigma { t}_{})}{{ s}_{}{ t}_{} ({ s}_{}+{ t})}\right]_{\substack{ s\rightarrow\ {\bf s}  \\ t\rightarrow\ {\bf t} \\ u\rightarrow\ {\bf u}}}
\end{align}
where the term in between $[\ldots]$ is supposed to be simplified first, as usual for the flat VS amplitude, and the result, which is now a polynomial of degree $h$, 
covariantised w.r.t.\! the bold font variables of $AdS_5\times S^5$, which are such that ${\bf s}+{\bf t}+{\bf u}=-4$. This is also what 
we recover, independently, by looking at the $\zeta_5$ contributions of the full $AdS_5\times S^5$ amplitude in \eqref{zeta_5}, rewritten under the large $p$ stratification. 

At the saddle point ${\bf s}_{cl}+{\bf t}_{cl}+{\bf u}_{cl}=0$, the series in the large $p$ limit 
can now resummed, in the very same way as for $\mathcal{A}_{VS}$. We obtain the nice result in terms of $\Gamma$ functions
\begin{align}
\!\!\lim_{p\rightarrow \infty} \mathcal{M}(s_{cl},t_{cl},\tilde s_{cl},\tilde t_{cl}) \Big|_{genus\,=\,0} \!&=\! 
\frac{1}{ {\bf s}_{cl} {\bf t}_{cl} {\bf u}_{cl}}  
\frac{ \Gamma[1- \frac{\Sigma}{\sqrt{\lambda}} {\bf s}_{cl} ]\Gamma[1-\frac{\Sigma}{\sqrt{\lambda}} {\bf t}_{cl} ]\Gamma[1-\frac{\Sigma}{\sqrt{\lambda}}{\bf u}_{cl} ] }{
\Gamma[1+\frac{\Sigma}{\sqrt{\lambda}} {\bf s}_{cl} ] \Gamma[1+\frac{\Sigma}{\sqrt{\lambda}}{\bf t}_{cl} ]\Gamma[1+\frac{\Sigma}{\sqrt{\lambda}} {\bf u}_{cl} ]} 
\end{align}
which is quoted in section in \ref{VSlargepequal10d}.

\section{Leading discontinuities from $\widehat{\mathcal{D}}(\vec{p} )$ and $\Delta^{(8)}$}\label{app_delta8}

%
In section \ref{one_loop_sec} we constructed the generic prepotential 
$\mathcal{P}_{p_1p_2p_3p_4}$ by acting on $\mathcal{P}_{2222}$ 
with the operators $\widehat{\cal D}_{\tilde s , \tilde t}(\vec{p}\,)$.  
For completeness we repeat $\mathcal{P}_{2222}$ in our conventions. 
In position space, 
 \begin{align}
 \mathcal{P}_{2222}=& \frac{ \mathcal{P}_{2222}^{2,2^-} }{(x_1-x_2)^{7} } (\Li_1^2(x_1)-\Li_1^2(x_2) ) +   \frac{ \mathcal{P}_{2222}^{1,2^-} }{(x_1-x_2)^{7} } (\Li_2(x_1)-\Li_2(x_2) ) \notag\\
 &\frac{ \mathcal{P}_{2222}^{1^-} }{(x_1-x_2)^{7} } (\Li_1(x_1)-\Li_1(x_2) ) +   \frac{ \mathcal{P}_{2222}^{1^+} }{(x_1-x_2)^{6} } \log(v)  + \frac{ \mathcal{P}_{2222}^{0} }{(x_1-x_2)^{6} } 
 \end{align}
 where the various polynomials $ \mathcal{P}_{2222}^{basis\, element, weight}$ are
\begin{align}
&
\mathcal{P}_{2222}^{2,2^-}= +\tfrac{1}{4}U^4 V^2 ( 1-U-V)\\[.15cm]
&
\mathcal{P}_{2222}^{1,2^-} =-\tfrac{1}{2}  U^4 ((1-V)^2(1+V)+U(1+V^2)) \notag \\[.15cm]
&
\mathcal{P}_{2222}^{1^-} =
-\tfrac{1}{24} U^3( U^4 + (1-V)^4-4U^3 (1+V)+8 U(1-V)^2 (1+V)-2 U^2 (3-8V+3V^2) ) \notag \\[.15cm]
&
\mathcal{P}_{2222}^{1^+}=
+\tfrac{1}{24} U^3(1-V)(U^2+(1-V)^2+10U(1+V) )\notag \\[.15cm]
&
\mathcal{P}_{2222}^{0}=-\tfrac{1}{12}U^4(2U^2-7(1-V)^2-7U(1+V))
\end{align}
This is the prepotential for the amplitude (normalised as in previous section). 

Mellinising the above result we find 
\begin{align}
\!\!\!
\mathcal{P}_{2222}(U,V)&= -\iint ds dt\ (-U)^{s+4}V^t\ \frac{\Gamma[-s]}{\Gamma[s+1]} \Gamma[-t]^2 \Gamma[-{u}]^2 \times \mathcal{N}_{2222}(s,t)
\end{align}
where the amplitude $\mathcal{N}_{2222}(s,t)$ is \eqref{N2222}. 
The result we gave next, in \eqref{Mellin_p1234},
follows from a pencil and paper computation, as we show here below. 

The operators $\widehat{\cal D}_{\tilde s ,\tilde t}(\vec{p}\,)$ in \eqref{dpqrs}-\eqref{mostgeneralDpqrs}, 
when acting on the Mellin integrand of $\mathcal{P}_{2222}$, only act on the monomial $U^{4+s} V^t$, 
returning a combination of Pochhammers in $s$ and $t$, which can be straightforwardly 
turned into $\Gamma$ functions.  Since the denominator of $\widehat{\cal D}_{\tilde s ,\tilde t}(\vec{p}\,)$ 
is immediately  recognisable as the denominator of $\Gamma_{\otimes}$, 
we shall focus on the numerator, by defining 
\begin{align}
\widehat{\cal D}_{\tilde s ,\tilde t}(\vec{p}\,)=\frac{(-)^{c_t+c_s}\widetilde{\cal D}_{\tilde s ,\tilde t}(\vec{p}\,) }{ 
				\Gamma[1+\tilde s] \Gamma[1+\tilde t]\Gamma[1+\tilde u] \Gamma[1+\tilde s + c_s]\Gamma[1+\tilde t + c_t] \Gamma[1+\tilde u + c_u] }
\end{align}
Then, acting on the integrand of $\mathcal{P}_{2222}$ we find.
\begin{align}
&
U^{-4-s} V^{-t} \ \widetilde{\cal D}_{\tilde s,\tilde t}(\vec{p}\,) \left( U^{4+s} V^t\right) = \notag \\[.2cm] 
&
( s+1-\tilde s)_{\tilde s} ( s+1-\tilde s-c_s)_{\tilde s+c_s }( t+1-\tilde t)_{\tilde t} ( t+1-\tilde t+c_t)_{\tilde t+c_t }( s+t+4)_{\tilde u} ( s+t+4)_{\tilde u+c_u } =\notag \\[.2cm]
&
\rule{1.8cm}{0pt}
\frac{ \Gamma[s+1]^2 \Gamma[t+1]^2  \Gamma[s+t+4+\tilde u]\Gamma[s+t+4+\tilde u +c_u] } {\Gamma[s+1-\tilde s]\Gamma[s+1-\tilde s-c_s] \Gamma[t+1-\tilde t]\Gamma[t+1-\tilde t-c_t]\Gamma[s+t+4]^2}
\end{align}
The next step is to put together $\widetilde{\cal D}(\vec{p}\,) \left( U^{4+s} V^t\right)$ with
$\Gamma[-s] \Gamma[-t]^2\Gamma[s+t+4]^2/\Gamma[s+1]$, 
where the latter is part of the integrand of $\mathcal{P}_{2222}$. Let us split the 
computation in the three channels. 

{\bf u-channel}. This is the simplest, 
\begin{align}
\frac{ \Gamma[s+t+4+\tilde u]\Gamma[s+t+4+\tilde u +c_u] }{ 
			\Gamma[s+t+4]^2}  \times \Gamma[s+t+4]^2=  \Gamma[s+t+4+\tilde u]\Gamma[s+t+4+\tilde u +c_u] 
\end{align}
Notice that $s+t+4+\tilde u= (s-\tilde s) +(t-\tilde t) +p_3+2$. 

{\bf t-channel}. In this case we will need a reflection identity, 
\begin{align}
(-)^{c_t} V^{t-\tilde t} \frac{ \Gamma[t+1]^2  }{ \ 
				\Gamma[t+1-\tilde t]\Gamma[t+1-\tilde t-c_t]}  \times \Gamma[-t]^2=  V^{t-\tilde t}\Gamma[\tilde t-t]\Gamma[\tilde t +c_t-t]
\end{align}
where on the r.h.s we used $\Gamma[t-n_i] = (-)^{n_i-1} \Gamma[-t]\Gamma[t+1]/\Gamma[n_i+1-t]$ 
with $n_1=\tilde t -1$ and $n_2=\tilde t+c_t-1$ (assumed to be integers, as it should).

{\bf s-channel}. Similarly to the previous case,
\begin{align}
&
(-)^{c_s} U^{2+s-\tilde s} \frac{ \Gamma[s+1]^2  }{ \ \Gamma[s+1-\tilde s]\Gamma[s+1-\tilde s-c_s]}  \times \frac{  \Gamma[-s]}{\Gamma[s+1]}=  
U^{2+s-\tilde s}\frac{ \Gamma[\tilde s-s]\Gamma[\tilde s +c_s-s] }{\Gamma[-s]\Gamma[s+1]}\notag \\[.2cm]
\end{align}
The gamma function kernel on the r.h.s.~should only have simple poles, 
whose residue would be such that it reproduces that of $\Gamma_{\otimes}$. 
The idea is that $\tilde s$ and $\tilde s+c_s$ are both integers, therefore we 
can turn around the reflection identity, in two way
\begin{align}
\Gamma[-s]\Gamma[s+1]&=(-)^{\tilde s+ c_s-1} \Gamma[\tilde s+c_s-s]\Gamma[s-c_s-\tilde s+1]\\
				&=(-)^{\tilde s-1}\Gamma[\tilde s-s]\Gamma[s-\tilde s+1]
\end{align}
depending on the situation. What fixes this is the leading twist of the correlator $U^{max(0,c_s)}$.
Therefore, if $c_s<0$ we consider the first line, otherwise the second line. 
To consider both cases at once it is useful to introduce
$max(0,c_s)$. 

Let us illustrate the final result in the case $c_s<0$ 
just for concreteness, 
\begin{align}
\mathcal{P}_{p_1p_2p_3p_4}=&(U\newsigma)^2 \sum_{\it T}\int U^{s-\tilde s} V^{t-\tilde t}\  \tilde U^{\tilde s} \tilde V^{\tilde t} \times \mathcal{N}_{2222}(s,t) \times (-)^{s+1} 
\frac{\Gamma[\tilde s-s]  }{ (-)^{\tilde s+c_s-1}\Gamma[s-\tilde s -c_s+1]}
\times \notag\\
&
\quad
\frac{ \Gamma[\tilde t-t]\Gamma[\tilde t -t+c_t]  \Gamma[ s-\tilde s +t-\tilde t +p_3+2]\Gamma[ s-\tilde s +t-\tilde t +p_3+2+c_u] }{  \Gamma[1+\tilde s] \Gamma[1+\tilde t]\Gamma[1+\tilde u] \Gamma[1+\tilde s + c_s]\Gamma[1+\tilde t + c_t] \Gamma[1+\tilde u + c_u] } 
\end{align}
Finally, by changing variables $s\rightarrow s+\tilde s$ and $t\rightarrow t+\tilde t$ and defining $u=-s-t-p_3-2$, we get
\begin{align}
\mathcal{P}_{p_1p_2p_3p_4}=(-)^{c_s}(U\newsigma)^2 &\sum_{\it T}\int (-U)^{s} V^{t}\  \tilde U^{\tilde s} \tilde V^{\tilde t} \times \mathcal{N}_{2222}({\bf s},{\bf t})
\frac{\Gamma[-s]  }{ \Gamma[s -c_s+1]}
\times \notag\\
&
\frac{ \Gamma[-t]\Gamma[-t+c_t]  \Gamma[ -u]\Gamma[ -u+c_u] }{  \Gamma[1+\tilde s] \Gamma[1+\tilde t]\Gamma[1+\tilde u] \Gamma[1+\tilde s + c_s]\Gamma[1+\tilde t + c_t] \Gamma[1+\tilde u + c_u] } 
\end{align}
Notice, $\mathcal{N}_{2222}$ is evaluated in ${\bf s}=s+\tilde s$ and ${\bf t}=t+\tilde s$ 
and  $\Gamma_{\otimes}$ is essentially manifest,  we still have simple poles in $s$ 
but the residue is the one $\Gamma_{\otimes}$ would induce.

Let us move now to $\Delta^{(8)}$.
The eight-order differential operator $\Delta^{(8)}$ is best written in the form  \cite{Caron-Huot:2018kta}
\begin{align} \label{delta8}
\delta^{(8)}_{[p_{21},p_{43} ]} &=  \frac{x_1 { x_2} y_1 { y_2 }}{(x_1-{x_2 })(y_1-{ y_2})}
\prod_{i,j=1}^{2} \left(\mathbf{C}_{x_i}^{[+\frac{p_{21}}{2} ,+\frac{p_{43}}{2},0]} -\mathbf{C}_{y_j}^{[-\frac{p_{21}}{2} ,-\frac{p_{43}}{2},0]}\right) 
\frac{(x_1-{ x_2})(y_1-{ y_2})} {x_1 { x_2} y_1 { y_2}}  \notag \\
\Delta^{(8)}_{[p_{21},p_{43}]}&=   
\left( \frac{U}{\newsigma}\right)^{ -\frac{p_4+p_3}{2} }\  \ \delta^{(8)}_{[p_{21},p_{43} ]} \ \  \left( \frac{U}{\newsigma}\right)^{ \frac{p_4+p_3}{2} }
\end{align}
where $\mathbf{C}_{z}^{[\alpha,\beta,\gamma]}$ is the elementary $2d$ casimir
$\mathbf{C}_{z}^{[\alpha,\beta,\gamma]}= z^2(1-z)\partial_{z}^2 +z(\gamma-(1+\alpha+\beta)z)\partial_{z} - \alpha\beta z$.
$\Delta^{(8)}$ depends non trivially on the external charges. It is invariant 
under the `flip' symmetry $x_i\rightarrow x_i/(x_i-1)$ and $y_i \rightarrow y_i/(y_i-1)$
which is a symmetry of the correlator when the external charges satisfy 
$p_{21}=0 || p_{43}=0$.

Continuing with the amplitude we find  
\begin{align}
\mathcal{A}_{p_1p_2p_3p_4}={U^{-2} \tilde U^{-2} }\Delta^{(8)}  \mathcal{P}_{p_1p_2p_3p_4} 
\end{align}
The action of $\Delta^{(8)}$ is localised on $U^{2+s} V^t \tilde U^{\tilde s+2} \tilde V^{\tilde t}$, 
which appear in the Mellin representation of $\mathcal{P}_{p_1p_2p_3p_4}$. It returns
\begin{align}
&
{U^{-2} \tilde U^{-2} }\Delta^{(8)}\left[ U^{2+s} V^t \tilde U^{\tilde s+2} \tilde V^{\tilde t}\right]= \notag\\
&
U^{s }{ V }^{t } \newsigma^{\tilde s} \newtau^{\tilde t} 
\left[  \sum_{0 \leq i+j \leq 4} \left( \tfrac12 p_{21}\right)^i\left( \tfrac12 p_{34} \right)^j  \Omega_{ij}^{}( U,V,\newsigma,\newtau, s, t, \tilde s, \tilde t, p_3) \right]
\end{align}
where in the second line we also used $\frac{p_4+p_3}{2}=-\frac{1}{2}p_{34}+p_3$.
The $\Omega_{ij}^{}$ can be expanded as 
\begin{align}
\Omega_{ij}^{}( U,V,\newsigma,\newtau, s, t, \tilde s, \tilde t, p_3)= 
 \sum_{ \substack{ 0\leq m+n\leq 4 \\[.1cm] 0\leq \tilde m+\tilde n\leq 4} } A_{ij}[m,n,\tilde m,\tilde n](s,t,\tilde s,\tilde t,p_3) U^m V^{n-2} \tilde U^{\tilde n} \tilde V^{\tilde m-2}
\end{align}
and the $A_{ij}[m,n,\tilde m,\tilde n](s,t,\tilde s,\tilde t,p_3)$ are attached in 
an ancillary. Notice the overall $V^{-2} \tilde V^{-2}$. 
For each $i,j$, i.e.~powers of $p_{21}$ and $p_{34}$,  there are 225 terms. 
This means that when we Mellinise a given $\Omega_{ij}^{}$ we find a shift-operator with 225 terms. 
These shifts act on the Mellin variables ${\bf s}$ and ${\bf t}$, i.e. those in $ \mathcal{N}_{2222}({\bf s},{\bf t})$, as we showed above. 
However, there are only $25$ different shifts and these are the ones we collect.
Out of an $\Omega_{ij}(U,\ldots ,p_3)$ we obtain a shift operator
\begin{align}
\widehat{\Omega}_{ij}(a,b)\qquad;\qquad -4 \leq a \leq 0 \quad;\quad -3-a \leq b\leq 3
\end{align}
where $a$ and $b$ label the shifts. 

There is something interesting to point out about the $\widehat{\Omega}_{ij}(a,b)$. 
The coefficients of these shift-operators are rational functions, since they come 
directly from rearranging ``$\Gamma_{\otimes}$", which has numerator and denominator.
However, when we sum over $i,j$ we find polynomials. In section 
\ref{big_mellinoneloop_section} we used, 
\begin{align}
{\delta}_{p_1p_2p_3p_4}(a,b)=\sum_{i,j} \widehat{\Omega}_{ij}(a,b) 
\end{align}
which are indeed polynomials in $s,t,\tilde s, \tilde t, p_{i=1,2,3,4}$, 
and are attached in an ancillary file for illustration.\footnote{In the ancillary file we 
use $S$ and $T$ instead of $\tilde s$ and $\tilde t$, just for convenience.} 

The relation between $\Omega^{(8)}(U,\ldots p_3)$ and $\Omega^{(8)}(U,\ldots ,p_3)$ 
evaluated at $\{U,\tilde U,V,\tilde V\} \rightarrow$ $\{$\,$U/V$, $\tilde U/\tilde V$, $1/V$, $1/\tilde V\}$, 
can be seen as follows, 
\begin{align}
\!\!\!\!\!A_{0j}(m,n,\tilde m,\tilde n,s,t,\tilde s,\tilde t,p_3)=+A_{0j}(m,-n-m+4,\tilde m,-\tilde n-\tilde m+4,s,u,\tilde s,\tilde u,p_3)
\end{align}
and also 
\begin{align}
\!\!A_{ij\neq0}(m,n,\tilde m,\tilde n,s,t,\tilde s,\tilde t,p_3)=-A_{ij\neq0}(m,-n-m+4,\tilde m,-\tilde n-\tilde m+4,s,u,\tilde s,\tilde u,p_3)
\end{align}

\subsection{Leading log at two-loop}\label{two_loop_exercise}

The scaling with $p$ of the Mellin amplitude, generalised to the $\ell$-loop,
gives the following scheme:
\begin{align}
&
 \lim_{p\rightarrow \infty} \mathcal{M}_{\vec{p}}^{\ell-loop} ( s_{cl} ,t_{cl}, \tilde s_{cl},\tilde t_{cl}) =  
 \frac{1}{ {\bf s}_{cl} {\bf t}_{cl} {\bf u}_{cl}  }
 \times  
 \left( \frac{\Sigma^4 {\bf s}_{cl}^4 }{N^2} \right)^{\!\!\ell}  
 \Bigg[  \notag\\
&
\rule{.5cm}{0pt}
 \sum_{0\leq d\leq 2\ell} \texttt{lim-Transcendetal}_{\ell,d}({\bf s}_{cl} ,{\bf t}_{cl} ) \times 
 					\texttt{Rational}_{\ell,d}\left( \frac{ {\bf t}_{cl} }{ {\bf s}_{cl} }\right)\, +\, {\rm crossing} \Bigg]
\end{align}
as we explained around \eqref{scheme_all_loops}.

We can test our scheme at two-loops, for the top-weigh function $\texttt{Rational}_{2,6}$, 
since this is given only by the triple discontinuity, which we know \cite{Aprile:2018efk,Caron-Huot:2018kta}. 
In fact, we can also illustrate with a short computation how to organise the 
top-weight construction of the full two-loop amplitude.  At two-loops in position space, 
the top-weight term is accompanied by a weight six function, and there are six independent 
such pure functions  \cite{Drummond:2012bg}, i.e.~no symmetry. None of them has 
the $\log^3 u\log^3 v$ contribution, rather we find at most a $\log^3 u \log^2 v f_1(u,v)$ 
in one orientation. We pick this one, and even though we don't know all the details of 
the full amplitude, we can just focus on the Mellin transform of its coefficient function, 
which is a rational function and will give us $\texttt{Rational}_{2,6}$ in the end. 
We Mellinise that coefficient function for the stress-energy tensor correlator and 
we obtain $\mathcal{T}^{(6)}(s,t)$. Then, we adapt the same procedure that led us 
from $\mathcal{P}_{2222}$ to $\mathcal{P}_{p_1p_2p_3p_4}$ at one-loop. Thus
we obtain a Mellin amplitude which includes $\Gamma_{\otimes}$ with two $sin$ flips, 
because we are studying a rational function in position space, and $\mathcal{T}^{(6)}({\bf s},{\bf t})$.
In sum, 
\begin{gather}
\mathcal{T}^{(6)}(s,t)=-\frac{1}{21600}\frac{120 +99s + 21 s^2 + 29 t + 12 s t + t^2 }{(s+1)(s+2)(t+1)(t+2)(t+3)}\qquad\quad \\[.2cm]
\texttt{Rational}_{2,6}={\bf s}_{cl}{\bf t}_{cl} {\bf u}_{cl} \lim_{p\rightarrow\infty}\mathcal{T}^{(6)}({\bf s}_{cl} ,{\bf t}_{cl} ) 
				= -\frac{21 ({\bf s}_{cl} )^2 {\bf u}_{cl} +({\bf t}_{cl})^2 {\bf u}_{cl}+ 12 {\bf s}_{cl} {\bf t}_{cl} {\bf u}_{cl}  }{ 21600({\bf s}_{cl}) ({\bf t}_{cl})^2}
				\label{something_2loops}
\end{gather}
Indeed we find that $\texttt{Rational}_{2,6}$ has degree zero under the large $p$ limit, 
and we expect the action of $\Delta^{(8)}\Delta^{(8)}$ to bring $\Sigma^8 {\bf s}_{cl}^8$. 
The matching with the ten-dimensional double boxes at top-weight starts from 
\eqref{something_2loops} and the Mellin transform of the weight six function  \cite{Drummond:2012bg}, 
which we postpone to a future work. Our strategy complements the infinite sum discussion of~\cite{Bissi:2020wtv}.

\end{document}